\documentstyle[12pt,aasms4]{article}

\def\Lymana{Lyman-$\alpha$}

\def\Lya{Ly$\alpha$}
\def\Lyb{Ly$\beta$}

\def\Lyd{Ly$\delta$}

\def\kms{km s$^{-1}$}
\def\kmsmpc{km s$^{-1}$ Mpc$^{-1}$}
\def\h0{${h_{0}^{-1}}$}
\def\h100{${h_{100}^{-1}}$}
\def\h75{${h_{75}^{-1}}$}
\def\H0{$H_{0}$}

\def\gamres{$\Gamma_{res}$}
\def\nhi{N$_{HI}$}

\newcounter{ntab}
\newcounter{nfig}
\slugcomment{Submitted to the Astrophysical Journal}

\lefthead{Impey et al.}
\righthead{A Study of Lyman-Alpha Quasar Absorbers in the Nearby Universe}

\begin{document}

\title{A STUDY OF LYMAN-ALPHA QUASAR ABSORBERS \\
  IN THE NEARBY UNIVERSE\footnote{Based on Observations made 
  with the NASA/ESA Hubble Space Telescope, obtained at the Space Telescope 
  Science Institute, which is operated by AURA, Inc., under NASA contract 
  NAS 5-26555. } }

\vskip 2truecm

\author{C. D. Impey}
\affil{Steward Observatory, University of Arizona, Tucson, AZ 85721}
\affil{email: cimpey@as.arizona.edu}

\author{C. E. Petry}
\affil{Steward Observatory, University of Arizona, Tucson, AZ 85721}
\affil{email: cpetry@as.arizona.edu}

\and

\author{K. P. Flint}
\affil{Department of Astronomy and Astrophysics, University of California, 
  Santa Cruz, CA 95064}
\affil{email: flint@ucolick.org}

\begin{abstract}
Spectroscopy of ten quasars obtained with the Goddard High Resolution 
Spectrograph (GHRS) of the Hubble Space Telescope (HST) is presented.
We detect 357 absorption lines above a significance level of $3\sigma$ 
in the ten sightlines, and 272 lines above a significance level of 
$4.5\sigma$. Automated software is used to detect and identify the
lines, almost all of which are unresolved at the GHRS G140L resolution 
of 200 kms$^{-1}$. After identifying galactic lines, intervening metal 
lines, and higher order Lyman lines, we are left with 139 \Lya\
absorbers in the redshift range $0 < z < 0.22$ (lines within 900 km
s$^{-1}$ of geocoronal \Lya\ are not selected). These diffuse
hydrogen absorbers have column densities that are mostly in the range
10$^{13}$ to 10$^{15}$ cm$^{-2}$ for an assumed Doppler parameter of
30 kms$^{-1}$. The number density of lines above a rest equivalent 
width of 0.24 \AA, $dN/dz = 38.3 \pm 5.3$, agrees well with the the 
measurement from the Quasar Absorption Line Key Project. There is
marginal evidence for cosmic variance in the number of absorbers
detected among the ten sightlines. A clustering analysis reveals an 
excess of nearest neighbor line pairs on velocity scales of 250-750 
km s$^{-1}$ at a 95-98\% confidence level. The hypothesis that the 
absorbers are randomly distributed in velocity space can be ruled 
out at the 99.8\% confidence level. No two-point correlation power 
is detected ($\xi < 1$ with 95\% confidence). \Lya\ absorbers 
have correlation amplitudes on scales of 250-500 kms$^{-1}$ at least 
4-5 times smaller than the correlation amplitude of bright galaxies.
A detailed comparison between absorbers in nearby galaxies is carried
out on a limited subset of 11 \Lya\ absorbers where the galaxy
sample in a large contiguous volume is complete to $M_B = -16$. 
Absorbers lie preferentially in regions of intermediate galaxy 
density but it is often not possible to uniquely assign a galaxy
counterpart to an absorber. This sample provides no explicit support
for the hypothesis that absorbers are preferentially associated with
the halos of luminous galaxies. We have made a preliminary comparison
of the absorption line properties and environments with the results
of hydrodynamic simulations. The results suggest that the
\Lya\ absorbers represent diffuse or shocked gas in the IGM
that traces the cosmic web of large scale structure.

\end{abstract}


\section{INTRODUCTION}

The systematic study of quasar absorption is a powerful cosmological
tool.  Given a bright enough illuminating source and a combination of
observations from the ground and space, the properties of the
absorbers can be studied over 90-95\% of the Hubble time. Sharp
intervening absorption features are used to locate cold, diffuse and
dark components of the universe --- the traditional view is that
\ion{C}{4} and \ion{Mg}{2} doublets are tracers of the halos of
luminous galaxies (\cite{wey79}) and \Lya\ lines are tracers
of intergalactic hydrogen (\cite{sar80}). Recent work has blurred the
distinction between the types of absorbers, and has given us a much
more sophisticated and complex view of the intergalactic medium.  The
rapid evolution in the subject over a ten year span is amply conveyed
by the contents of the two conference proceedings edited by Blades,
Turnshek \& Norman (1988) and Petitjean \& Charlot (1997). 

The study of quasar absorbers is an important complement to galaxy surveys 
which catalog the luminous content of the universe. For suitable background 
sources, quasar absorbers can be detected over the range $0 < z < 5$ with an 
efficiency that is almost independent of redshift. Galaxy surveys are 
inevitably affected by Malmquist bias, surface brightness selection effects, 
cosmological dimming, and k-corrections. On the other hand, absorbers can 
only be surveyed along lines of sight with a suitable quasar, so most measures
of large scale structure must use the one dimensional redshift distribution
of absorbers.

High redshift quasars show a dense ``forest'' of \Lya\ absorption
lines, first recognized to be discrete intervening absorbers by Lynds (1972).
The observational situation at $z \ga 2$ has been transformed by the high
resolution and sensitivity of the HIRES spectrograph on the Keck telescope.
Since the distribution of H I column density is a power law, the demarcation
of the \Lya\  forest is somewhat arbitrary --- we adopt $N_{HI} < 
10^{17}$ cm$^{-2}$, where the absorbers are optically thin in the Lyman
continuum. Surveys for the C IV doublet show that the metallicity of the 
hydrogen absorbers is a few percent of solar from 10$^{17}$ cm$^{-2}$ down 
to 10$^{14}$ cm$^{-2}$ (\cite{cow95}; \cite{tyt95}), but the metal
abundance drops sharply by an order of magnitude below 10$^{14}$ cm$^{-2}$
(\cite{lul98}). The clustering properties also depend on column density.
A two-point velocity correlation is detectable above 10$^{14}$ cm$^{-2}$
(although well below the level of galaxy-galaxy correlations) and is much 
weaker or absent at lower column densities (\cite{cri97}).

Both of these observations can be understood in the context of cosmological
simulations that incorporate gas dynamics. These supercomputer simulations
show that the \Lya\  absorbers trace a filamentary network of
highly ionized gas (\cite{cen94}; \cite{her96}). At $z > 2$, 
a majority of the baryons in the universe are contained in the absorbers
of the \Lya\  forest (\cite{mir96}). At column
densities above 10$^{15}$ cm$^{-2}$, the absorbers are roughly spherical and 
trace the skeleton of the large scale structure defined by collapsed 
objects. At column densities below 10$^{13}$ cm$^{-2}$, the absorbers 
are underdense and form a web of filaments and sheets
(\cite{cen97}). A column density of 10$^{14}$ cm$^{-2}$ corresponds
approximately to the transition between these two regimes. 

The insights from simulations affect the interpretation of quasar spectra. 
It is clear that the idea of a spherical cloud or even a ``characteristic'' 
size is naive --- the absorbers trace a complex topology. The low column 
density absorbers are particularly interesting for cosmological studies, 
because they accurately trace the underlying dark matter potential and 
may be primitive enough to retain a memory of initial conditions, in 
contrast to highly non-linear objects like galaxies. Croft et al. 
(1998) have shown that the shape and amplitude of the power spectrum 
of mass fluctuations can be recovered directly from observations of the 
\Lya\ forest (see also \cite{gne96}; \cite{bih97}).

The nature of the hydrogen absorbers at low redshift is not clear. At $z 
< 1.6$, the \Lya\  line shifts below the atmospheric cutoff and
quasar spectra can only be obtained with the relatively modest aperture 
of the Hubble Space Telescope. Also, the number density of absorbers 
drops rapidly with redshift so the line samples are relatively small at 
low redshift. The evolution with redshift shows an inflection at $z 
\sim 1.5$; data from the HST Absorption Line Key Project show strong 
evolution at high 
redshift and much weaker evolution for the 2/3 of a Hubble time since 
$z = 1.5$ (Jannuzi 1997). In detail, there is differential evolution at 
low redshift --- strong lines evolve, and lines near a rest equivalent
width of 0.24 \AA\ show no evolution (Dobrzycki \& Bechtold 1997).

The strong lines at low redshift ($z<1.3$) appear clustered in velocity space 
with an amplitude similar to that of galaxy-galaxy correlations (Ulmer 1996). 
Additional evidence for clustering comes from the HST Key Project, where 
\Lya\ absorbers are clumped around metal line systems (Bahcall et al.
1996; Jannuzi 1997). Nothing is known about the clustering of the unevolving
weak lines, but a few high sensitivity spectra show that there are a large 
number of lines below 0.24 \AA\ (which corresponds to a column density of
10$^{14}$ cm$^{-2}$ for a Doppler parameter of 30 \kms). At $z \sim 
0$ in the local universe, the number density rises from $dN/dz \approx 20$ 
above 10$^{14}$ cm$^{-2}$ to $dN/dz \approx 250$ above 10$^{12.6}$ cm$^{-2}$
(Shull 1997).

Low redshift absorbers offer the great advantage that galaxy counterparts
can be detected directly. If a single galaxy is responsible, the most 
plausible counterpart is a luminous galaxy with a small impact parameter 
to the line of sight and a small velocity separation from the absorber 
(Lanzetta et al. 1995; Chen et al. 1998, CLWB hereafter). However, it is 
difficult to identify a unique counterpart since galaxies cluster in space 
and there are many faint galaxies for each luminous one. There is an 
ambiguity between a luminous galaxy and an invisible dwarf at a smaller
impact parameter (Linder 1998). Moreover, the velocity resolution of 
most published HST spectroscopy is only 200-300 \kms, leading to an 
ambiguity between an absorber that samples the velocity dispersion of a
halo or the rotation of a massive disk, and an absorber that is part 
of a quiescent structure like a loose group of galaxies. This issue is
highlighted by the study of quasar pairs, which show common \Lya\ 
absorption at intermediate redshift ($0.5 < z < 0.9$) with zero velocity
difference on transverse scales far larger than a galaxy halo (Dinshaw et 
al. 1995). In addition to looking for a single counterpart, pencil-beam 
redshift surveys are used to statistically relate the one-dimensional 
absorber distribution to the three-dimensional galaxy distribution.
 
Morris et al. (1993) made the first detailed study of \Lya\  
absorbers and galaxies along the single line of sight toward 3C~273. 
They concluded that the absorbers were more clustered than a random
population but less clustered than galaxies were with each other. 
Different studies have disagreed on the strength of the relationship
that would point to bright galaxy counterparts --- an anticorrelation
between \Lya\  equivalent width and galaxy impact parameter
(Lanzetta et al. 1995; Le Brun, Bergeron, \& Boiss\'{e} 1996; Bowen, 
Blades \& Pettini 1996). Using HST data sensitive to column densities 
above 10$^{14}$ cm$^{-2}$, these authors find that the fraction of 
absorbers associated with galaxies (either within a halo or in a 
correlated structure) is $f_{\rm gal}$ = 0.3-0.7. The story is quite 
different when using HST data that is sensitive to lines of lower column 
density. At $z < 0.1$, where galaxy surveys are sensitive and relatively 
complete, the fraction of weak absorbers that are associated with galaxies 
is $f_{\rm gal}$ = 0-0.2 (Mo \& Morris 1994; Shull, Stocke, \& Penton 1996; 
Grogin \& Geller 1998). Low column density absorbers appear to be 
unclustered and uncorrelated with galaxies.

Many questions about the low redshift \Lya\  absorbers remain
unanswered. Are they kinematically linked to galaxies or are they merely 
tracers of large, unrelaxed structures? Is there a sharp transition 
in properties such as metallicity and ionization at a column density of
10$^{14}$ cm$^{-2}$? How are they related to the rapidly evolving 
population of absorbers at higher redshift? Some insights have been provided
by the first hydrodynamic simulations to predict absorber properties at
$z=0$.  
For example, \cite{dav98p} find that the \Lya\ forest arises primarily 
from shock-heated gas associated with the large scale structures surrounding
the galaxies.
The evolution of the absorber is governed by the trade-off between the 
declining recombination rate due to the expansion of the universe and the
photoionization rate, which declines sharply due to the fading ultraviolet
background at $z<2$ (see also Riediger, Petitjean \& M$\ddot{{\rm u}}$cket 
1998; Theuns, Leonard, \&  Efstathiou 1998).
Absorbers with column densities above 10$^{14}$ cm$^{-2}$ may 
sample a population of absorbers that is rapidly evolving as the gas
drains onto galaxies and filaments. At low redshifts, the residue 
of this gas would display much of the clustering power of galaxies. 
Lower column density absorbers may sample gas in void regions, and 
consequently these slowly evolving absorbers would be less chemically
enriched and less clustered. 
  
This paper presents new observations of \Lya\  absorbers at low
redshift ($z \lesssim 0.2$). The approach is to use multiple lines of sight 
in a single region of sky to thread a large, contiguous volume. In this 
way, absorbers can be compared with individual galaxies down to a low
luminosity limit.  The target area is the Virgo region, chosen because it
contains a significant number of background probes and because the galaxy
distribution is reasonably well sampled --- in addition to the Virgo 
cluster and the southern extension of the Coma cluster at $z = 0.02$, 
there is a sheet of galaxies at least 150 Mpc in extent at $z = 0.08$ 
(Flint \& Impey 1996). We have used the Goddard High Resolution Spectrograph
(GHRS) to detect 139 \Lya\ absorbers in the redshift range $0.003 < z < 
0.225$. The total volume threaded by the 10 pencil beams is 3 $\times 
10^6$ Mpc$^3$. Several other studies have presented low redshift \Lya\ 
absorbers along widely separated, and therefore unrelated, sightlines. 
The primary comparison sample comes from 
the HST  Key Project (Bahcall et al. 1996; Jannuzi et al. 
1998). There have been several other studies of multiple sightlines 
(Stocke et al. 1995; Shull, Stocke \& Penton 1996; Grogin \& Geller 
1998) and a couple of sensitive surveys of individual sightlines 
(Morris et al. 1993; Tripp, Lu, \& Savage 1998). 

The goal of this paper is to present the new sample of low redshift
\Lya\ absorbers, summarize their statistical properties, and 
relate them to the individual galaxies. A deeper spectroscopic
survey is underway to measure galaxy redshift in cones around each of 
the Virgo sightlines. Another eventual goal is to compare the spatial
distribution of absorbers and galaxies to the results of hydrodynamic
simulations of the local universe. In \S 2, we discuss the new HST
observations and data reduction procedures. The line selection and
identification process is discussed in \S 3. Following that we describe 
in \S 4 the statistical properties of the \Lya\ absorbers and 
compare the data to other published samples. In \S 5 we relate the
absorbers to the luminous matter distribution defined by galaxies.
The paper ends with a brief discussion of the nature of the low redshift 
hydrogen absorbers.

\section{OBSERVATIONS}

\subsection{Target Selection}

Most of what we know about \Lya\ absorbers at $z < 2$ comes from 
studies of single lines of sight. This information can be combined to 
produce absorber samples with great statistical power; this approach is
exemplified by the HST Key Project (Jannuzi et al. 1998, and references 
therein). The transverse scale of the absorbers can be measured by looking 
for common absorption along adjacent lines of sight. Experiments using
gravitational lenses and quasar pairs probe scales from 100 pc up to 1 
Mpc (e.g. Weymann \& Foltz 1983; Fang et al. 1996; Petry, Impey, \& Foltz 
1998; Dinshaw et al. 1998). However, the low optical depth to lensing and 
the low surface density of bright quasars mean that these asterisms are 
rare. The connection between galaxies and absorbers can be established with
galaxy redshift surveys along individual lines of sight. But the field of view 
of multi-object spectrographs is too small to relate absorbers to large scale
structure in this way.

We favored a hybrid strategy in this study of \Lya\ absorbers at 
very low redshift, $z \lesssim 0.2$. The well-sampled galaxy distribution 
in the direction of Virgo provides an excellent opportunity to study the 
relationship of \Lya\ absorbers not only to individual bright galaxies but 
also to the large scale structure traced by those galaxies. The Virgo region 
is covered by the Large Bright Quasar Survey, ensuring a suitable grid of 
probes (\cite{hew95}). Target quasars were selected by their location  
on the sky and in redshift, by their estimated 1300 \AA\ flux,
and by the number of galaxies detected along the line of sight.  

Target quasars were chosen to span a large region centered on the Virgo 
cluster on the sky ($\alpha$: 12$^{\rm h}$ to 13$^{\rm h}$; $\delta:
-5\arcdeg\  \rm to +20\arcdeg$). We adopted a lower redshift bound of 
$z = 0.1$ for the target quasar, 
to give a pathlength of $\Delta z > 0.1$ for galaxy-absorber comparison, and 
an upper redshift bound of $z = 0.9$, to avoid the likelihood of Lyman limit
absorption in quasars without previous ultraviolet photometry or spectroscopy.
The final target list contains one exception of a quasar at $z = 0.08$ whose 
ultraviolet brightness offset the increased amount of time required to 
detect the necessary number of \Lya\ lines in the smaller redshift
pathlength. Quasars were selected from the  V\'eron catalog (\cite{ver93}),
with preference given to targets with at least 40 galaxies 
($N_{\rm gal} \ge 40$) within a radius of 2 degrees out to $z \sim 0.1$.
This generous criterion chose lines of sight that had a minimum sample of 
detected galaxies (i.e. with or without a measured redshift) 
within a large volume around each line of sight, taking 
galaxies from an early version of the CfA Redshift Catalog (ZCAT)
ca. 1994 (\cite{zcat}). $N_{\rm gal} \ge 40$ was a conservative limit
to ensure data existed in the literature, and in fact, using our
current galaxy sample from the Virgo region (a combination of ZCAT
version November 1998, and NED\footnote{The NASA/IPAC Extragalactic Database 
(NED) is operated by the Jet Propulsion Laboratory, California Institute 
of Technology, under contract with the National Aeronautics and Space
Administration.}, defined in \S 5.1), only 12/100 randomly generated
lines of sight within our overall region would have $N_{\rm gal} < 40$ 
out to $z \sim 0.1$. The two exceptions are quasars in the direction of 
the southern extension of the Virgo cluster, where $N_{\rm gal} \ge 20$. 

The candidate target list was further refined to exclude radio loud quasars 
which could prove to be variable, as well as those quasars expected to have
low ultraviolet flux. Nine of the ten remaining quasars had no UV observations
in the literature. The tenth object, PKS~1217+1804, had been observed with 
IUE (\cite{lan93}). Eight of the nine objects were observed optically with
the Multiple Mirror Telescope in March and April of 1995 to measure an
individual spectral index for each object, which was subsequently used 
to extrapolate to a 1300 \AA\ flux ($S_\nu \propto \nu^{\alpha}$). The
remaining unobserved object, Q1214+1804, is an optically selected quasar 
and had a high probability of having a reliable extrapolated flux calculated
from an average of our optical spectral indices ($\alpha = -0.71$). We 
required the quasars to have an expected 1300 \AA\ flux greater than 
$5.0\times10^{-15}$ erg s$^{-1}$ cm$^{-2}$ \AA$^{-1}$,\ which was found 
to be the minimum flux needed to achieve the prescribed data quality. With
UV flux level as an overall limitation, the HST observations were planned
to yield a significant number of \Lya\ absorbers along each line of sight. 
The expected number of \Lya\ lines, $N_{\rm exp}$, was evaluated
using the absorber density relation from the maximum likelihood model of
\cite{bah93p}, assuming a  $4.5 \sigma$ limiting equivalent width and using 
a SNR calculated from the 1300 \AA\ flux. Integration times were adjusted to 
maximize observing efficiency versus $N_{\rm exp}$, yielding an average 
$N_{exp}$ of 4 for $\Delta z = 0.1$, and an average $N_{exp}$ of 9 over 
the whole accessible range $0 < z < 0.22$. 
The actual yield was an average of $\sim 10$ $4.5\sigma$ lines per quasar 
over the 10 lines of sight.

Information on the final list of 10 target quasars and details of the
observations is summarized in Table \arabic{ntab}. 
The SNR of the GHRS spectra agree well with the
predictions, except in two cases, Mark~1320 and Q~1228+1116, which have very 
low SNR --- either the 1300 \AA\ flux for these objects was underestimated 
or they are variable sources. These two objects are not included in the
analysis.  A search of the HST Archive yielded two additional 
targets, 3C~273 and J~1230.8+0115, which were observed using the same
instrumental configuration. The SNR for these two targets is higher 
than for the other 10 objects
and so they are included in our analysis to enhance the statistics. The 
details of these observations are also included in Table \arabic{ntab}.

\subsection{Observations}

Spectroscopy of 10/12 quasars listed in Table \arabic{ntab} was
obtained with the Hubble Space Telescope GHRS (post-COSTAR)
using the Side 1 digicon 
detector with the Large Science Aperture (LSA) and the G140L grating (see 
Table \arabic{ntab} for the observational details). This configuration
yields a wavelength coverage of 1200--1480 \AA, which is sensitive to
\Lya\ absorption from $z = 0$ to $z = 0.22$. Because the Side 1
acquisition mirror of the GHRS only reflects far-ultraviolet light,
the targets were too faint to accumulate enough counts over the maximum
acquisition integration time, and so the Faint Object Spectrograph
(FOS) blue-side mirror was used to acquire the objects. Acquisitions
were made with the 4.3\arcsec\ FOS aperture then followed with a blind
offset to the GHRS 2\arcsec\ LSA for observations. Such FOS-assisted
GHRS acquisitions have a pointing uncertainty of 0\farcs1 (Leitherer
et al. 1994).

The G140L grating produces a dispersion of 0.57 \AA\ $\rm diode^{-1}$
and the instrumental FWHM (\gamres) is 1.40 diodes (GHRS Instrument 
Handbook v6.0). To obtain
full Nyquist sampling, the observations are substepped into quarter-diode 
steps, providing 4 pixels per diode and thus a dispersion of $\sim 0.143$ 
\AA\ pixel$^{-1}$, and spectral resolution of $\sim 6$ pixels or 0.80 \AA.
Furthermore, to account for the granularity of the diodes and increase 
the SNR, the observations were split into 4 subexposures, rotating the 
grating carousel by $\sim 5$ diodes per subexposure. The reduced spectra
are shown in Figure \arabic{nfig}. 

\subsection{Data Reduction}

The data were re-reduced with the standard GHRS data pipeline,
implementing updated calibration files from July 1997. In particular,
the grating sensitivity and the LSA incidence angles have been
recently recalibrated for the G140L, so the newest references were
used. The GHRS reduction pipeline includes a correction to the
wavelength scale for heliocentric velocities. 
The default wavelength scale, which has proven to be very stable
(\cite{lan97}), has a maximum RMS dispersion of 55 m\AA\ for
this grating. The largest source of wavelength error was the thermal
variation in the spectrograph, with the G140L showing the greatest
temperature sensitivity of all the GHRS gratings.  These variations
resulted in significant zero-point shifts, which typically were
corrected with intermediate CzPtNe wavelength calibration exposures.
The cross-correlation, however, between these calibration exposures
and an artificially created CzPtNe spectrum yielded unsatisfactory
offsets and large errors. The offsets can be calculated independently
from the Galactic absorption lines present in the spectra using the
algorithm described in \S 3.2. This method assumes that the gas
causing the Galactic absorption is at rest within the LSR ($v_{\rm
LSR}= 0$ \kms).  Although some lines of sight may pierce high-velocity
clouds, inducing potential variation in $v_{\rm LSR}$ on the order of
$\pm100$ \kms, average LSR velocities measured from HI
emission by the HST Key Project (\cite{sav93}; \cite{loc95}) are
typically on the order of $|v_{\rm LSR}| \la 10$ \kms. 
However, this is much smaller than the instrumental resolution 
(0.8 \AA, or 195 \kms\ at 1230 \AA), in 
addition to being smaller than the match window used in the line
identification process (see \S 3.2 for details).
The combined $1\sigma$ errors in the wavelength solution are well
represented by the dispersion in the zero-point offsets for each
spectrum, with a typical value of 18 \kms.  These $1\sigma$ errors
(rms) of the offset for each individual spectrum are included in Table
1.

\section{SELECTING THE ABSORBERS}

Line-profile fitting is the simplest and most direct way to detect and 
measure quasar absorption lines. Line-profile fitting implicitly assumes 
that the regions causing the absorption are discrete structures in 
thermodynamic equilibrium which are well described by the chosen profile.
However, supercomputer simulations have shown that the structure of the
absorbing regions is complex and filamentary, and the gas is subject to
a wide variety of dynamical processes, each of which has an influence on
the resultant spectral profile (\cite{cen94}, \cite{her96}, \cite{mir96}).  
In fact, the entire notion of a ``cloud'' is inappropriate; at the lowest
column densities the hydrogen distribution tends towards a diffuse and
smoothly fluctuating intergalactic medium (Gunn \& Peterson 1965;
Kirkman \& Tytler 1997). 
Absorption features studied in higher resolution GHRS G160M data 
(\cite{wey95}) are well fit by Voigt profiles and so their Doppler
parameters may be inferred.
However, at the resolution of the GHRS G140L data 
($\Gamma_{\rm res} = 0.80$ \AA), 
any thermal or turbulent imprint on the line profiles will 
not be resolved. This assumes that the Doppler parameter distribution at 
low redshift is similar to that found found at high redshift using very 
high resolution spectra (e.g. \cite{hu95}; \cite{wom96}). At low redshift,
the number density of absorbers is low enough that the spectral features 
are isolated and deblending is not an issue. A key feature of our analysis
is the use of an automated line selection and fitting process that is
reproducible and quantifiable.

\subsection{Selection and Measurement of the Absorbers}

Line-profile fitting requires identification of the continuum for the 
observed flux. Typically, an accurate estimate of the continuum is limited
by the cumulative effect of the increasing number of low column density 
lines which act to depress the continuum. However, at very low redshift 
this effect is negligible because the line density is low and the continuum
can readily be located adjacent to each spectral feature. A continuum was
fit for each of the 12 spectra using software designed for this purpose 
as well as for fitting line profiles.  The software is a significant
elaboration and modification of the algorithm of \cite{ald93p},
which produces a self-consistent and repeatable result.
For details, see \cite{pet98p}. The continuum is fit by-hand in the 
region of the damped \Lya\ absorption and 
geocoronal \Lya\ emission features; no subtraction of these features
was attempted and adjacent regions ($\pm 900$ \kms) were omitted from the 
analysis. The final continuum fits are overplotted on the reduced spectra 
in Figure \arabic{nfig}.

\addtocounter{nfig}{1}

The limiting equivalent width, $\sigma_{lim}$, of each spectrum was 
computed as a function of wavelength in order to assess the quality 
of the data and to set limits for inclusion of lines in the subsequent 
analysis.  The computation of $\sigma_{lim}$ is described in \S 3.2.
The 4.5$\sigma_{lim}$ detection limit is shown for each spectrum in
Figure \arabic{nfig} for the wavelength range corresponding to $0.003<z<0.225$.
For comparison, the completeness level of 0.24 \AA\ used by \cite{jan98p}
is overplotted and the tickmarks schematically indicate the location of 
\Lya\ lines.  Note that the data for Mark~1320 and Q~1228+116 
have detection limits that are too high to use in this study and, although 
line lists were developed, they were excluded from the analysis.

In order to select and measure the absorption features, we assume that the 
observed flux profiles are well represented by the convolution of a Voigt 
profile with the line spread function of the GHRS G140L grating.
To verify this, subroutines from the program AutoVP (\cite{dav97})
were used to generate flux profiles for \Lya\ absorption lines 
with lower and upper limits for the expected Doppler parameter, $b$, 
and for a range of column densities, \nhi. 
The convolution of this intrinsic line profile with the instrumental
line spread function is the expected line profile.  
If the distribution of Doppler parameters at low redshift
is similar to that at high redshift, the respective lower and upper 
limits are approximately 20 \kms\ and 80 \kms\ (\cite{hu95}).
The line spread function is essentially a Gaussian distribution
with $\rm FWHM = 0.80$ \AA\ (\cite{isr63}; \cite{hea95}).
By inspection, none of the absorption features in the 12 quasars
in our sample had a central flux lower than $\sim 10$\% of the 
continuum level, so we examined profiles computed for values of
$b$ and \nhi\ that resulted in this value for the central flux.
We then compared them to a Gaussian fit to the profile and found that
the difference
between the actual and fitted profiles was very small. In other words, 
given the the column densities and Doppler widths of the absorbers and 
the resolution of the spectrograph, the instrumental profile dominates the 
intrinsic line profile in the resultant flux profile. We conclude that the 
use of a Gaussian profile in fitting absorption features is appropriate for 
our purposes.

Line-profile fitting was performed by software based on the \cite{ald93p}
code. New algorithms for selecting and fitting lines as well as deblending
were implemented, completely automating the process and eliminating 
``by-hand'' intervention.
\cite{pet98p} used this software on a high redshift lensed quasar, 
where the line density was much higher and the width of the
instrumental profile dominated the distribution of Doppler parameters,
so the FWHM was held constant (all of the absorption lines are unresolved). 
In this work, the intervening
\Lya\ lines are expected to be unresolved but some high ionization
Galactic lines may be resolved due to inflow and outflow processes (Savage,
Sembach \& Lu 1997). We allow for resolved lines but restrict the minimum
allowable FWHM to be \gamres, following the HST Absorption Line 
Key Project (\cite{bah93}). Even though our methodology differs slightly
from that of the Key Project, similar results are produced in a direct
comparison of line lists for the three objects in common. 

In the simultaneous fitting phase, the algorithm allowed variation of all 
three parameters which describe the Gaussian. 
After fitting a particular combination of lines, the program examined
the FWHM for each component, and if any value for the  
FWHM fell below \gamres, the FWHM for 
that component was reset to \gamres. The fit was then performed again.
This algorithm prevents fits to noise spikes, and sets a minimum allowable
FWHM for real absorption lines which cannot be narrower than the instrumental
resolution.  
Inspection of the distribution of velocity widths shows that a small fraction
of the total number of lines have FWHM larger than 375 \kms\ --- 13 lines or
3.6\%.  Six of these are strong lines identified with Galactic and 
extragalactic metal line systems.
The remaining seven lines yield an unphysically broad FWHM most likely due
unresolved, blended components or because of the uncertainty in the continuum 
fit and noise. This small number of lines has a negligible impact
on the analysis. Given the average line density, the probability that two 
lines will fall close enough by chance to appear as a blend is only 2.6\%.
This is evidence that some of the lines with FWHM larger than instrumental 
resolution are truly resolved and are not the result of individual blended
components.  
\addtocounter{ntab}{1}

Parameters fit for lines selected in each spectrum are listed in Table
\arabic{ntab}. Blended lines which were fit simultaneously to a feature 
have identical $\chi^2_{\nu}$ values. Lines for which the quoted error 
in the FWHM is exactly zero are considered to be unresolved and were not 
varied in the final fit.
Five lines with significance lower than 3$\sigma_{lim}$ were removed
from Table \arabic{ntab}. Since a significance level of 3$\sigma$ is low, we made a
line by line comparison in the case of 3C~273, the only object in our sample
where a higher resolution spectrum is available. The only lines in the list
of Morris et al. (1991) that do not appear in our line list are either very 
weak lines ($W<75$ m\AA), or they are very close blends that our G140L data 
could not separate. Therefore, we recover lines as well as would be expected 
given the signal to noise and resolution.

The lines that appear in our list that do not appear in the Morris et al.
list are used to estimate the false detection rate for very weak lines. 
In 3C~273, our software recovers 29 lines above three times the 1$\sigma$ 
limiting equivalent width). Adopting the Morris et al. spectrum as a ``truth''
spectrum, five of these lines are false detections. This is a conservative 
estimate of our ``false'' detection rate, since these are all weak lines 
where the exact choice of continuum fit makes a substantial difference to 
the detectability (and significance level) of the line. Using these numbers, 
we estimate that $\sim$10\% stronger than 4.5$\sigma_{lim}$ might be false 
detections and $\sim$50\% of the lines between 3$\sigma_{lim}$ and 
4.5$\sigma_{lim}$ might be false. As we will see, this projects to no
more than 16\% possibly false lines in the \Lya\ sample, a level of 
contamination that cannot affect the main scientific conclusions of 
the paper. We include all lines in Table 2 in the identification procedure.
The total number of lines above 3$\sigma_{lim}$ is 357, and the number above 
4.5$\sigma_{lim}$ is 272.

\subsection{Identification of the Absorption Lines} 

A list of \Lya\ lines for each quasar was created by removing
lines from the observed lists that could be otherwise
identified. Because the spectra span the redshift range down to $z=0$, a
significant number of features are due to absorption by metal species
in the Galaxy --- these lines were used to give an independent
measure of the wavelength calibration zero-point and error.
Metal-line absorption systems due to extragalactic sources were identified
using previously published redshifts, and a search was made for new systems.  
We distinguish metal line systems, which have strong associated \Lya\ 
absorption, from much weaker metal lines that have been found to be associated 
with most \Lya\ absorbers down to the limits of detection.  
Lastly, we search for higher order Lyman lines in systems which may or may 
not have associated metal lines.

Candidate identifications for absorption lines were made by searching
the line lists for matches to the comparison lines. A match
was declared when the absolute value of the difference between the
comparison and observed wavelengths was less than some multiple of
$\sigma_{res}$, which is related to the instrumental resolution, \gamres.
The comparison line list is a compilation of the strongest transitions
of the most abundant elements from \cite{bah93p} and \cite{mor88p}.
Some more recent measurements of wavelengths and oscillator strengths are 
taken from \cite{mor91p} and \cite{sav96p}.  

Tentative identifications intially
selected by proximity to the predicted wavelength were then subjected
to a series of tests designed to check consistency with atomic
physics. These have been defined by \cite{bah92p}.
First, \Lya\ must have the greatest equivalent width.
Second, doublets tenatativly
identified as \ion{O}{6} $\lambda\lambda$1031/1037, \ion{Si}{2} $\lambda\lambda$1190/1193,
\ion{N}{5} $\lambda\lambda$1238/1242, or 
\ion{Si}{4} $\lambda\lambda$1393/1402 must have the correct separation within 
a tolerance of 3$\sigma_{res}$ or about 180 \kms\ (although 70\% of 
these doublets have separations correct to within 1$\sigma_{res}$).
Third, the doublets as well as lines identified as transitions of 
\ion{N}{1}, \ion{S}{2}, and \ion{Si}{2} must also meet as set of
criteria based on line strength.  If the weaker component is
tentatively identified but the stronger one is not, the identification
is not accepted.  If only the stronger component is identified, the
minimum expected equivalent width of the weaker component,
$W_{w}^{min}$, must be below the detection threshhold, which we define
to be 3.5$\sigma_{lim}$, to be accepted.
Here 1$\sigma_{lim}$ is the 1$\sigma$ limiting equivalent width
computed by convolving the 1$\sigma$ flux error array, where the regions 
occupied by absorption features have been replaced by values from the 
adjacent continuum regions, with a Gaussian
having FWHM equal to the instrumental resolution
\begin{equation}
W_{w}^{min} = \frac{f_{w}}{f_{s}} (W_{s} - 2 \sigma_{s}),
\end{equation}  
Here $f_{w}$ and $f_{s}$ are the oscillator strengths for the weaker
and stronger components, and $W_{s}$ and $\sigma_{s}$
are the measured equivalent width and error for the stronger
component.  If both components are tentatively identified, the value
of the equivalent width for the stronger component must be at least
$W_s^{min}$, where
\begin{equation}
W_{s}^{min} = W_{w} - \sigma_{m}.
\end{equation}
Here $W_{w}$ is the equivalent width for the weaker component, and
$\sigma_{m}^{2} = \sigma_{w}^{2} + \sigma_{s}^{2}$, or the errors in
the measured equivalent width added in quadrature.  In all cases if
either component is identified and the other is not, but its predicted
location is outside the observed spectrum, it is accepted as a final
identification.  Finally, if any absorption line can be identified
with more than one system, preference for identification is given by
the following order: interstellar line, extra-galactic line, isolated Lyman
line.  
For competing identifications within an extra-galactic system, the
closer match with a higher expected strength based on oscillator
strength is chosen. 
If one is closer and the other has a larger expected strength, an
alternate identification is noted with the closer match listed in 
Table \arabic{ntab} and the second identification indicated by a footnote.
For competing identifications between extra-galactic systems, the
closer match is chosen. 

We determined the zero-point offset for the wavelength calibration
by identifying strong interstellar lines in each spectrum.
This procedure assumes that the Galactic ISM is at rest, and that mean
deviations from 0 \kms\ due to high-velocity clouds along the line of
sight are negligible in comparison to our resolution and errors (as
described in \S 2.3). Candidate
identifications for galactic absorption lines were made by searching
the observed line lists for matches to the comparison lines; the match
window was set to be $4.5\sigma_{res}$.  Final identifications were 
assigned after verifying they are consistent with atomic physics as
itemized by the rules above.  
At least 3 lines (for the two poorest SNR spectra), but typically 5 or 6 lines
were used to measure the zero-point offset for each spectrum. Generally, the
transitions used were the \ion{Si}{2} $\lambda\lambda$1190/1193 doublet, 
\ion{Si}{3} $\lambda$1206, \ion{Si}{2}$\lambda$1260, \ion{O}{1} 
$\lambda$1302, \ion{C}{2} $\lambda$1334, and the doublet \ion{Si}{4} 
$\lambda\lambda$1393/1402.
The mean residual weighted by the line significance, S$\sigma_{W}$,
is the zero-point offset, and the rms, $\sigma_{\lambda}$, is a measure 
of the total uncertainty in the wavelength calibrations. Both quantities
are listed for each spectrum in Table \arabic{ntab}. 
The average of these rms values results in a number
that characterizes the uncertainty in the wavelength calibration for
the sample as a whole and is 0.072 \AA\ or 18 \kms.  The maximum value
for any quasar used in the subsequent analysis is 0.11 \AA\ or 27 \kms.  
Although the match window is
$4.5\sigma_{res}$ ($5\sigma_{res}$ for J1230.8+0115) all the lines used
to determine the zero-point offset have a maximum absolute residual of
0.36 \AA\ ($\sim 1\sigma_{res}$), with a more typical value of 0.16
\AA\ ($\sim 0.5\sigma_{res}$), after the offset is applied.

After the zero-point correction was made to the spectra and line lists,
we searched for interstellar lines using the complete
comparison list, which not only included the strong lines used to calculate
the zero-point correction but also additional weaker features.
Candidate identifications were initially chosen as lines with a match
window of $3\sigma_{res}$, and finalized after being tested for
consistency with atomic physics.  The final identifications for
the interstellar absorbers along with their residuals, 
$\Delta \lambda = \lambda_{meas} - \lambda_{pred}$,
are listed in Table \arabic{ntab}.

Following the search for Galactic lines, absorbers associated with 
extragalactic sources were identified by first searching for lines
associated with published heavy element systems, which are more commonly
termed ``metal-line systems''. Then a search is made for new systems.

\subsection{Comments on Newly Identified Systems}

Three absorption line systems have been identified in an FOS spectrum of
PG 1216+069, presented by \cite{jan98p}, at redshifts 0.0063,
0.1247, and 0.2822.  Systematic redshifts were redetermined from the
strongest associated lines in our GHRS spectum and were found to be  
0.0063$\pm$0.0001, 0.1250$\pm$0.0005, and 0.2923$\pm$0.0001 (the quoted
errors do not include systematic errors). 
We identify all lines as tabulated by \cite{jan98p}.
As noted by and in agreement with \cite{jan98np}, we find the \Lya\ absorption 
at $z_{abs}=0.0063$ to be unusually strong, and we do not resolve \Lya\ into 
components.  However, we do detect metal-line absorption associated with this
system. Metal lines \ion{C}{2} $\lambda$1334 and \ion{Si}{4}
$\lambda$1402 have been identified as members of this system; 
\ion{Si}{2} $\lambda$1260 was also a candidate identification with
this system, but it was superceded by a closer match to an identification
with \ion{O}{6} 1037 for $z_{abs}=0.2221$ and could possibly be a blend. 
The automatic line finding software did not find a line at the predicted
location of the stronger component of the \ion{Si}{4} $\lambda\lambda$
1393/1402 doublet; however, there is a feature at this location which
when measured by hand has a marginal significance.
Additionally, the \ion{Mg}{2} $\lambda$2796
line was identified in the incomplete sample of \cite{jan98p}, so
\ion{C}{4} $\lambda$1402 is identified and this system is considered 
confirmed.  Four higher order Lyman lines were
identified with the $z_{abs}=0.2882$ system.  \Lyd\ is not listed because
although an absorption feature corresponds to its predicted location,
it lies in the wavelength region which was omitted because of the 
geocoronal \Lya\ feature.
Two heavy element lines are found: 
the stronger component of the
\ion{O}{6} doublet (the expected strength of the weaker component is
below the detection threshold) 
and \ion{C}{3} $\lambda$977  (which was superceded by 
identification as Galactic \ion{S}{2} but may possibly be a blend.) 

A search was made for new metal line systems in all of the
spectra by assuming in
turn each as yet unidentified line to be \Lya\, and looking
for matches to the expected location of the strongest lines in the
comparison list.  Lines that fall within $3\sigma_{res}$ are
considered candidate identifications.  
In order for a new system to be accepted either \Lya\
and both components of one of the four doublets mentioned in Rule 2
above, or \Lya\ and three other strong lines must be identified and be
in compliance with the rules specified above.  
These lines are then
used to redetermine the redshift of the system (by taking the
average of the redshift weighted by the significance of each line), 
and a second pass was made with
the complete comparison list to look for additional associated lines
(which must also meet the consistency criteria).  This search also found 
higher order Lyman lines for systems which may or may not have
associated metals.  All candidate \Lyb\ lines were preferentially
identified as metals associated with the new metal-line systems, 
and so no higher Lyman lines are listed in Table 2, except for the 
strong Lyman series at $z_{abs}=0.2823$ in PG~1216$+$069. 
Ten new metal systems are found in 5 of
the 12 quasar spectra and are listed along with their identified
\addtocounter{ntab}{1}
lines in Table \arabic{ntab}.  

There are  a total of 11 \Lya\ lines found to have associated metal-line 
absorption, and these plus the remaining 128 unidentified lines in the 
wavelength region corresponding \Lya\ at $0.003<z_{abs}<0.225$ 
are assumed to be \Lya\ absorbers. These 139 lines comprise the sample 
which will be examined in the subsequent analysis. All of these have 
S$\sigma_{lim}\ge 3$, and 108 have S$\sigma_{lim}\ge 4.5$. Based on
the comparison with a single higher resolution spectrum of 3C~273
(Morris et al. 1991), we estimate that no more than 16\% of these
lines are potentially false detections due to details in the line
selection process. The lines used in the detailed comparison with 
galaxies are all strong enough that the analysis in \S 5 is not 
affected by this issue.

\section{PROPERTIES OF THE ABSORBERS}

This dataset provides a unique opportunity to examine the properties
of the \Lya\ absorbers in the local universe.
If these absorbers can be characterized by a random distribution, this 
would suggest that they have maintained their ``primeval'' state, and have not
evolved gravitationally from their higher redshift counterparts.
If they are clustered, then the gas may have collapsed into
structures that are in some way related to galaxies. 
In this section we describe the 
general properties of the \Lya\ absorbers, such as their number density and 
their distribution of equivalent widths. We also check for consistency with 
values measured from larger samples of data.  
The scale and amplitude of the clustering of the absorbers, 
compared to similar statistics for galaxies,
can give clues to the origin and evolution of the structures.  
We use two statistics to address the hypothesis that the 
\Lya\ absorbers are randomly distributed:  the nearest neighbor distribution 
and the two-point correlation function (TPCF).  We then 
test the hypothesis that the \Lya\ absorbers are clustered in the same
way that galaxies are clustered by comparing the \Lya\ TPCF to
the TPCF measured for galaxies.

\subsection{The Statistical Properties of the \Lya\ Absorbers}

To check that our sample of \Lya\ absorbers is representative of its
parent population, the number of lines per redshift interval and the 
number distribution of rest equivalent widths is compared with values 
derived from a much larger sample of data by \cite{wey98p}.  
The range in wavelength to be included in the analysis is determined
at the blue end by obscuration due to the geocoronal \Lya\ line, $z=0.003$, 
and at the red end by the limit of the data, $z=0.225$.
The evolution in the number density of lines
is undetectably small over this range, so we assume it 
to be constant. We compare to the \cite{wey98p} sample,
which has a uniform detection limit of 0.24 \AA\ and counts both \Lya-only
lines as well as \Lya\ lines with associated metals. 
We count lines in our sample which are located in regions of the spectra 
which are complete to 0.24 \AA\ for 4.5$\sigma$ lines, and compute
$dN/dz = 38.3 \pm 5.3$.  The mean is an unweighted average, and the error 
is computed by combining in quadrature
the Poisson error in $dN/dz$ from each line of sight.
This is considered to be the internal error obtained
by treating each line of sight as an independent measurement. 
\addtocounter{nfig}{1}
The values for $dN/dz$ computed for each line of sight individually are shown 
in Figure \arabic{nfig}.

Our number for $dN/dz$  agrees with the
predicted value from the fitted coefficients of \cite{wey98p} to within their
1$\sigma$ errorbars.   Also, as expected, the distribution of rest equivalent 
widths of this sample of lines is well fit by an exponential distribution.
The observed number of lines is compared to the number
expected for each line of sight with a $\chi^{2}$ test and
results in a probability of 15\% that
the $\chi^{2}$ would be larger than it is observed. This indicates that the 
scatter in the observed number of lines is greater than would be expected 
from an assumption of Poisson errors. We interpret this marginal evidence for 
cosmic variance in the number of absorbers among the lines of sight. The typical
transverse separation of any two sightlines is $\sim 40 h_{75}^{-1}$ Mpc.
Variations on such a large scale would be unprecedented for \Lya\ absorbers,
and this issue is worth revisiting with a larger data set. We note that
the simulations of Dav\'{e} et al. (1998) are not sensitive to structure
on this scale due to the limited box size.

Evaluation of the significance of the results of the nearest neighbor
distribution and the TPCF depends on computing a random distribution of 
absorbers using a Monte Carlo technique.  The number of lines chosen for 
each realization depends on the extrapolation of the fitted distribution of 
the number of lines per interval redshift per interval rest equivalent width, 
$d^{2}N/dWdz$, to the highest sensitivity limit, $w_{min}$, of each spectrum. 
We can compare the extrapolated values with the observed values for low 
redshift $dN/dz$ measured at higher sensitivity limits from \cite{shu97p} and 
\cite{tri98p}.
\addtocounter{nfig}{1}
Their points are presented as a function of sensitivity limit, $w_{min}$, in 
Figure \arabic{nfig} by solid symbols. 
Overplotted as a straight line with dashed $1\sigma$ errorbars is the Quasar
Absorption Line Key Project distribution from \cite{wey98p}, 
\begin{equation}
\frac{d^{2}N}{dzdw_{min}} = 
\left(\frac{dN}{dz}\right)_0(1+z)^{ \gamma} \exp\left[\frac{-(w_{min}-0.24)}{w{*}}\right],
\end{equation}
where $(dN/dz)_0=32.7$, $\gamma=0.26$, and $w_{*}=0.283$. 
Also plotted is our computed value of $dN/dz$ for a completeness limit of
0.24 \AA.  
Note that the \cite{shu97p} and \cite{tri98p} measurements (solid symbols) 
are slightly higher than the Key Project extrapolation.
To evaluate whether the extrapolation with $w_{min}$ breaks down at
lower equivalent width thresholds,
we compute a second point at a higher sensitivity limit from a subset of our
data which has slightly larger errorbars (open symbol).  
We also plot points at lower sensitivity limits quoted by 
\cite{shu97p} and \cite{tri98p}.  The results suggest that if there
is a real increase in the number density of lines in excess of the 
extrapolation, it occurs only among the very weakest lines.  It is also
possible that the \cite{shu97p} point samples a line of sight with an 
unusually high number of absorbers.  We 
chose to use the extrapolation of \cite{wey98p} in performing the Monte 
Carlo simulations.

\subsection{Nearest Neighbor Distribution}

The nearest neighbor distribution is computed for the observed sample 
of 139 lines by finding the nearest neighbor in velocity space for every 
\Lya\ absorption line along each line of sight, and plotting the frequency 
distribution of velocity splittings for all the lines of sight combined.  
The expected number of pairs in each bin due to a random distribution
of \Lya\ absorbers is determined by Monte Carlo simulation. Even though 
the number of lines per unit redshift for the sample agrees with that
of \cite{wey98p} for all lines of sight, there is a variance in $dN/dz$ 
among the lines of sight as shown in 
\addtocounter{nfig}{-1}
Figure \arabic{nfig}. 
\addtocounter{nfig}{1}
Therefore, to obtain the random distribution of velocity separations,
the number of expected lines for each simulated line of sight must 
be drawn from a Poisson distribution with a mean given by $d^{2}N/dzdW$ using
fitted coefficients from \cite{wey98p}, instead of using the observed number
of lines per line of sight. 
This turns out to affect the amplitude of the nearest neighbor distribution 
in the smallest bins by about 25\%.  

The nearest neighbor distribution expected for a random distribution of
absorbers is computed 1000 times for
a simulated set of 10 quasars having the measured 3$\sigma_{lim}$ detection
threshhold.  The number of lines per quasar is initially chosen by scaling
$dN/dz$ using fitted coefficients by \cite{wey98p} to the most
sensitive part of each spectrum, $w_{min}$.
The finite resolution of the spectrograph is accounted for by not allowing
any two lines to be closer than 2.5$\sigma_{res}$.  This value was chosen
based on simulations performed on the software to quantify the recovery 
reliability of input line parameters.
We have previously performed this test for lower resolution higher redshift 
quasar spectra (\cite{pet98}), where the recovery rate for input central 
wavelengths,
FWHMs and equivalent widths as a function of line strength, separation and 
SNR was evaluated using a Monte Carlo technique.  For a separation of
2.5$\sigma_{res}$, the central wavelengths of the input lines were recovered
to within $1\sigma_{res}$ 99\% of the time, and the equivalent widths were
recovered to within 20\% of the input value 95\% of the time.

In order to use the maximum number of lines from the sample, we account
for the varying sensitivity of each spectrum in the simulation.  
Each line was assigned a wavelength corresponding to a random location in 
space, then assigned an equivalent width drawn from an exponential
distribution, again using the fitted coefficients of \cite{wey98p}.
For each simulated absorption feature, the assigned equivalent width was
compared to the detection limit at its location in the spectrum, and
was removed from the list if it was below the detection limit.  This procedure
simulates the entire observed line list with randomized locations.
The distribution of velocity 
separations for the nearest neighbor pairs along each line of sight were
computed for every realization
and the mean number of pairs for each bin is the expected number for that bin.  
Confidence intervals were evaluated by summing over the distribution of pairs
in each bin.  The bin size is set to be 250 \kms, and the first bin is
not meaningful because the resolution of the spectrograph limits sensitivity
to about 210 \kms.  
\addtocounter{nfig}{1}
The results are shown in Figure \arabic{nfig}$a$.  
The first two bins, corresponding to a velocity splitting of
250-750 \kms, each show a clustering signal with greater than 95\% 
confidence level. The amplitude of this clustering signal may have
been underestimated by as much as $\sim$15\% due to the contamination
of the weakest lines with (randomly distributed) false detections.

Another way to evaluate the significance of this signal is to compute
the probability that the observed and expected nearest neighbor distributions
as a whole are drawn from the same random parent distribution. 
This can be estimated by forming
the distribution of the variance between the mean expected distribution and 
each realization and is shown in Figure \arabic{nfig}$b$.  The 
variance for the observed distribution and the expected mean 
distribution is shown as a dotted line.
Only two out of the 1000 random realizations have a larger variance.
We conclude that the probability that
the observed distribution of nearest neighbor velocity separations is
obtained from a random distribution of absorbers is very small.

\subsection{Two-Point Correlation Function}

The TPCF along the line of sight is computed for the observed sample 
of lines with significance greater than $3\sigma_{lim}$.
The TPCF, $\xi(\Delta v)$, is defined as
\begin{equation}
\xi(\Delta v) = \frac{N_{obs}}{N_{exp}} -1,
\end{equation}
where $N_{obs}$ is the frequency distribution of observed velocity splittings
of all pairs of absorbers along the lines of sight, and $N_{exp}$ is the expected
number of velocity splittings in each bin and is determined 
using a Monte Carlo technique.
This process for computing $N_{obs}$ is similar to the nearest neighbor 
distribution computation except instead of
only the nearest line contributing a velocity splitting, all possible
pairings of lines in a line of sight are computed.  The results from
all 10 lines of sight are combined to form $N_{obs}$.  The distribution
expected from a random population of absorbers is computed by Monte Carlo
simulation in the same manner as for the nearest neighbor distribution.
Because the number of velocity pairings goes
as $N^{2} - N$, instead of with $N$ as with the nearest neighbor distribution,
a slight difference in the normalization of the number of lines per
line of sight makes a very large difference in the normalization of $N_{exp}$.
Since we are interested in the relative shapes of the distributions,
the random distribution of velocity splittings, $N_{exp}$, is scaled so
both distributions have equal numbers of velocity pairs. 
Because of the finite length of the spectrum, the distribution $N_{exp}$ 
has a slope due to the fall-off of pairs with larger separations.  To
account for this aliasing effect in the normalization, we sum over a velocity
range corresponding to half the redshift range under study.  This 
also corresponds to the velocity splitting where the number
of observed pairs is zero in some bins.

\addtocounter{nfig}{1}
The TPCF is shown in Figure \arabic{nfig}. The 68\% and 95\%, confidence intervals 
are overplotted and were computed as for the nearest neighbor distribution. 
This statistic is, in principle, sensitive to clustering at all scales.
But because all pairings are used, in practice it is not as sensitive as the
nearest neighbor test to clustering at the smallest scales. Added pairs 
produce added noise to all bins and any small scale clustering signal is 
diluted.

In order to test the hypothesis that galaxies are clustered
like \Lya\ absorbers are clustered, we compare our TPCF to that determined 
for bright galaxies. The measured TPCF for galaxies
from \cite{dav83p} is represented in Figure \arabic{nfig} for the smallest 
bins by black dots.  
We use a parameter choice of $r_p \sim 500$ \h75\ kpc in this comparison,
appropriate to the observed coherence length of the absorbers at $z<1$
(\cite{din95}).
If galaxies are clustered like \Lya\ absorbers, they should have the same
amplitude.
Figure \arabic{nfig} indicates that while \Lya\ absorbers have a marginal clustering
signal for small velocity splittings (almost 95\% confidence level),
galaxies clearly cluster much more strongly.

The only two previous studies of the clustering of
\Lya\ absorbers at low redshift both use  data obtained by
the HST Key Project (\cite{bah93}) over the range 
$0\lesssim z \lesssim1.3$. \cite{bah93p} analyzed line lists from 
the first set of quasar 
spectra obtained by the Key Project with the HST's FOS, and found
no evidence for a strong correlation in the TPCF.
Subsequently, \cite{ulm96p} used the line lists from these inital 
observations plus a second set of line lists (\cite{bah96}), for a total
sample of 100 lines, to 
look for a clustering signal with the expanded set of data.
He found a clustering signal that is similar in strength to that of the
galaxy-galaxy correlation function, $\xi(\Delta \rm v) = 1.8^{+1.6}_{-1.2}$,
90\% confidence level, for separations of 250-500 \kms.
This work is the first observational study of clustering to focus on 
the local universe, $0.003 < z < 0.225$, where the mean redshift
corresponds to 15\% of the lookback time compared to $\sim 55$\% ($q_{0} = 0.5$)
in previous work; our total sample contains 139 lines compared to 15 
found in this redshift range in the sample studied by \cite{ulm96p}.

\section{COMPARING GALAXIES AND ABSORBERS}
\subsection{The Virgo Region}

All ten of these lines of sight (LOS) were chosen for their position 
behind the Virgo
cluster region which provides an excellent opportunity to explore the
galaxy-absorber connection in a well-studied region of varying density
environments containing numerous surveys complete to faint limits.  We
constructed a sample of galaxies from the literature, using NED
(ca. October 1998), supplemented with ZCAT (version November 1998;
\cite{zcat}), with RA from $12^{\rm h}$ to $13^{\rm h}$, and
declination from  $-4$\arcdeg\
to 19\arcdeg, and radial velocity less than 3000 \kms, which we will
call the Virgo galaxy sample although it encompasses more than just
the Virgo cluster proper.  Fitting a
Schechter luminosity function to the sample with a flat faint-end
slope  ($\alpha = -1.0$) and $M_B^* = -20$,   we
find it to be complete to $M_B  = -16$, containing  galaxies as faint
as SMC-type dwarfs ($L \gtrsim 0.04L^*$, adopting the more standard
$M_B^* = -19.5$ from \cite{lov92}). 
Extending this sample further in redshift, sampling incompleteness
sets in quickly, and is only complete to $\sim L^*$\ for $v \le 4000$\
\kms. 
\addtocounter{nfig}{1}

The galaxy distribution can be seen in Figure \arabic{nfig}, where the
Virgo sample 
galaxies are plotted in units of galaxy per unit magnitude and the
Schechter function is overplotted in the same units. The function was
arbitrarily normalized to fit the turnover, and the error bars
indicated are Poisson.
Galaxy absolute
magnitudes were calculated assuming pure Hubble flow, with
$H_0$ = 75 \kmsmpc, with no Virgo-centric infall model 
applied, since the Virgo triple-value problem (e.g. \cite{ton81})
introduces scatter at all magnitudes and so will not 
greatly affect the shape of the luminosity function and thus the
completeness limit.  The quasar path lengths
were likewise limited to \Lya\ in the range   
$900 < v < $~3000 \kms, using the same upper limit as the Virgo sample,
and excluding all possible lines below 900 \kms\ due to interference by the
geocoronal \Lya\ emission (see \S 3.1).  This yields 11 \Lya\ lines
amongst the ten  
 LOS, satisfying the 3$\sigma$\ limiting equivalent width
criterion (although they are all at least 4$\sigma$\ lines).  
\addtocounter{nfig}{1}

The distribution of galaxies and absorbers can be seen in the pieplots
in Figure \arabic{nfig}, where the declination range has been split up
into three slices of $\sim8$\arcdeg\ each.  The Virgo sample of galaxies, as
defined above, are plotted for $L \gtrsim 0.04L^*$ out to $v = 3000$\ \kms.
The ten lines of sight 
are also plotted with the open circles indicating the absorber
positions; the large circles are the 4.5$\sigma$\ lines,
and the small circles are the remaining 3$\sigma$\ lines.
\addtocounter{nfig}{1}
The one-dimensional galaxy distributions along the lines of sight are
indicated in Figure \arabic{nfig}.  The distribution of $L \ge
0.04L^*$\ 
galaxies within impact parameters $\rho \le$~ 1\h75\ Mpc of each of the ten
lines of sight can be seen as the unshaded histograms, and the 
galaxies falling within $\rho \le$~ 250\h75\ kpc as the shaded histograms.
The \Lya\ lines in this 
range are also plotted, with the longer vertical bars representing
$4.5\sigma$\ lines and the shorter vertical bars $3\sigma$\ lines.  

While limiting our comparison galaxy volume to the Virgo 
region and to a shortened redshift range greatly diminishes our
available spectral path length, it significantly increases 
the contiguous volume within which to compare to individual
galaxies with a uniform luminosity limit.   
The typical distance between any two lines of 
sight within this volume is about 5 degrees, or 2\h75\ Mpc, and the
total volume probed is $\sim 10^{6} h^{-3}_{75}$ Mpc$^{3}$. 
Moreover, in addition to probing
primarily the field galaxy population, we have the
opportunity to probe a galaxy cluster environment down to very
faint completeness levels.  

In comparing absorbers and galaxies, we make no corrections for
peculiar velocities, assuming they
will share the same velocity field.  However, the Virgo cluster itself
($v_c = 1050 \pm 35$; \cite{bin93})
presents a special case, having a large velocity dispersion 
($\sigma
\simeq 700$ \kms; \cite{bin93}), and a possibly non-virialized
structure (c.f. \cite{fuk93}, \cite{bin93}).  This makes identifying an
absorber with any galaxy in the Virgo core ambiguous.  However, only
3/10 lines of sight intersect the 6\arcdeg\ Virgo core
(following the definition of \cite{tul84}), and have no absorbers within
the included 1$\sigma$ velocity range of 900--1700 \kms.  Of these
three LOS, only PG1211+143 has an absorber with  $v < 3000$ \kms,
one which falls in the 2$\sigma$ tail of the Virgo
velocity distribution at 2160 \kms.
If absorbers follow the galaxies, one might expect a number of
absorbers in the dense Virgo core, but the small number statistics of
this analysis and the exclusion of the low-velocity end of the core
(namely, 500 - 900 km s$^{-1}$) make the lack of absorbers less
compelling.   The remaining lines of sight all fall well beyond the
Virgo core, but within the maximum angle of influence in the \cite{tul84p} 
Virgocentric infall model (28\arcdeg), the majority falling within 
11\arcdeg\ of the core.  According to the model, the extrema of
galaxy peculiar velocities caused by infall within 11\arcdeg\ 
have a dispersion of roughly 350 - 400 \kms.
At the lowest
velocities, this will only affect comparisons to
one absorber (3C~273: 1012 \kms),
and although this dispersion is on the order of the velocity-space
window in the later galaxy-absorber pair analysis, our techniques may
not give a reliable result for this one \Lymana\ line.

\subsection{Absorbers and Local Galaxy Density}

To pursue the relationship of absorbers to galaxy density, we compared
the distribution of 
galaxy density at the absorber positions to the same distribution 
for randomly distributed absorbers.  The galaxy
number densities 
were counted in 2\h75 Mpc-radius spheres centered on the actual absorber
positions.  Virgo sample galaxies were placed in
three-dimensional space assuming pure Hubble flow.  The 2\h75 Mpc
radius, roughly the Abell radius, serves to smooth the small-scale
galaxy distribution, although the counts in the spheres are still
subject to some shot-noise.  This size of sphere is smaller, but
roughly comparable to 
the Gaussian smoothing length of 5$h^{-1}_{100}$\ Mpc used by \cite{gro98p} to smooth the
CfA2 galaxies around 3C~273, where their simulations demonstrate their
density contours are not sensitive to smoothing lengths varied between
2$h^{-1}_{100}$\ to 10$h^{-1}_{100}$\  Mpc. 

Artificial absorbers were randomly generated according to a Poisson
distribution with a mean equal to the mean number of $3\sigma$\ lines
found along the path length: 1.1 absorber per LOS.  We found constant
$dN/dz$\  at these low redshifts (see also \cite{wey98}).  The galaxy
density 
distribution at the random absorber positions was then
determined in the same way, and this was repeated for 50  
 trials.  We then compared the distributions of galaxy density with
both a KS test and a $\chi^2$\ test.
The advantage of these tests  that they assume no {\em a priori} model
for 
galaxy-absorber correlation, and so are sensitive to a wider range of
scenarios. 

\addtocounter{nfig}{1}
The distributions of galaxy density can be seen in Figure \arabic{nfig}, where
the distributions for real and simulated absorbers (for all 50 trials)
are plotted together, each individually normalized to the total number of
absorbers.  Figure  \arabic{nfig} suggests
that the real absorbers seem to correspond to typically higher 
galaxy densities than the randomly distributed absorbers. 
A KS test between the two density distributions yields only a 12\%
chance the distributions are the same.   However, 
the reduced $\chi^2$\ is 1.04, implying the distributions are a good
match, but with only a 59\% certainty.  Although statistically
well-motivated, this test cannot distinguish between the case where
the absorbers trace the galaxy density and where the absorbers are
independent of the galaxies.

\subsection{Individual Galaxy-Absorber Pairs}

We tried two different methods to associate absorbers with individual
galaxies.  The first method matched 
galaxy-absorber pairs by finding the nearest galaxy three-dimensionally
that was of any luminosity down to our completeness level,
assuming pure Hubble flow (referred to as the 
$r^{3D}_{min}$\ method).  The second method
allowed for a velocity window around the absorber to account for the
uncertainty in mapping radial velocity into distance, and took the
galaxy with 
the smallest impact parameter that fell within that velocity range,
$\Delta v$,  
around the absorber (referred to as the $\rho^{\Delta v}_{min}$ 
method).  If 
no galaxy was found within $\Delta v$, the matching galaxy was then
chosen as the galaxy with the smallest three-dimensional distance,
$r^{3D}_{min}$\ (although, this was not necessary for any of the 11 observed
absorbers in this redshift range).
The velocity window $\Delta v$~=~300~\kms\ was chosen to 
allow for cosmic virial scatter and for small-scale peculiar motions,
being the approximate velocity dispersion of a poor group of galaxies.  
With this
value of $\Delta v$, an absorber counterpart was found for each of the
11 absorbers from $600 < v < 3000$\ \kms.

Other groups have chosen a wide range of methods for associating
absorbers with galaxies.  Variants of the
$\rho^{\Delta v}_{min}$\ method seem to be the most popular.  This is
probably due to the 
velocity-space uncertainties mentioned above,  to
the inherent velocity errors 
when working at high redshift, and to the fact that no
assumption is required beyond some degree of symmetry of the absorbing
object.  We report the results of both tests.  The $\Delta v$ chosen
by various groups varies enormously.  \cite{mor93p} considered each absorber more
individually, generally considering a galaxy ``associated'' for 
$\Delta v \la $\ 400 \kms. Le Brun et al. (1996) adopted a higher value
of $\Delta v = $\ 750 \kms, claiming this falls between galaxy
rotation and internal 
velocity dispersions of 100--200 \kms\ and emission-line region
velocity variations of up to 900 \kms.  \cite{lan95p} initially favored 
$\Delta v = 1000$ \kms, 
but that group now relies upon $v$\ and $\rho$\ parameters
from their galaxy-absorber cross-correlation function, and only
consider galaxy-absorber pairs with $\Delta v \la $\ 500, and $\rho <
270$\h75 kpc (\cite{che98}).  We adopt  $\Delta v = $\ 300 \kms, similar to
\cite{tri98p}, since it encompasses the velocity dispersions of
massive galaxy halos, and since dispersions in this region roughly
correspond to poor group dispersions of 300 \kms. With the exception
of the Virgo cluster itself ($\sigma_v = 700$  \kms), the volume
contains no Coma-cluster-like dispersions of $\sim 1000$ \kms.

\addtocounter{ntab}{1}

For each of the two methods, the absorbers were paired to the Virgo
galaxy sample (as defined in \S 5.1), but the limiting luminosity of 
the sample was varied to simulate survey selection effects.  
Surface brightness selection effects, which could also affect
galaxy-absorber pairing, were neglected (c.f. \cite{lin98}, \cite{rau96}).
First, absorbers were matched to the closest $L \ge
0.04L^*$\ galaxy, then matched to the closest $L \ge 0.25L^*$\ galaxy,
and lastly matched to the closest $L \ge L^*$\ galaxy.  The results of
these three pairings are listed in Tables \arabic{ntab}a and \arabic{ntab}b
for the $r^{3D}_{min}$\ and $\rho^{\Delta v}_{min}$\ methods,
respectively.  The wavelength, velocity and rest equivalent width of
the absorbers for each line of sight are listed with 
the three-dimensional distance to the partner galaxy in kpc, the
impact parameter of the galaxy to the LOS in kpc, the galaxy name,
position and velocity, and the absolute magnitude (calculated according
to a distance from pure Hubble flow), recessional velocity in \kms, and
velocity reference code.
Velocity reference codes are described in Table \arabic{ntab}c.
The  pairings using
the two methods  were not unique, and in fact multiple absorbers
along the same line of sight chose the same galaxy as the closest
match. 
For the $L \ge 0.04L^*$\ sample, 7/11 pairs were different between the
two methods, for $L \ge 0.25L^*$, 5/11 were different, and for 
$L \ge L^*$, 7/11.  In addition, for the $\rho^{\Delta v}_{min}$\ method,
each luminosity cut had 2 absorbers in one LOS with the same galaxy
as a match.  This degeneracy
of pairing  demonstrates the inherent difficulties in choosing
a single method for pairing up an absorber with an individual galaxy.  
It also suggests that there may be no unique and physically reasonable
way to identify a galaxy responsible for any particular absorption
line.

The impact parameters for the galaxy-absorber pairs can be compared to
the impact 
parameters for galaxy-absorber pairs found in the same way for
randomly-distributed, artificial absorbers.  The artificial absorber
redshifts were generated by the same method as the previous KS test,
consistent with a Poisson distribution with a mean of 1.1 absorbers per
LOS, then 
matched with real galaxies in the Virgo sample according to both the 
$r^{3D}_{min}$\ method and the $\rho^{\Delta v}_{min}$\ method. For each
pairing method, a KS test was performed, comparing the distribution of
impact parameters for the real galaxy-absorber pairs and the
artificial pairs.   This was repeated for 50 trials of artificial
absorbers for each method, and the $D$ values were again averaged
over those trials, and the probability that the
distributions are the same, $P(\langle D\rangle)$, was calculated.  
This test was repeated for the two extremes of the luminosity cuts in
Tables \arabic{ntab}a and \arabic{ntab}b, $L \ge 0.04L^*$\ and $L \ge
L^*$. 

\addtocounter{nfig}{1}
In Figure \arabic{nfig}, the distributions of impact parameter for
real and simulated absorbers are plotted
together, where
the simulated absorbers are presented for the sum of
50 trials,  normalized to the total number of
absorbers. Panels a and b show the $L \ge 0.04L^*$\ pairs for the two
methods, and panels c and d show the $L \ge L^*$\ pairs.
 In the upper panels, the differences between the two pairing methods can be
seen in the fact that the $\rho^{\Delta v}_{min}$\ method is slightly skewed
towards smaller $\rho$\ than the $r^{3D}_{min}$\ method, due to the fact
that the $r^{3D}_{min}$\ method chooses the closest galaxy in three-dimensions, which is
not necessarily the galaxy with the smallest impact parameter.  However, both
methods produce similar results for this test.  While in both methods
the real and random distributions appear to be very similar, 
the real absorbers in both cases tend towards smaller impact
parameters and do not have the same high $\rho$\ tail as in the random
distributions.  This can be seen in the resultant KS probabilities
where for the $r^{3D}_{min}$\ method, there is a 36\% probability the real
and random distributions are the same, and for the $\rho^{\Delta v}_{min}$\
method we find a 27\% probability.  In the lower panels, it is clear
that limiting the analysis to
only the most luminous galaxies introduces significant noise. The KS
probabilities bear out the visual impression that the impact parameter
distributions are both close to being random, with probabilities of
73\% and 60\% for the $r^{3D}_{min}$\ and the $\rho^{\Delta v}_{min}$\
methods, respectively.  The severity of the duplicity of
galaxy-absorber pairings, plus the tendency towards a random
distribution of impact parameters for more luminous galaxies highlights
the potential severity of survey selection effects, especially with
the high-redshift galaxy work. 

We then did the complementary experiment of looking for galaxies
that fall close to the line of sight but do not produce absorption
within the detection limit.  To do this, we selected
bright galaxies ($L^*$ or greater) that fell within $\rho
\le$~500\h75 kpc of a quasar LOS, and then searched for an absorber within
$\Delta v \le$~300 \kms\ of the galaxy velocity.  To ensure complete
velocity coverage for this search, the galaxy pathlength searched was
shortened to $1200 \le v \le 2700$ \kms.  If no absorber is
found, we can assign an upper limit to the equivalent width of the
possible absorption line using the 
3$\sigma$\ limiting equivalent width at that wavelength in the spectrum.
Of the five $L \ga L^*$\ galaxies falling within 
$\rho \le$~500\h75 kpc of the 10
lines of sight, only one of them matched an absorber within 300
\kms, suggesting a covering factor for $L^*$ of $\sim 20$\%.  For fainter
galaxies,  \cite{che98p} found a covering factor of 50\% for 
$L \ga 0.25L^*$\ for $\rho < 270$\h75\ kpc, and they suggest that for
fainter samples it should increase 
to 100\%.  For a more direct comparison, we consider galaxies in the
Virgo sample with $\rho < 270$\h75\ kpc and use $\Delta v$~ = 500 \kms\
which limits the search pathlength further to $1400 \le v \le 2500$ \kms.  
With these new constraints, we find 3/8 $L \ga 0.25L^*$\ galaxies to have
matching absorbers, giving a similar covering factor of 60\%.
However, for fainter galaxies, we find that only 4/18 $L \ga 0.04L^*$\
galaxies have a matching absorber, yielding a decrease in the covering
factor to 22\%. Despite the  small number statistics, we find a 
number of luminous galaxies in the Virgo region that do not cause
absorption, even when close enough to be considered a  ``physical
pair''.

\subsection{Galaxy-Absorber Correlations}
\addtocounter{nfig}{1}
One of the strongest pieces of evidence to associate \Lya\
absorbers with individual luminous galaxies is claimed to be the observed
 anticorrelation between rest
equivalent width and impact parameter ( e.g. \cite{tri98},
\cite{che98}).  In Figure \arabic{nfig},  we plot 
impact parameter, $\rho$, of the galaxy vs. rest equivalent width,
$W_r$,  of the absorber.  The top two panels (a \& b) show the absorbers
when matched to galaxies $L \ge 0.04L^*$, for the two methods,
$r^{3D}_{min}$\ and $\rho^{\Delta v}_{min}$, respectively, and the
lower two panels 
(c \& d) show the same for pairs matched to $L \ge 0.25L^*$\ 
galaxies. Our identified absorbers are indicated by
triangles, and the bright galaxies ($L > L^*$) with no detected
absorption within $\Delta v = 300$ \kms are
shown as upper limits in 
equivalent width.  The anticorrelation
relationship from \cite{che98p} is also plotted as the solid line, and
with that group's ``physical pair'' limit of $\rho = 270$\h75\ kpc
designated by the dashed line.

We note that for both pairing methods, the upper
panels which include fainter galaxy counterparts appear  consistent
with the anticorrelation 
line, whereas the lower panels of brighter galaxies are not.
This is interesting because the galaxies originally used to fit the
function, from \cite{che98p}, only extend to  $0.25L^*$ (with 3/35
exceptions, 2 of which are at the lowest redshifts).  If our sample
is similarly limited to $L \ge 0.25L^*$, as can be seen in Figures
\arabic{nfig}c and \arabic{nfig}d, for a given $W_r$ line, we identify
absorbers with galaxies at impact parameters  much larger than the
anticorrelation of \cite{che98p} would predict.

Removing the magnitude restrictions, our absorbers are invariably
identified with 
fainter galaxies at smaller impact parameters for both pairing methods
(seen in Figures \arabic{nfig}a and \arabic{nfig}b).  To some degree,
this can be expected for randomly distributed absorbers, which 
in all of our earlier random trials chose galaxies in the more
luminous galaxy sample at typically larger impact parameters than in
the fainter sample.  However, it is difficult to disentangle such random 
effects from the real physical association of the absorbers and galaxies.
\cite{che98p} define a model where the
absorption is caused by an extended gaseous halo around the galaxy,
and so a ``physical pair'' is an galaxy-absorber pair for which the
galaxy-absorber
cross-correlation function is greater than 1, and $\rho < 270$\h75\ kpc
(\cite{lan98}). In this scenario, the $W_r$--$\rho$\ anticorrelation
naturally arises for $\rho < 270$~kpc, and an absorber associated with
a galaxy at $\rho > 270$~\h75~kpc is caused by an undetected galaxy at
smaller impact parameter that is correlated with the detected galaxy.
In panels a and b of Figure \arabic{nfig}, 2/11 and 7/11 of our
absorbers fall within $\rho <$~ 270\h75 kpc, respectively, although
some of these galaxies still fall at impact parameters too large for
such low luminosity galaxies.  
The remaining pairs fall above this limit, which according to \cite{che98p}
means these galaxies are correlated with undetected (i.e. lower
luminosity) galaxies at smaller impact parameter.  If this physical
picture is correct, then roughly one-third to  half of our absorbers
are caused by galaxies somehow overlooked in our sample or by galaxies
falling below $0.04L^*$.

Combining data from the literature on the $W_r$--$\rho$\
anticorrelation in
\addtocounter{nfig}{1}
Figure \arabic{nfig}, we see that all data sets mostly find 
galaxy-absorber pairs at the highest impact parameters, with
the exception of \cite{che98p} which only find pairs for $\rho \lesssim$~
270\h75 kpc, by construction.  Here again we plot log
$W_r$\ vs. $\rho$\, with the solid line indicating the \cite{che98p}
best-fit and the large triangles are the data from the $\rho^{\Delta
v}_{min}$\ method from this
paper.  The open triangles are the galaxy-absorber pairs when
matching only to $L \ge 0.25L^*$\ galaxies, the filled triangles are
the pairs when matching to $L \ge 0.04L^*$\  galaxies, and the dotted line
connects the galaxy data points for the same absorber. The other data
included are from the literature, with filled symbols indicating
galaxies with $L < 0.25L^*$\ and open symbols galaxies with $L >
0.25L^*$.  

At high $\rho$, there is no measurable anticorrelation between $W_r$\
and $\rho$.  At high $\rho$, it is easy to select a bright galaxy counterpart
when there is in fact a fainter counterpart at smaller $\rho$, as observed for
about half of our absorbers.  However, of all the $L > 0.25L^*$ points
plotted, 8/11 from this paper, 4/5 from \cite{mor93p}, 5/5
from \cite{tri98}, and 0/3 from \cite{che98p} fall at $\rho >
270$\h75 kpc, which is too large to be caused by an extended
halo of such low luminosity galaxies.  
With the \cite{che98p} points removed, Figure \arabic{nfig}
would resemble more of a scatter plot.
As pointed out by \cite{tri98p}, there are a number of $W_r > 0.3$\AA\
 absorbers 
from \cite{che98p} sample with no counterpart galaxies, which could be
associated with galaxies beyond their search radius. 
If those absorbers fall at large $\rho$, the anticorrelation would be
further weakened.

Another feature of Figure \arabic{nfig} is upward trend of the $\rho
\la 200$\h75 kpc ($\simeq 160h_{100}^{-1}$\ kpc)  points,
 while the $\rho \ga 200$\h75 kpc points show no
real trend with $W_r$.  This division has been suggested as an
equivalent width effect ( c.f. \cite{sto95}), with weaker lines arising
from a different physical process.  However, the division in Figure
\arabic{nfig} does not correspond to any hard equivalent width cutoff,
but could correspond to the $W_r$,$\rho$\ position of a predominant
transition in gas phase as calculated from simulation (\cite{dav98};
see the next section of the paper.) 

In framing our data in the context of the current literature, we have
discovered significant problems with uniquely assigning an individual
galaxy as associated with an absorber.  In comparing two pairing
methods ( $r^{3D}_{min}$\ and $\rho^{\Delta v}_{min}$), 
which produce very different
galaxy-absorber pairings, we find there is no way to statistically
distinguish between the two methods as to which is a better
prescription.  Furthermore, each method is also very sensitive to 
magnitude completeness, since for three different absolute magnitude limits,
we could almost always find a fainter galaxy at smaller impact
parameters (an effect predicted by \cite{lin98}).  This is of
particular concern for high redshift \Lya\  
work, since the largest and brightest galaxies tell a different story than
going further down the luminosity function.  Moreover, these selection
effects in luminosity do not address further ambiguities due to
surface brightness selection effects (c.f. \cite{rau96}).
Our $\sim 60$\% covering factor for $L > 0.25L^*$\ galaxies
is consistent with previous estimates, but, contrary to some
predictions, yields smaller covering factors for fainter limits. 
This
is contrary to expectations ({\em c.f.} \cite{che98}) that by going to
faint enough magnitudes, every absorber can be reasonably associated
with a galaxy.  Limiting our data to galaxy-absorber pairs of limiting
magnitude 
similar to those in the literature ($L > 0.25L^*$), our data show no
anticorrelation between $W_r$\ and $\rho$.  Extending that limit to
intrinsically fainter galaxies does induce a $W_r$--$\rho$\
correlation, but these fainter galaxies have correspondingly
smaller halo sizes.  
Nothing in our data would specifically lead us to associate \Lya\
absorbers preferentially with luminous galaxies on halo size scales.

\section{SUMMARY}

The observation of low column density hydrogen absorbers has emerged as 
a powerful cosmological tool. Insights from theory and hydrodynamic
simulations give the basic picture: the \Lya\ forest at high 
redshift is the main repository of baryons in the universe and it is
a relatively unbiased tracer of the underlying dark matter distribution
(Rauch 1998). Diffuse and highly ionized hydrogen forms a ``cosmic web''
of large scale structure (Bond \& Wadsley 1998). As the universe expands
and evolves, much of the gas is heated and shocked or collapses into
galaxies and larger structures. The number density of absorbers declines
toward low redshift, and they can only be studied from space. However, at
low redshift it is possible to make direct comparisons with the galaxy
distribution.

We have studied \Lya\ absorption along ten sightlines in the 
direction of the Virgo cluster. The resulting sample of 139 lines
above a detection limit of 3$\sigma$ is the largest yet studied in
the local universe ($z \lesssim 0.2$). At the resolution of the GHRS
observations (200 \kms), essentially all of the absorption lines 
are unresolved. The number density of lines above a rest equivalent width
of 0.24 \AA, $dN/dz = 38.3 \pm 5.3$, agrees well with the the measurement from
the Quasar Absorption Line Key Project (Weymann et al. 1998). There is
marginal evidence for cosmic variance in the number of absorbers detected
among the ten sightlines. Down to a limit of 0.1 \AA, the line statistics 
are consistent with the study of two sightlines by Tripp et al. (1998) 
and with an extrapolation of the relationship for $dN/dW$ fitted to 
the Key Project data. The upturn in line density to $dN/dz = 250 \pm 40$ 
above 0.020 \AA\ observed by Shull (1997) must set in at column densities 
below 10$^{13}$ cm$^{-2}$.

We looked for clustering among the \Lya\ absorbers by carefully
modelling the varying sensitivity and redshift pathlength of the ten
different sightlines. Resolution and potential blending effects prohibit 
a search for clustering on velocity scales less than 250 \kms. We
detect an excess of nearest neighbor line pairs on velocity scales of
250-750 \kms\ at a 95-98\% confidence level. There is no significant
excess on larger scales that might correspond to the velocity dispersion
of a rich cluster. The hypothesis that the absorbers are randomly
distributed in velocity space can be ruled out at the 99.8\% confidence
level. No two-point correlation power is detected ($\xi < 1$ with 95\%
confidence), in marginal disagreement with Tripp et al. (1998). 
We do not have the resolution or line statistics to look for the
small scale clustering signal predicted by Cen et al. (1998).
We find \Lya\ absorbers to be less clustered than bright galaxies, in 
accord with Grogin \& Geller (1998). Absorber-absorber correlation
amplitude on scales of 250-500 \kms\ is 4--5 times smaller than
galaxy-galaxy correlation amplitude.

A detailed comparison between absorbers and nearby galaxies produces
results that are difficult to interpret. We restrict the comparison to
the eleven \Lya\ lines in the radial velocity range 900--3000 \kms.
Over the contiguous volume threaded by the ten sightlines, the galaxy
sample is complete to $M_B = -16$. Absorbers lie preferentially in regions 
of intermediate galaxy density. It is not possible to uniquely assign
a galaxy counterpart to each absorber, even if it is assumed that galaxies
are surrounded by spherical halos that can cause absorption
(CLWB). Ambiguities arise due to the uncertain mapping of redshift
into  
distance and due to the large number of low luminosity galaxies for every 
luminous galaxy. We find multiple or non-unique absorber counterparts in 
7/11 cases. The complete galaxy sampling allows us to do the converse 
experiment --- to look for absorbers at small impact parameters from 
luminous galaxies. A halo covering factor to \Lya\ absorption of 
20\% is deduced for galaxies of $L>L^{*}$ and impact parameters
$\rho<500$ kpc.
For somewhat fainter galaxies, $L>0.25L^{*}$, with $\rho<270$ kpc, the
covering factor is 60\%.
In general, there is no behavior in this sample that
specifically implicates luminous galaxy halos in causing the absorption.

Some insight into the physical state of the absorbers at low redshift
comes from a comparison with the recent hydrodynamic simulations of Dav\'{e} 
et al. (1998; see also Reidiger et al. 1988; Theuns et al. 1998). They 
found that the dynamical state of an absorber --- expanding or collapsing,
shocked or unshocked --- depends mainly on the overdensity of the gas,
$\rho_{gas}/\bar{\rho}_{gas}$. With decreasing redshift and universal 
expansion, 
a given column density selects absorbers which are increasingly overdense 
and which have progressively more advanced dynamical states. Figure 14$a$
shows the rest equivalent width distribution of our sample of absorbers.
In Figure 14$b$, the column density is estimated from rest equivalent width 
using a curve of growth and assuming a Doppler parameter of $b = 30$ \kms.
The vertical dashed line marks the column density below which the metal
abundance of the absorbers falls sharply (Lu et al. 1998). We can use the
relation between gas overdensity and column density from the simulations
(in a $\Lambda$-dominated CDM model) to estimate the overdensity of our
absorbers, as shown in Figure 14c (the conversion is relatively insensitive
to the assumed Doppler parameter). The association of overdensity with the 
phase of the gas --- diffuse IGM, shocked IGM, or condensed --- is crude,
because we have no estimate of the temperature of the gas and an increasing
fraction of the moderate overdensity gas is shocked at decreasing redshift.

Because only the weakest lines in the sample have equivalent widths and 
overdensities that correspond to overdensities of a few,
we infer that we are not in general probing the diffuse IGM. Tracing the
evolution of the most diffuse structures
from redshifts 3, 2, 1 and 0 means examining absorbers with column 
densities of $\log N_{HI} \approx 14.5$, 13.8, 13.2 and 12.7, respectively. 
For $b=30 $ \kms\ this corresponds to rest equivalent widths of about 0.39, 
0.21, 0.075, and 0.026 \AA. Possibly the sharp rise in $dN/dW$ seen by
Shull (1997) below 10$^{13}$ cm$^{-2}$ represents this intergalactic
population. Most of the absorbers in the Virgo sample have overdensities
ranging from a few up to $\sim 100$ and have not yet collapsed into 
galaxies. 

Figure 15 revisits the distribution of equivalent width and
galaxy impact parameter for our data (circled) and other studies in
the literature. The approximate regions of the three absorbing gas phases 
are superimposed (Dav\'{e} et al. 1998). An anticorrelation between $W_r$ 
and $\rho$ is anticipated because of the way gas traces the underlying mass 
distribution of large-scale structure. The strongest absorbers arise from 
the denser gas near galaxies, the majority of absorbers with 30\h75 $\la 
\rho \la$~270\h75 kpc arise from shock-heated gas near galaxies, and 
absorbers with $\rho \ga$~270\h75 kpc are associated with a cooler, 
diffuse gas component. We cannot be sure that the segregation of 
observations to the upper regions of this plot is physically
meaningful, since we have found ambiguities in the assignation of
absorber counterparts, and some absorbers identified with luminous
galaxies at large impact parameters might just as well be identified
with lower luminosity galaxies at smaller impact parameters. The
galaxy counterparts to our limited sub-sample of absorbers with radial
velocities of 900--3000 \kms\ are all at $\rho \ga 70$\h75 kpc, too
far to be bound to a halo potential. Our detection of weak clustering of
the absorbers is consistent with gas that loosely traces large scale
structure.

This study, and all others to this point, have been limited by the meager
statistics of absorbers at low redshift. Future observations will allow
us to increase the number of sightlines and revisit the issue of clustering.
It will be very interesting to study the relation between clustering
amplitude and gas overdensity. It may even be possible to identify a set 
of local absorbers at low column density that are primeval and completely
unrelated to the space distribution of bright galaxies. We also plan to 
make direct comparisons between observations and hydrodynamic simulations,
aiming to use \Lya\ absorbers for cosmological tests of increasing
sophistication. 

\acknowledgments
This research was supported by NASA and STScI under the G0 grant 
for program 5947. 
We are grateful for excellent advice and support from STScI staff.
We acknowledge useful discussions with Romeel 
Dav\'{e}, Craig Foltz, Buell Jannuzi, and John Stocke.
We thank the referee, Simon Morris, for his constructive
comments on this paper.



\clearpage

\begin{figure}
\epsscale{0.83}
\plotone{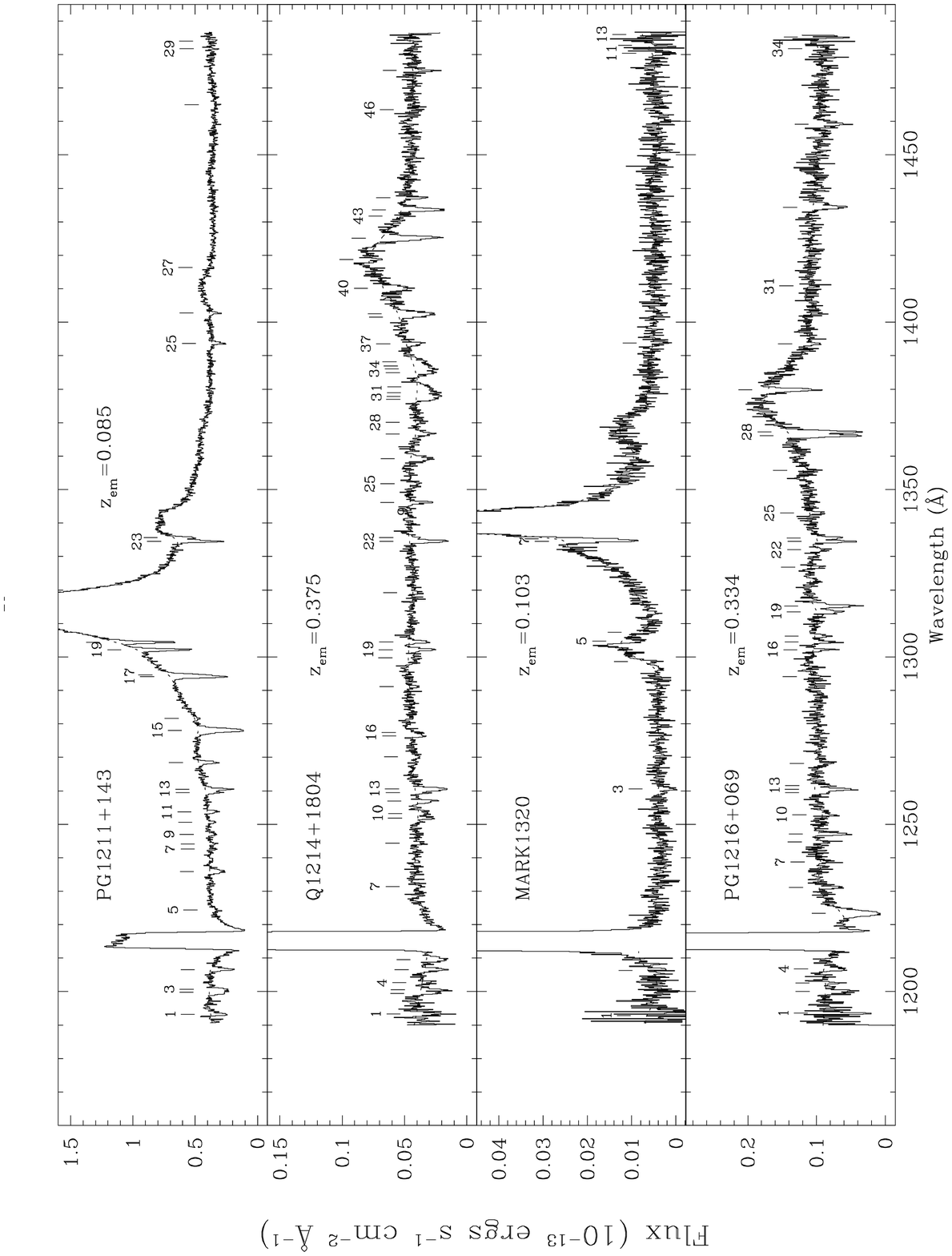}
\figcaption{Spectra of the 12 quasars in the sample are
displayed with the fitted continua (which omits the region near
the geocoronal \Lya\ line) overplotted.  
Absorption lines are indicated with tickmarks and are numbered
every two or three depending on the line density.
Each panel is labelled with the quasar name and its emission redshift.
Flux is plotted on the {\it y}-axis in units of 
$10^{13}\ \rm ergs\ s^{-1} cm^{-2} \AA^{-1}$}
\end{figure}
\epsscale{1.00}

\clearpage

\begin{figure}
\epsscale{0.80}
\plotone{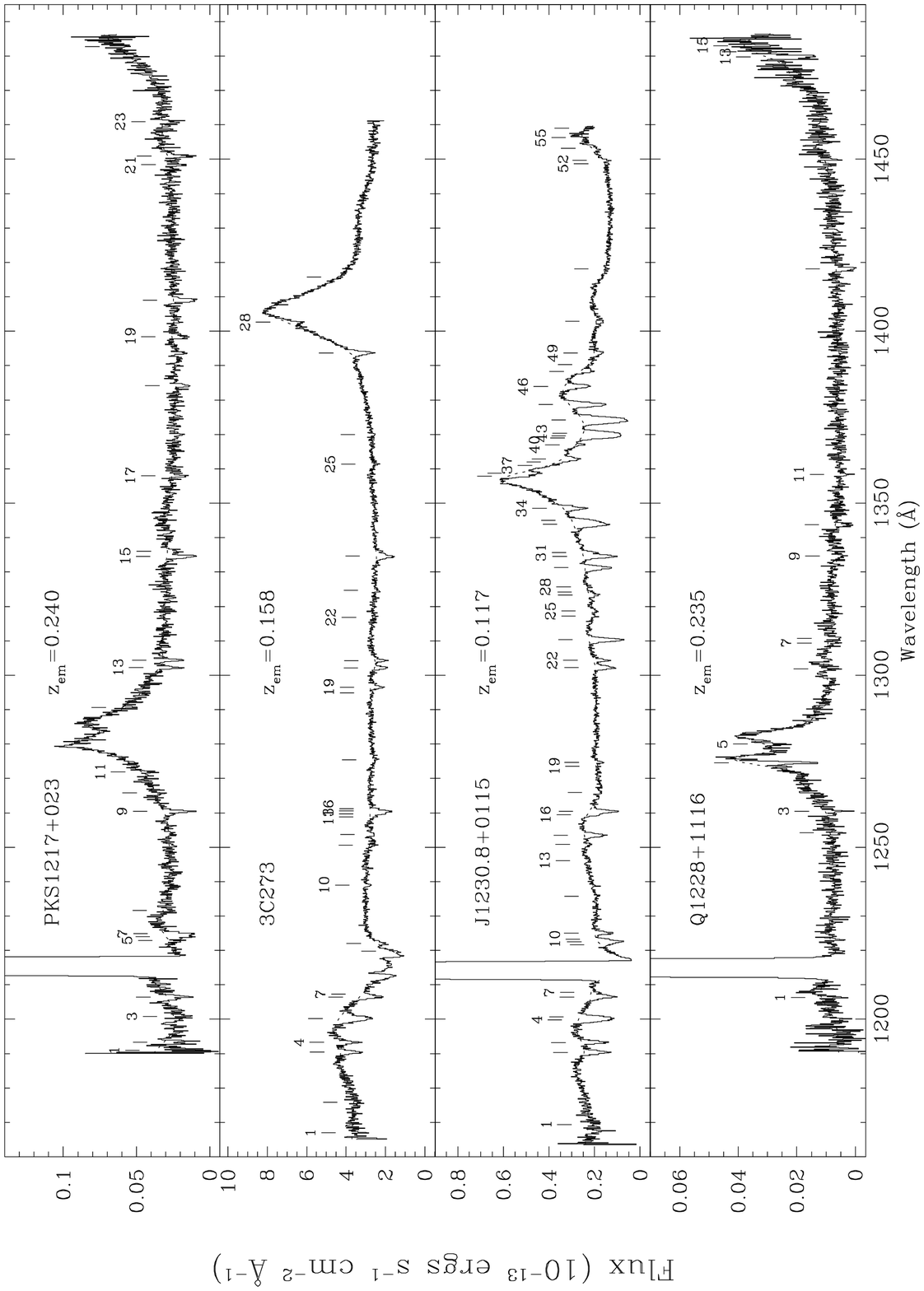}
\end{figure}
\begin{figure}
\epsscale{0.80}
\plotone{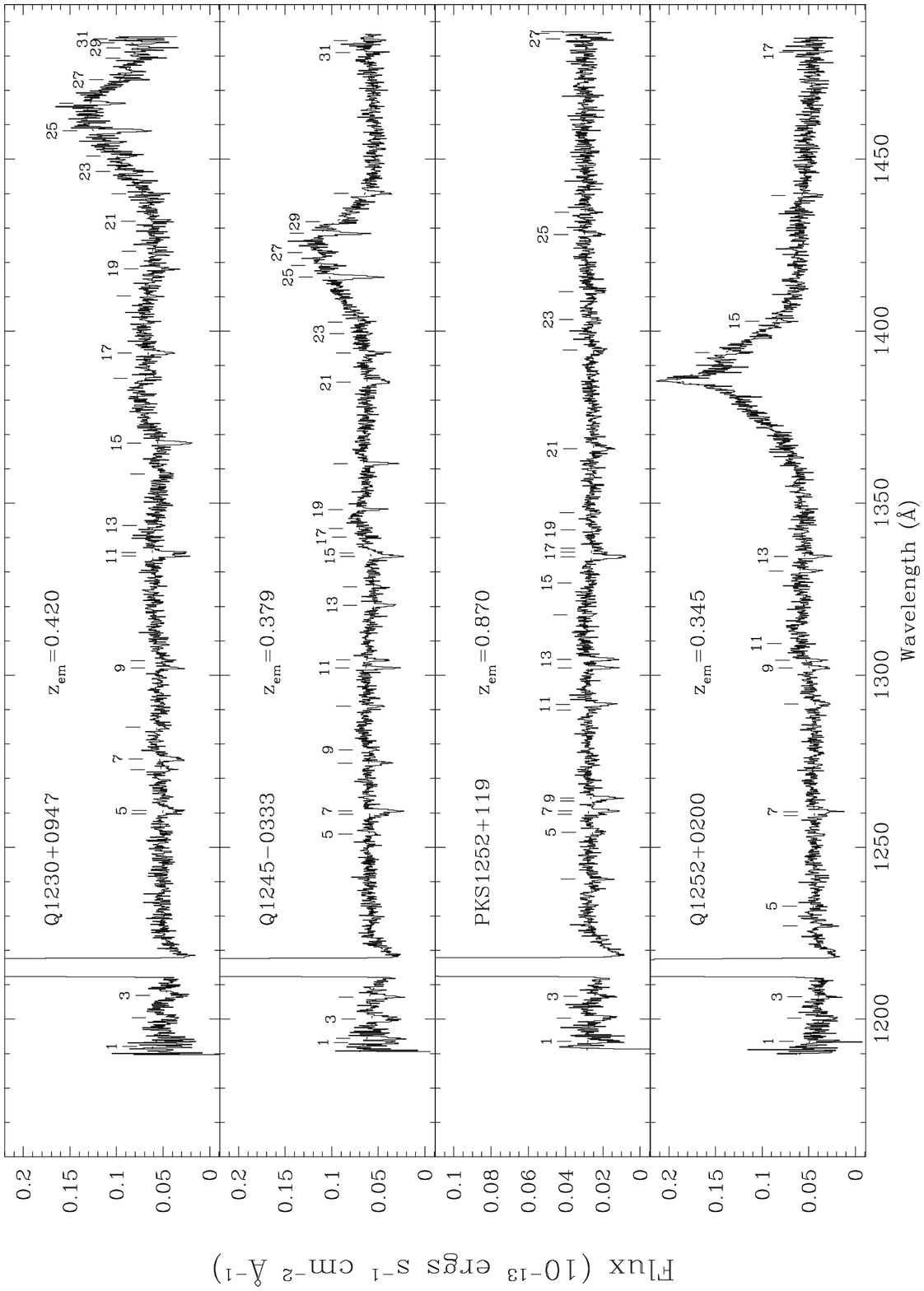}
\epsscale{1.00}
\end{figure}

\clearpage
\begin{figure}
\epsscale{0.75}
\plotone{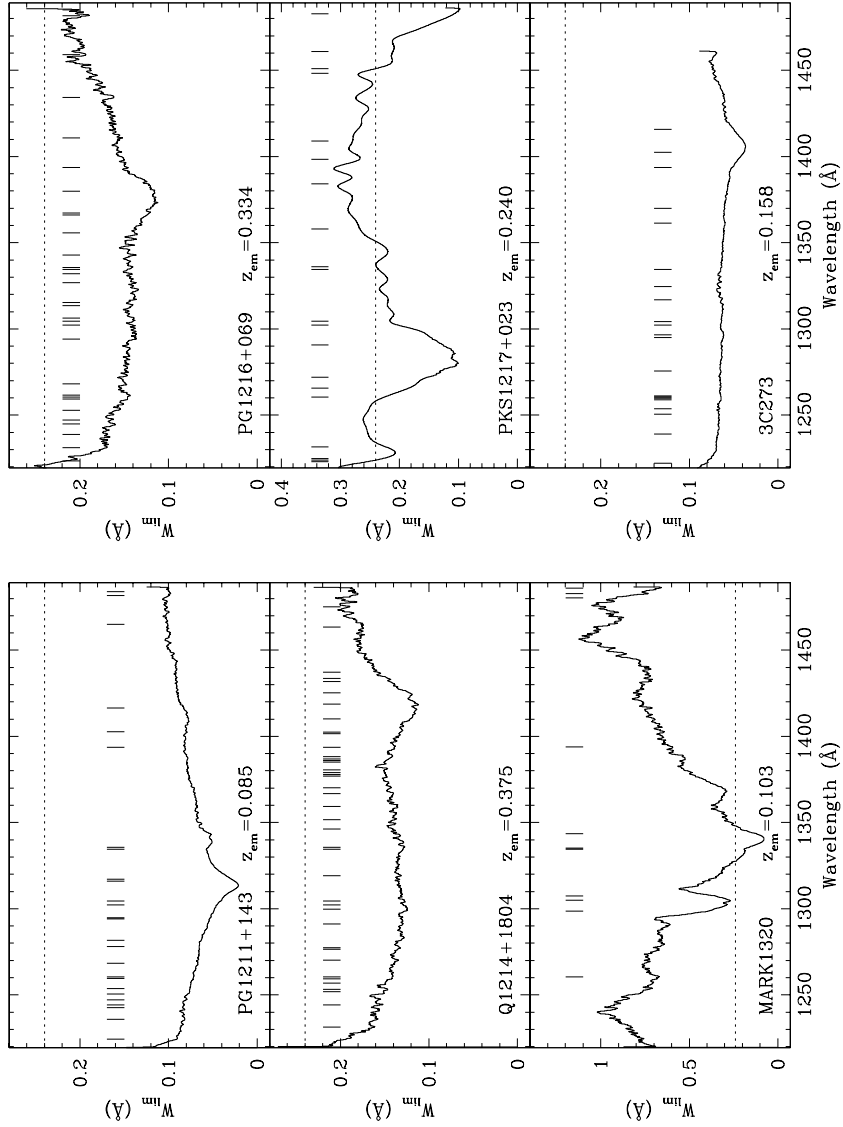}
\epsscale{1.00}
\figcaption{The 4.5$\sigma$ detection limit (rest equivalent width)
for each spectum over the wavelength range corresponding to $0.003<z<0.225$ is
shown by the solid curve.  The dotted line indicates the 4.5$\sigma$
completeness level of 0.24 \AA\ used by Jannuzi et al. (1998). The tickmarks
show the location of Ly$\alpha$ absorbers. }
\end{figure}

\clearpage
\begin{figure}
\epsscale{0.83}
\plotone{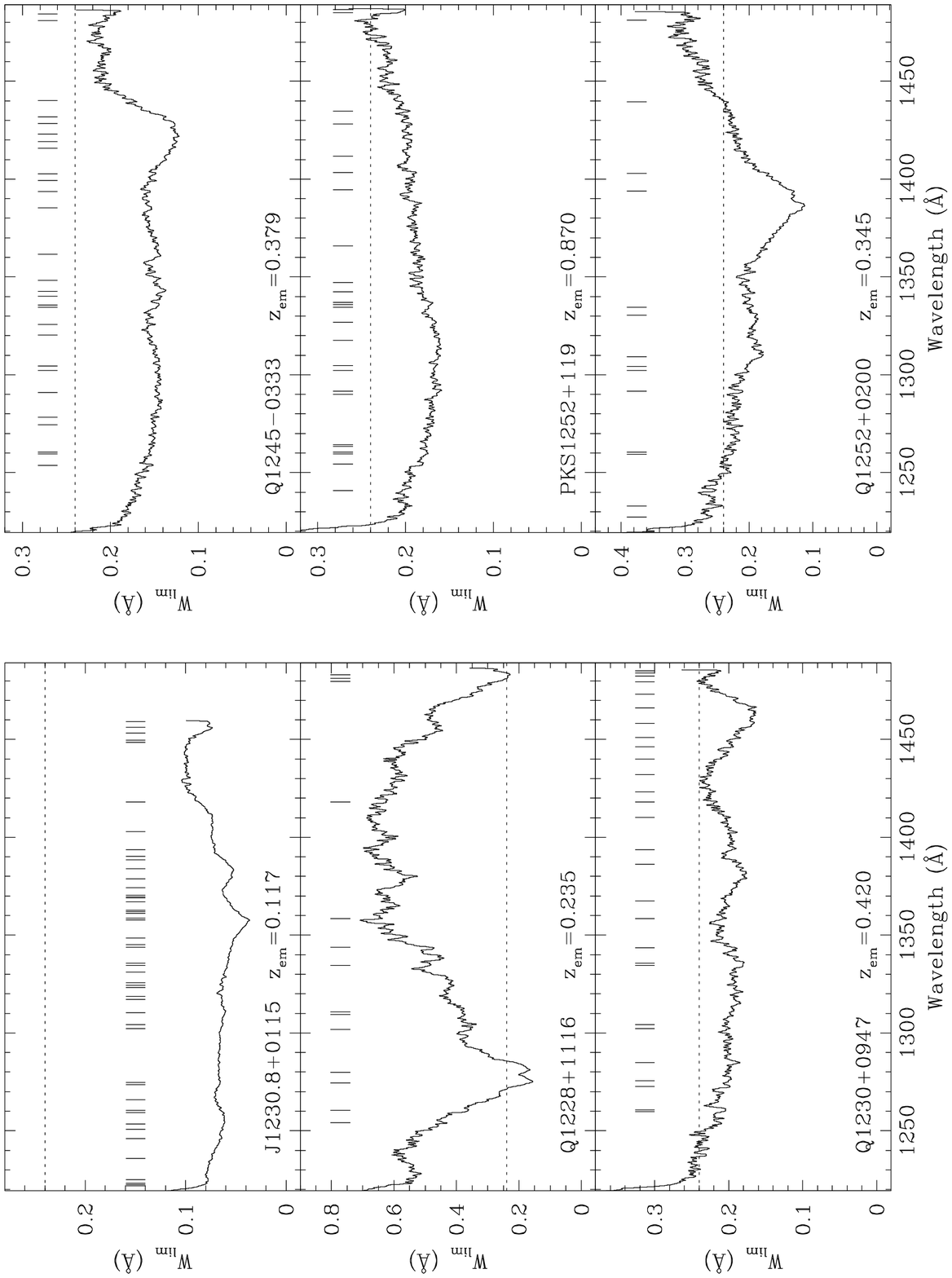}
\epsscale{1.00}
\end{figure}

\clearpage
\begin{figure}
\epsscale{0.83}
\plotone{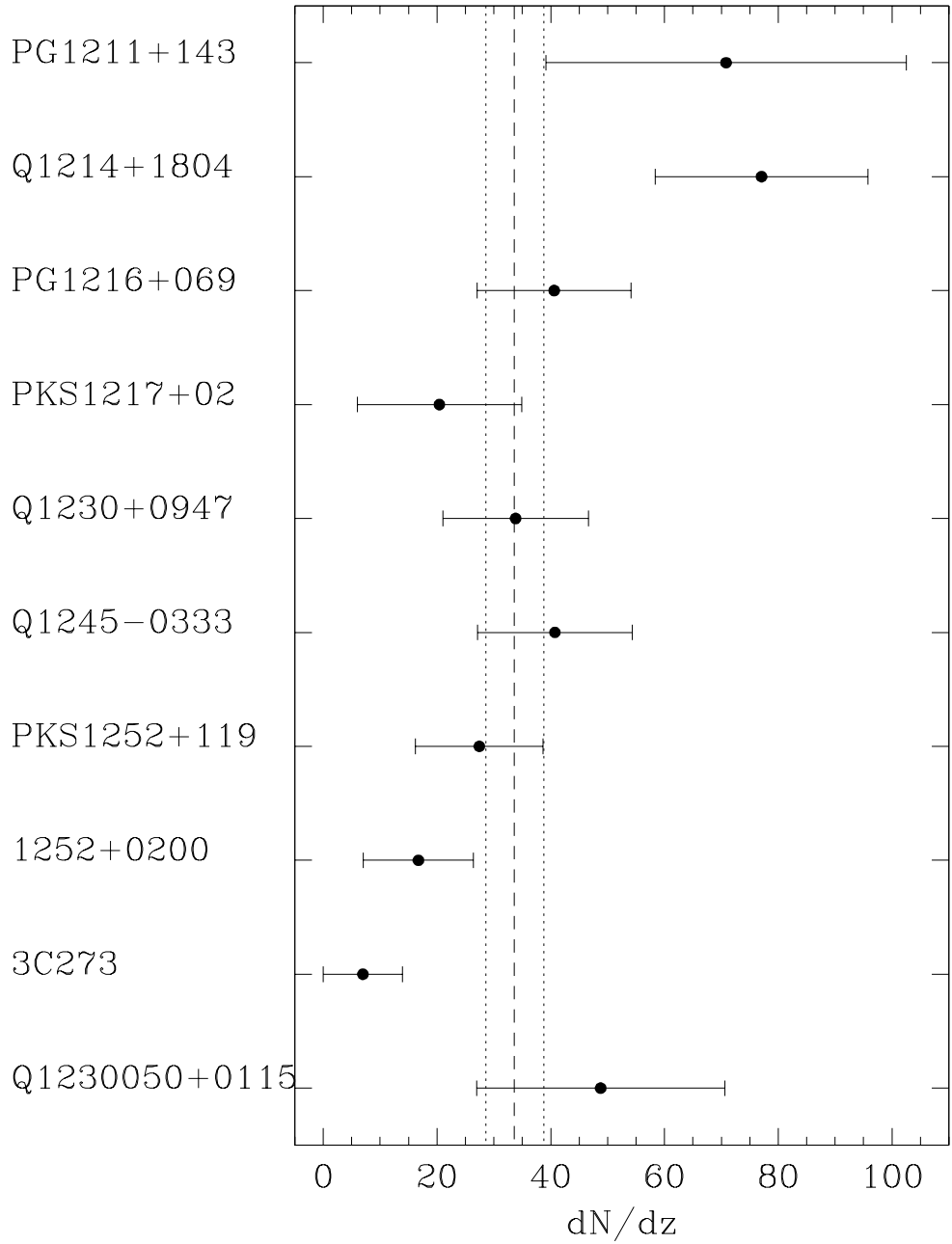}
\epsscale{1.00}
\figcaption{The number of absorbers per interval redshift for each
of the lines of sight individually.  As seen in Figure 4, the  average
value agrees closely with Weymann et al. (1998), although two lines of sight
differ by more than two standard deviations.}
\end{figure}

\clearpage
\begin{figure}
\plotone{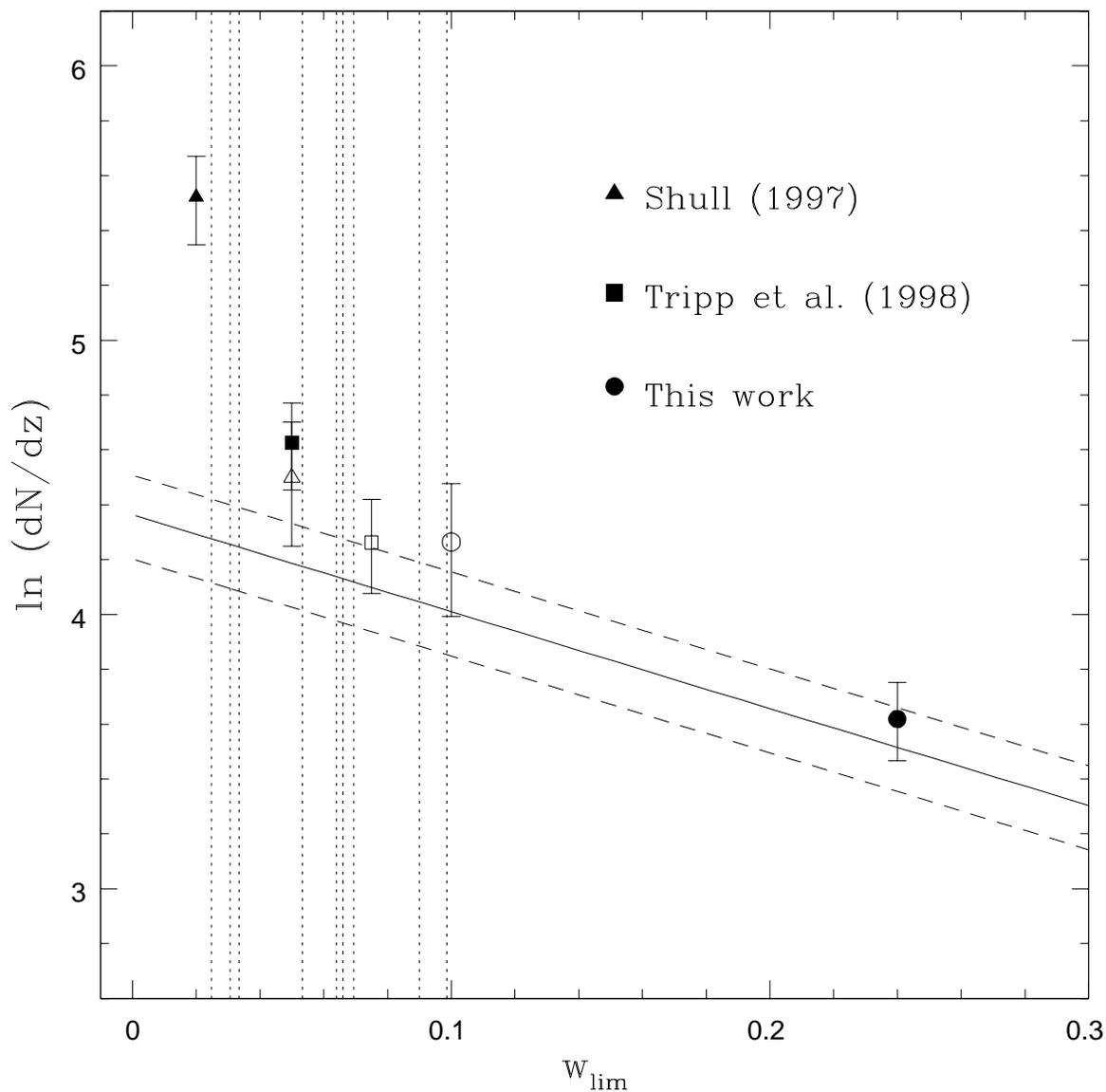}
\figcaption{The number of \Lya\ absorption lines per redshift interval
as a function of completeness limit.  The solid line is the fitted distribution
from the Quasar Absorption Line Key Project (Weymann et al. 1998), 
and the dashed 
lines are computed using the $1\sigma$ errorbars in $\gamma$ and $(dN/dz)_0$. 
The dotted lines represent the highest sensitivity of each GHRS spectrum.  The 
solid symbols are values quoted for each of the studies made as noted in the 
legend.  The open symbols are values quoted for some subset of the samples in 
each study, as described in the text.}
\end{figure}

\clearpage
\begin{figure}
\epsscale{.70}
\plotone{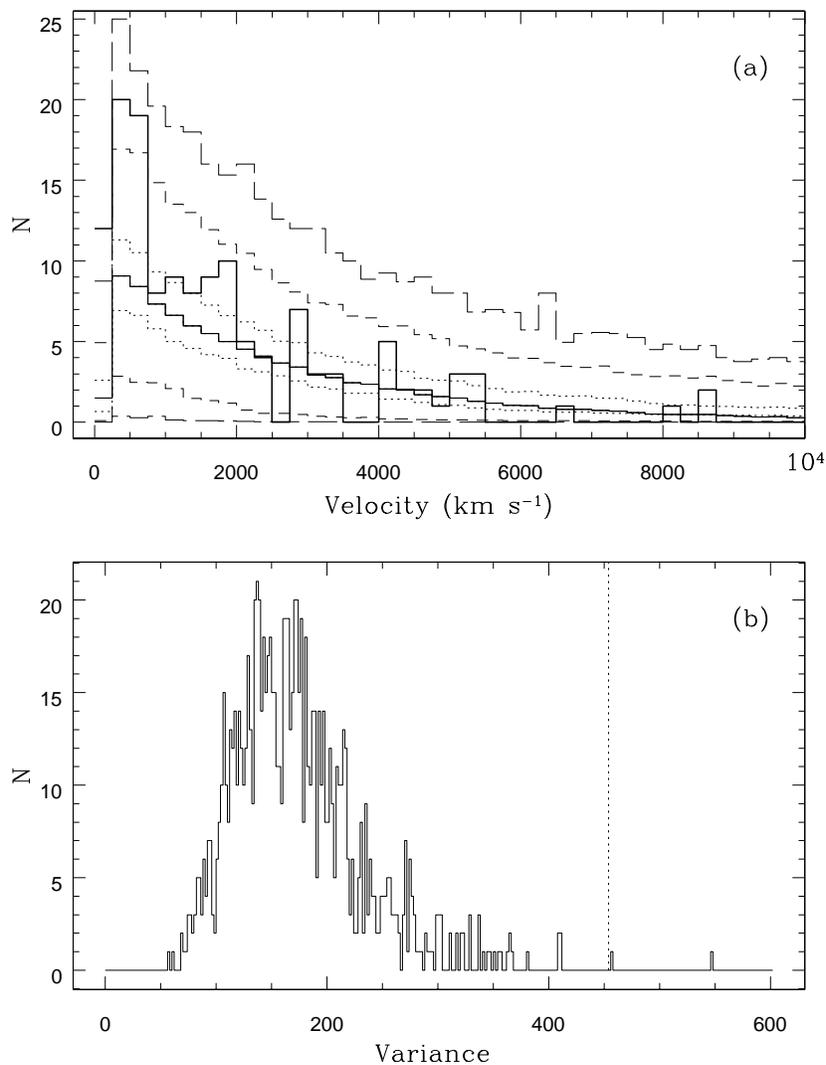}
\epsscale{1.00}
\figcaption{ {\bf (a)} The observed nearest neighbor distribution is shown
in the heavy solid line. The mean distribution expected from a random distribution
of absorbers is shown by the solid line.  The dotted and dashed lines indicate
the 68\%, 95\% and 99\% confidence intervals on the random distribution. Note that the strongest departures
from a random distribution are the two lowest velocity bins not affected by
resolution. {\bf (b)} The distribution of the variance of the mean expected
distribution and the expected distribution for each realization.  The dotted
line shows the variance of the observed and mean expected distributions.
This implies the observed distribution has a small probability ($\le 1$\%)
of having been drawn from a random velocity distribution.}
\end{figure}

\clearpage
\begin{figure}
\plotone{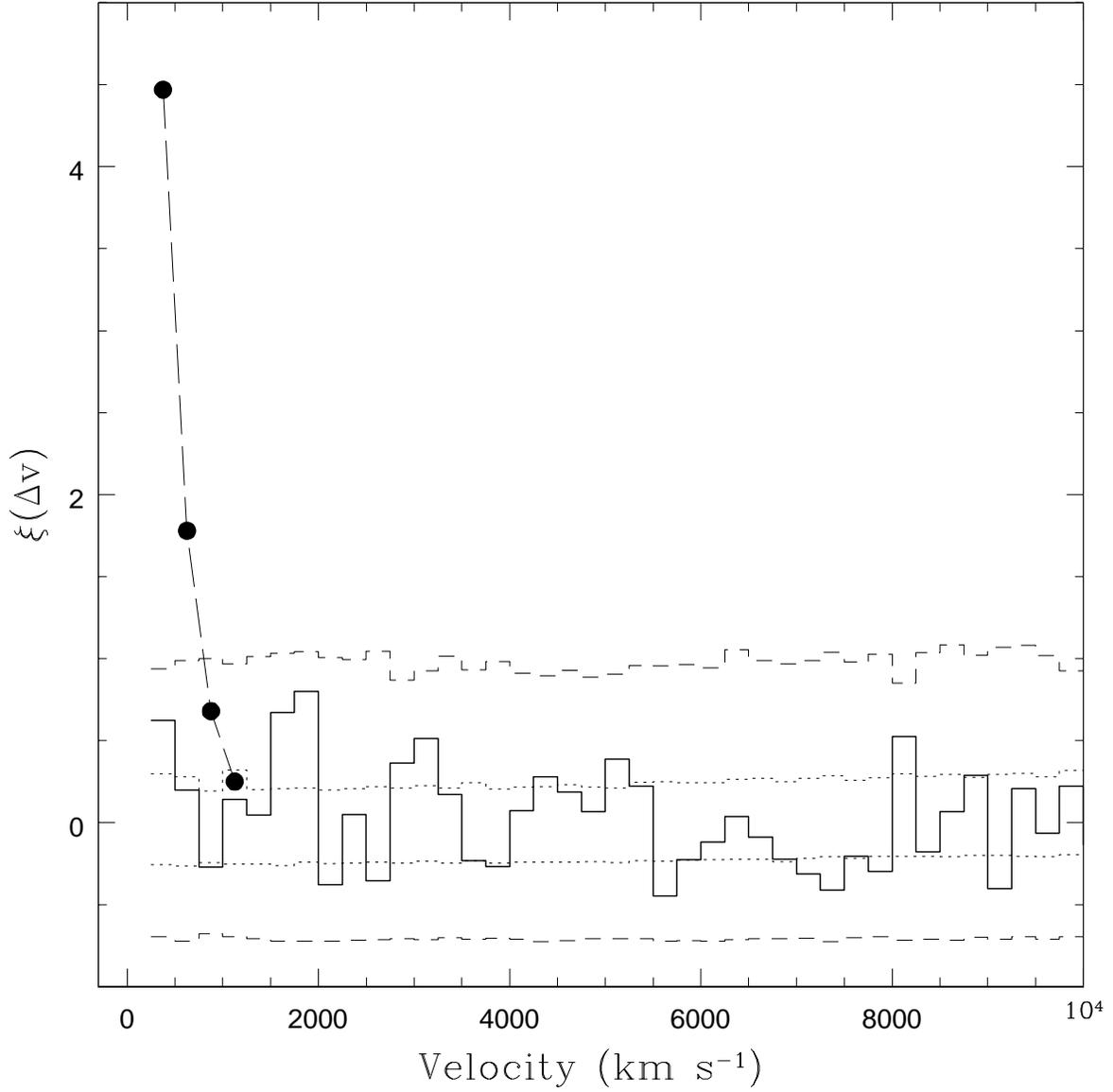}
\figcaption{The two-point velocity correlation function for the \Lya\ absorbers.
The dotted and dashed lines are the 68\% and 95\% confidence intervals for
a random distribution of absorbers.  The
black dots are the two-point correlation function for galaxies (Davis \&
Peebles 1983)
Data for the smallest bin ($<250$ \kms) is omitted because of resolution
limitations in the \Lya\ sample.}
\end{figure}

\clearpage
\begin{figure}
\plotone{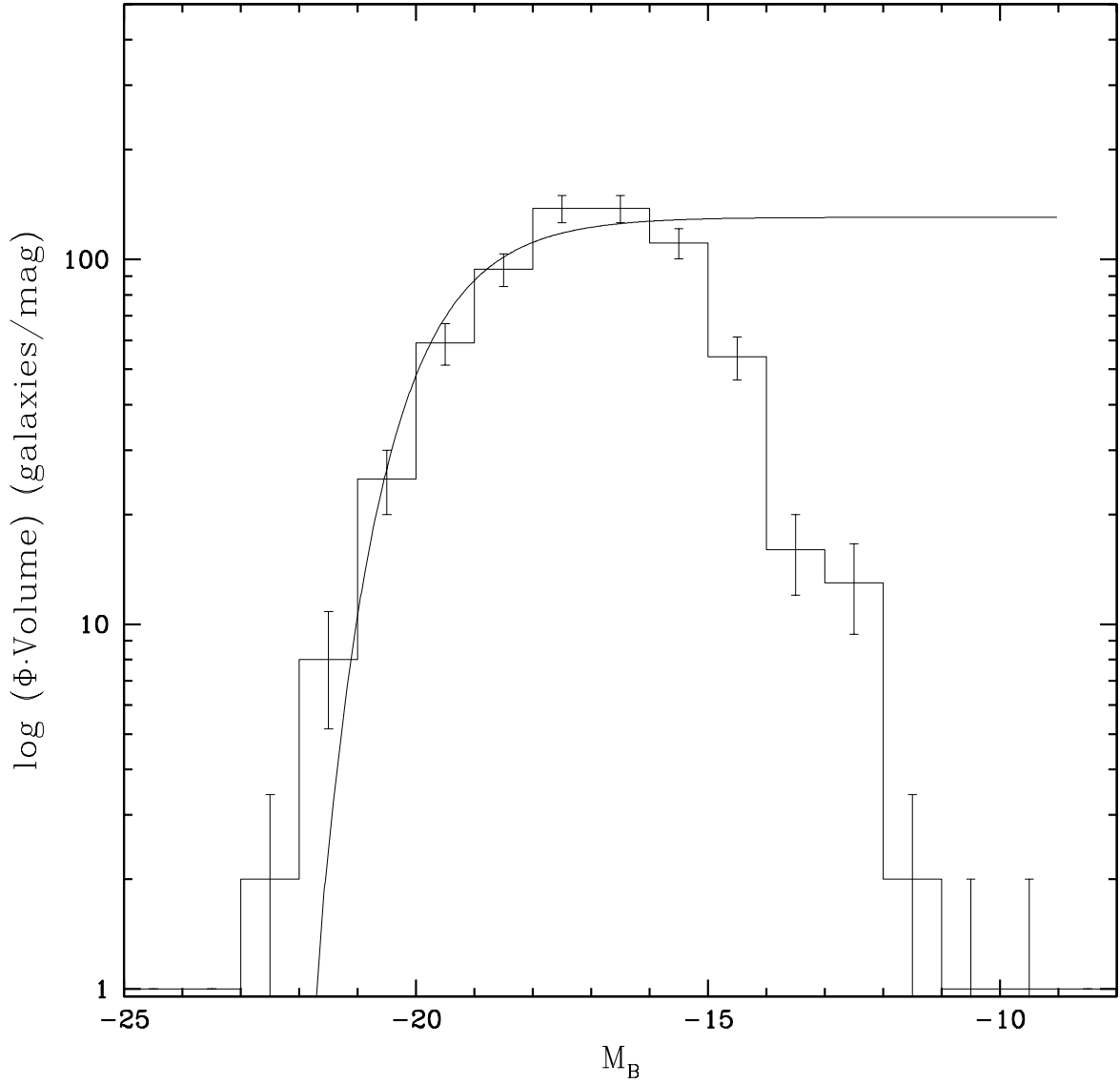}
\figcaption{Luminosity function of Virgo sample of galaxies,
spanning $600 < v < 3000$\ \kms, plotted in log($\Phi \cdot$\ Volume),
or counts/magnitude.  The error bars are Poisson.  The curve is
a Schechter luminosity function using average values for the Local
universe of $M_B^* = -20$, $\alpha = -1.0$, and arbitrarily
normalized  to fit the turnoff.}
\end{figure}

\clearpage
\begin{figure}
\epsscale{1.00}
\plottwo{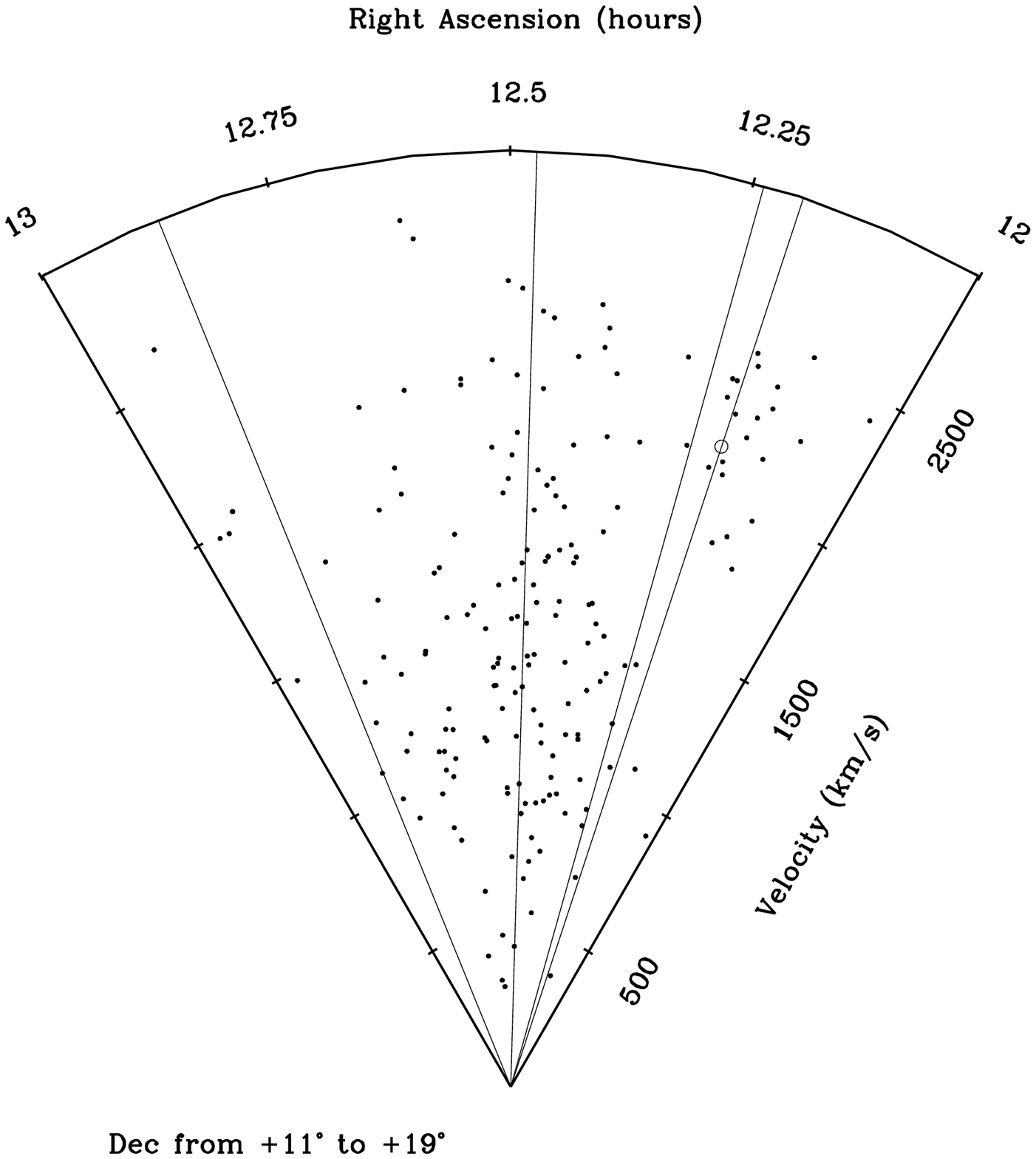}{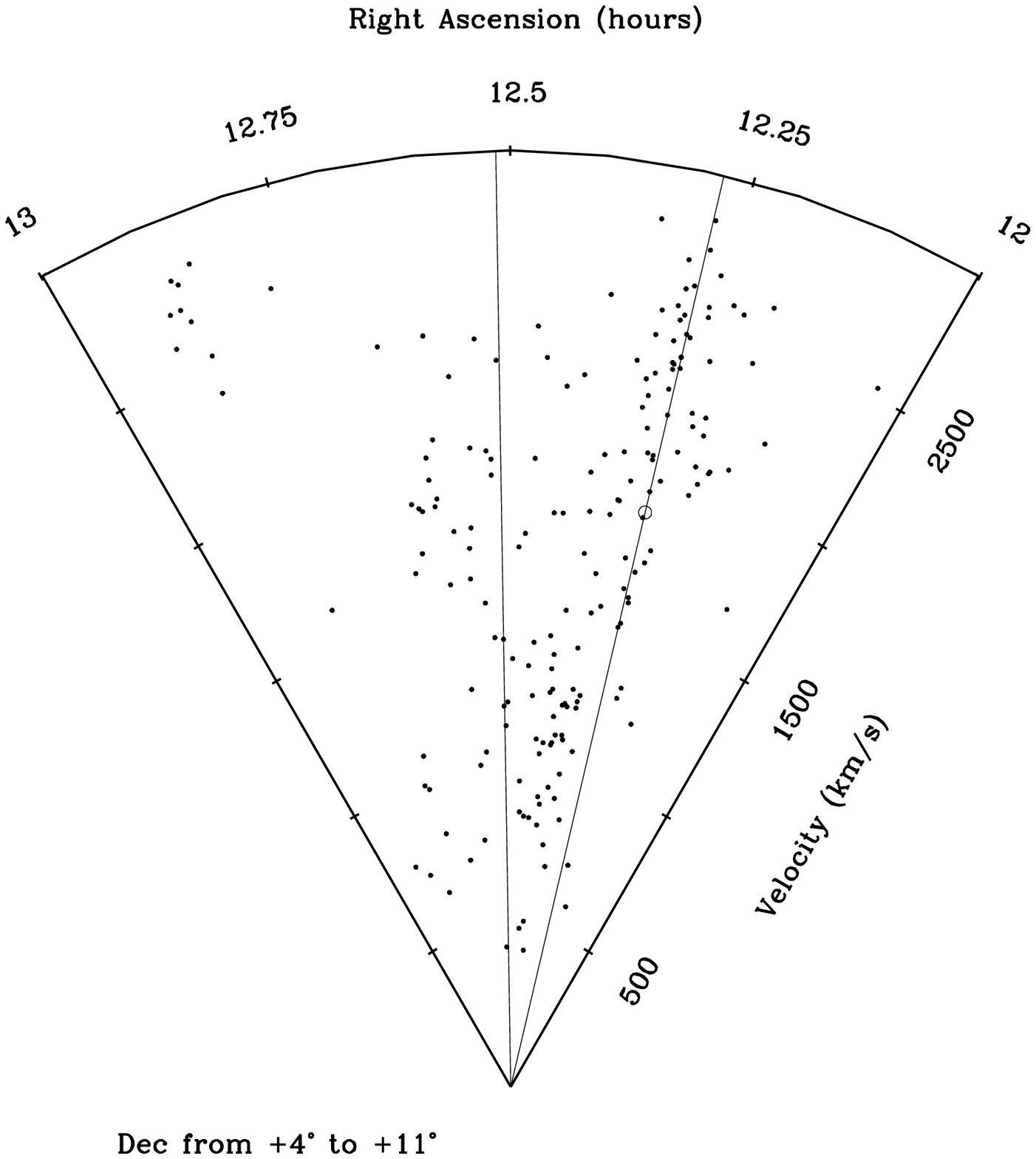}
\end{figure}
\begin{figure}
\epsscale{.45}
\plotone{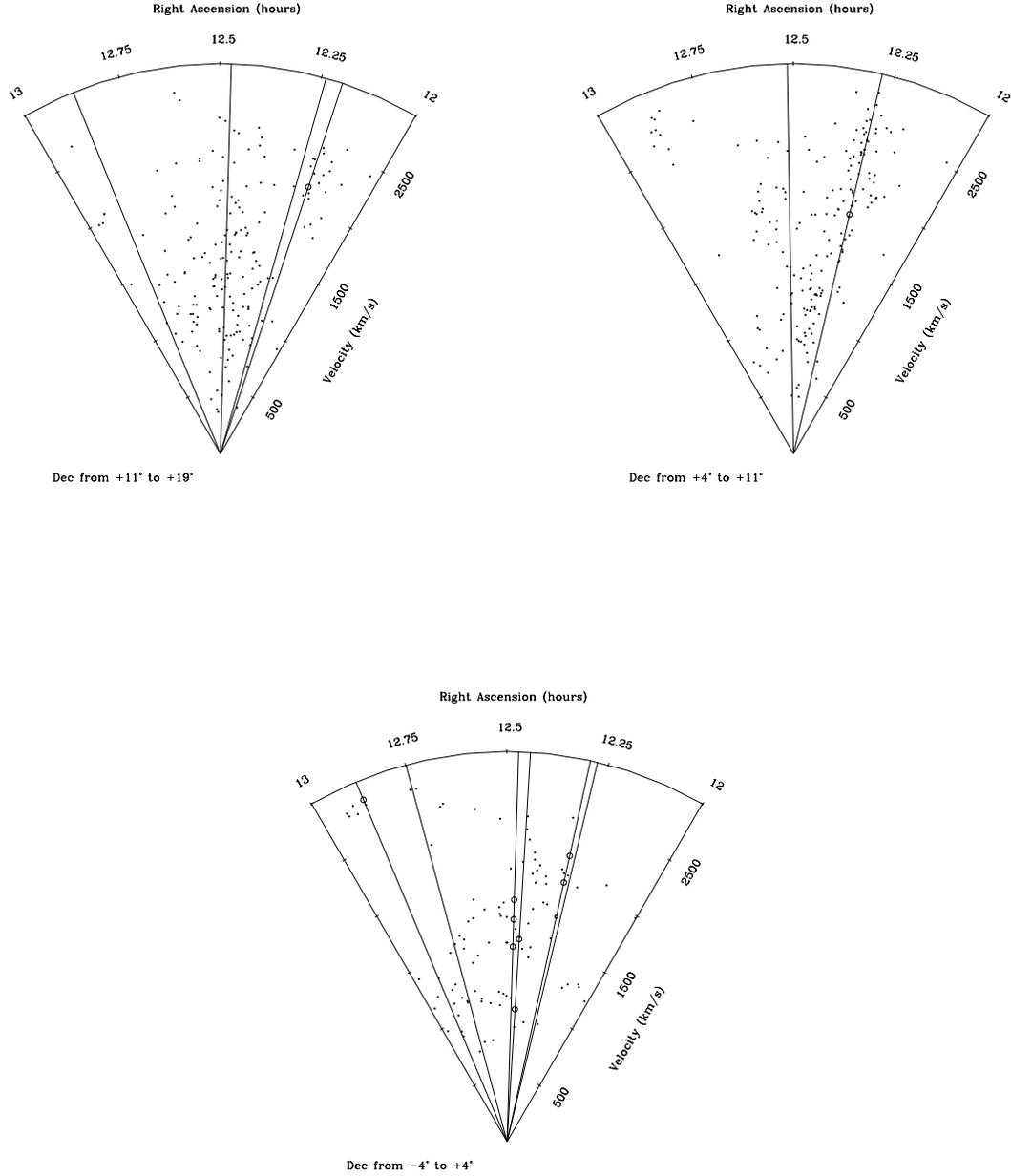}
\epsscale{1.00}
\figcaption{Pieplot distributions of galaxies in Virgo sample
out to $v = 3000$\ \kms.  Quasar lines of sight are plotted as lines
with absorbers indicated as circles.  The larger circles are
$4.5\sigma$\ lines, and the smaller circles are $3\sigma$\
lines. Each pieplot collapses $\sim 8$\arcdeg\ in declination, 
spanning ranges
{\bf (a)} 11\arcdeg\ to 19\arcdeg,
{\bf (b)} 4\arcdeg\ to 11\arcdeg, and
{\bf (c)}  $-$4\arcdeg\ to 4\arcdeg.}
\end{figure}

\clearpage
\begin{figure}
\epsscale{0.95}
\plotone{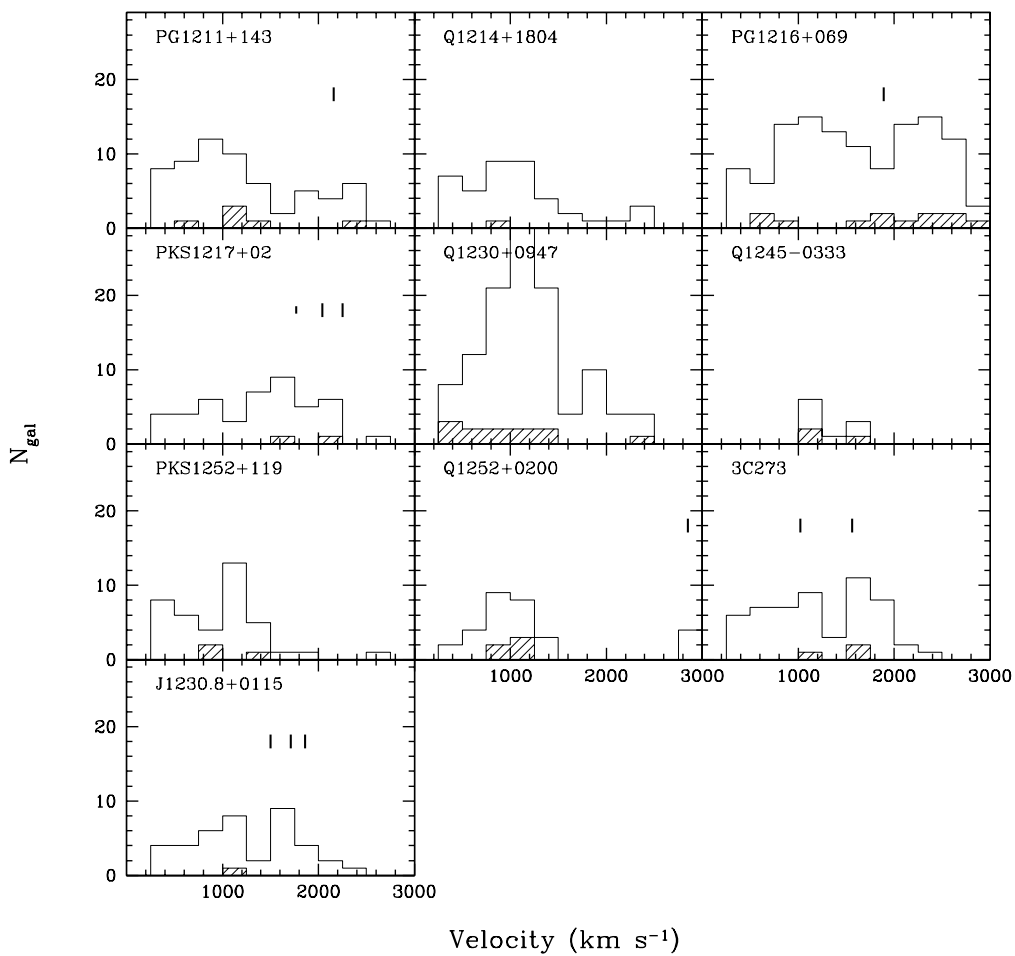}
\epsscale{1.00}
\figcaption{The one-dimensional galaxy distribution within
impact parameters, $\rho$, of 1 Mpc from the individual quasar lines of sight
are shown as the unshaded histogram.  The shaded histogram is the
galaxy distribution for  $\rho \le 250$\h75\ kpc.  The absorbers are
indicated by the vertical bars, with the longer bars for $4.5\sigma$\
lines, and shorter bars for $3\sigma$\ lines.}
\end{figure}

\clearpage
\begin{figure}
\epsscale{0.95}
\plotone{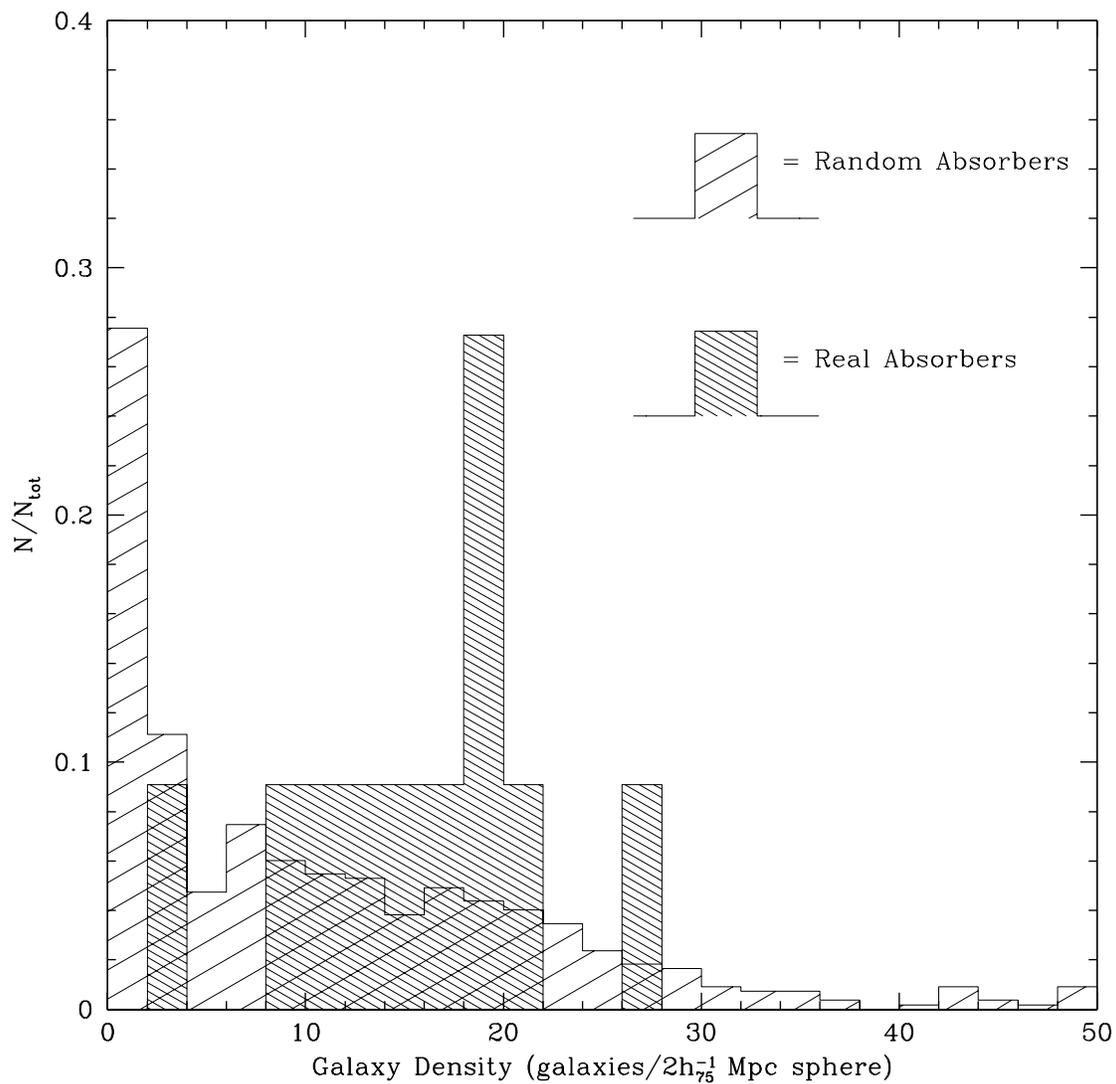}
\epsscale{1.00}
\figcaption{The distributions of galaxy densities around the
real absorber positions (darker histogram), compared to the galaxy
densities around artificial absorbers (lighter histogram).  Galaxy
densities are calculated in 2\h75\ Mpc spheres, assuming pure Hubble
flow.  Each histogram was individually normalized to the total number
of absorbers, and the random absorbers are presented for the sum of 50
trials.} 
\end{figure}

\clearpage
\begin{figure}
\epsscale{0.95}
\plotone{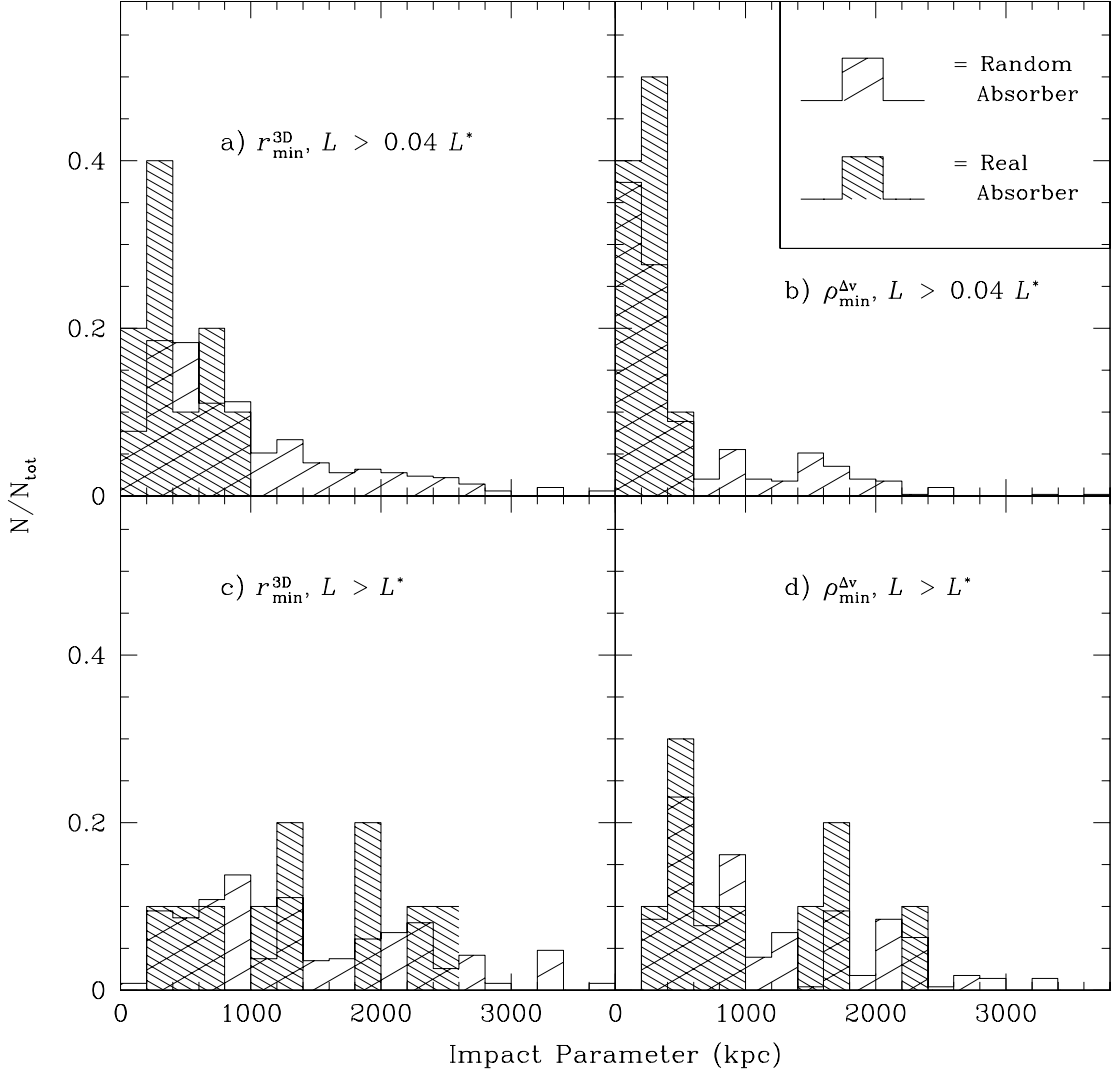}
\epsscale{1.00}
\figcaption{The distributions of impact parameters of the
galaxy counterparts for the different galaxy-absorber pairing methods
are plotted, with the darkly shaded histogram denoting the real
absorbers and the lighter histogram the random absorbers.  The
left-most panels, {\bf (a)} \& {\bf (c)}, are the pairings for 
the $r_{min}^{3D}$\ method, and the right-most panels, {\bf (b)} 
\& {\bf (d)}, are for the $\rho_{min}^{\Delta v}$ method.  
The upper panel absorbers are matched to $L \ge 0.04L^*$ galaxies, and the lower to $L \ge L^*$ galaxies.
The number of pairs is normalized to total number for each test, and
the random absorbers are presented for the sum of 50 trials.}
\end{figure}

\clearpage
\begin{figure}
\plotone{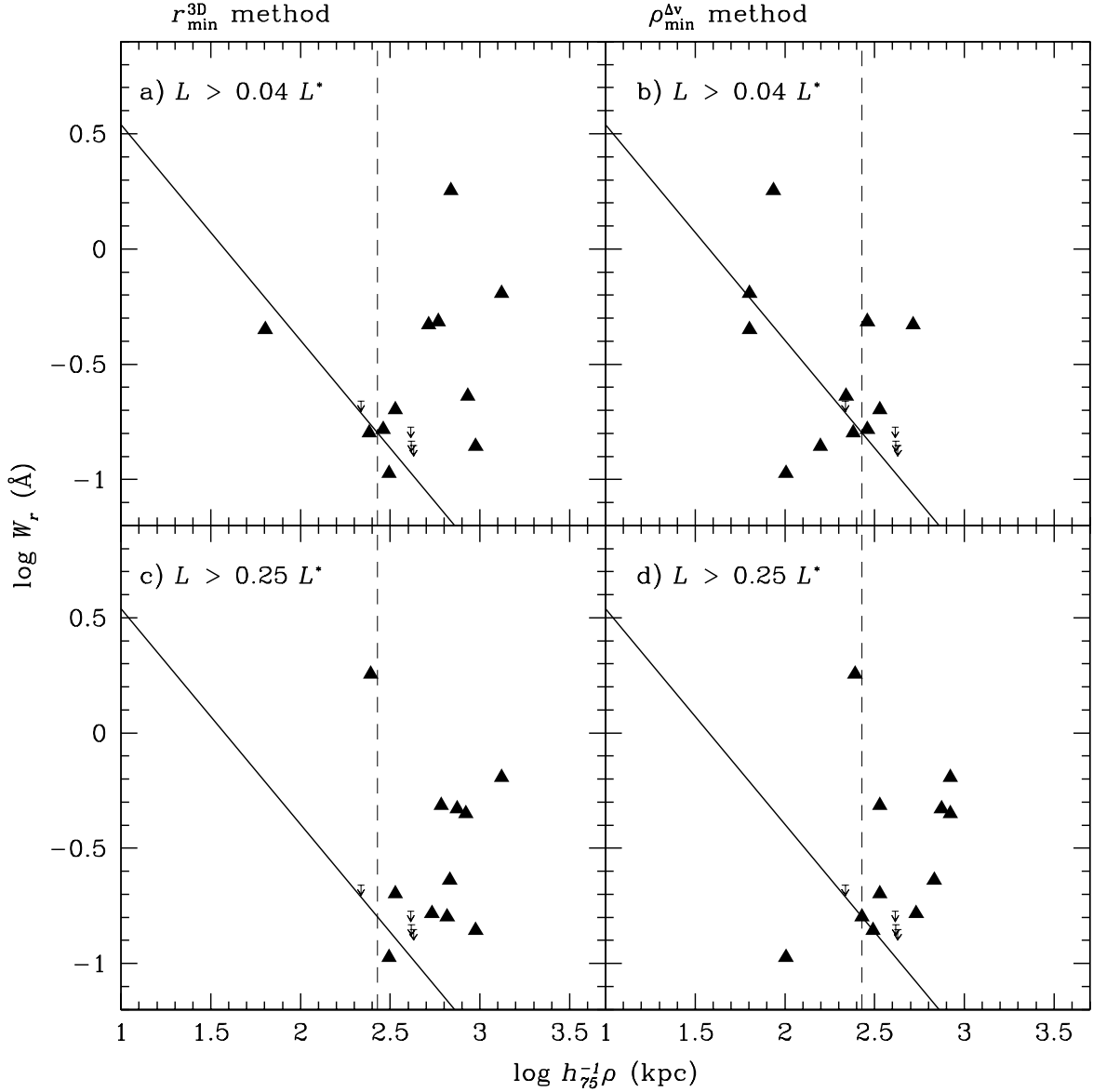}
\figcaption{The rest equivalent width ($W_r$) vs. impact
parameter ($\rho$) distribution is plotted for the two pairing
methods, with the upper panels corresponding to the $L \ge 0.04L^*$
galaxy counterparts, and the lower panels to the $L \ge 0.25L^*$
galaxy counterparts.  The solid line is the anticorrelation relation
from Chen et al. (1998), and the dotted line demarks that group's $\rho =
270$\h75\ kpc ``physical pair'' limit.  The triangles are the data
from this paper, and the limit signs indicate the $3\sigma$ $W_r$\ 
detection limits
for $L^*$ or brighter galaxies falling near the lines of sight that
have no absorber within $\Delta v = 300$ \kms.}
\end{figure}

\clearpage
\begin{figure}
\epsscale{0.83}
\plotone{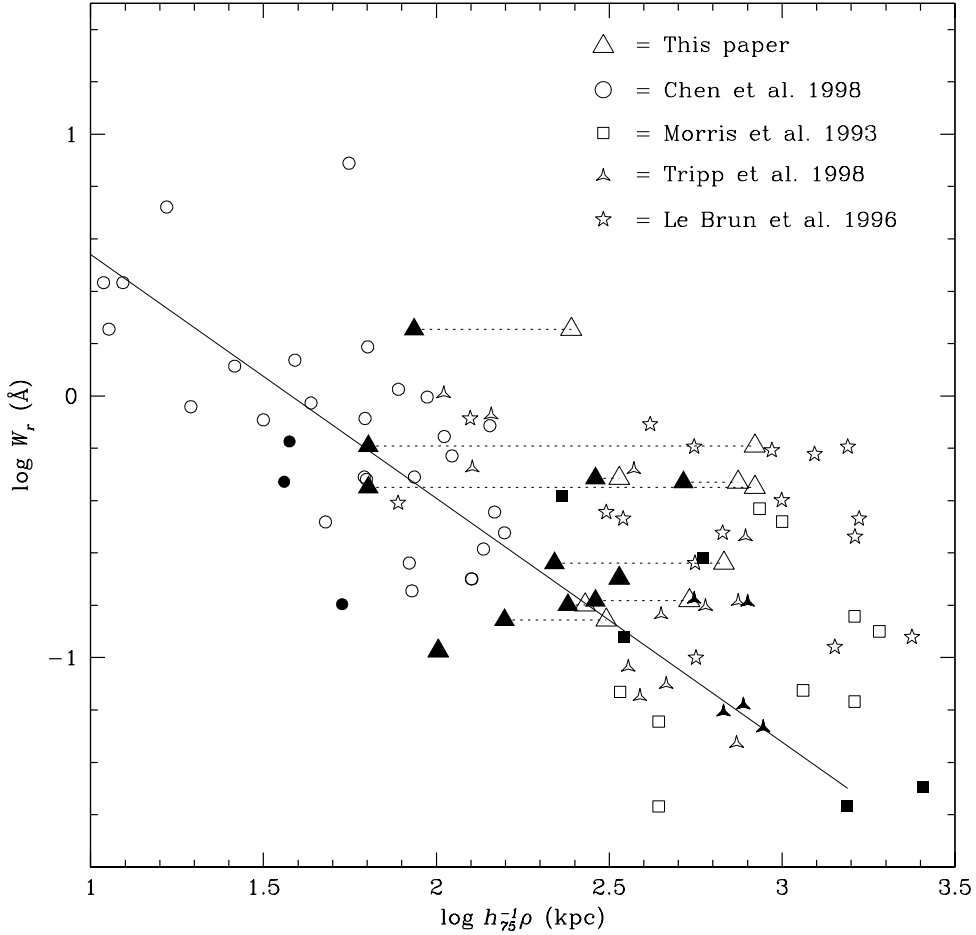}
\epsscale{1.00}
\figcaption{The $W_r$\ vs. $\rho$\ data from this paper are plotted with
data from the literature. Again, the solid line indicates the
Chen et al. (1998) best-fit and the large triangles are the data from the
$\rho_{min}^{\Delta v}$ method from this
paper.  The open triangles are the galaxy-absorber pairs when
matching only to $L \ge 0.25L^*$\  galaxies, the filled triangles are
the pairs when matching to $L \ge 0.04L^*$\  galaxies, and the dotted line
connects the galaxy data points for the same absorber. The other data
included are from Chen et al. (1998) [circles], 
Morris et al. (1993) [squares], 
Tripp et al. (1998) [3-pointed stars], 
and Le Brun et al. (1996)
[5-pointed stars], with filled symbols indicating galaxies with $L <
0.25L^*$\ and open symbols galaxies with $L \ge 0.25L^*$.}
\end{figure}

\clearpage
\begin{figure}
\epsscale{0.70}
\plotone{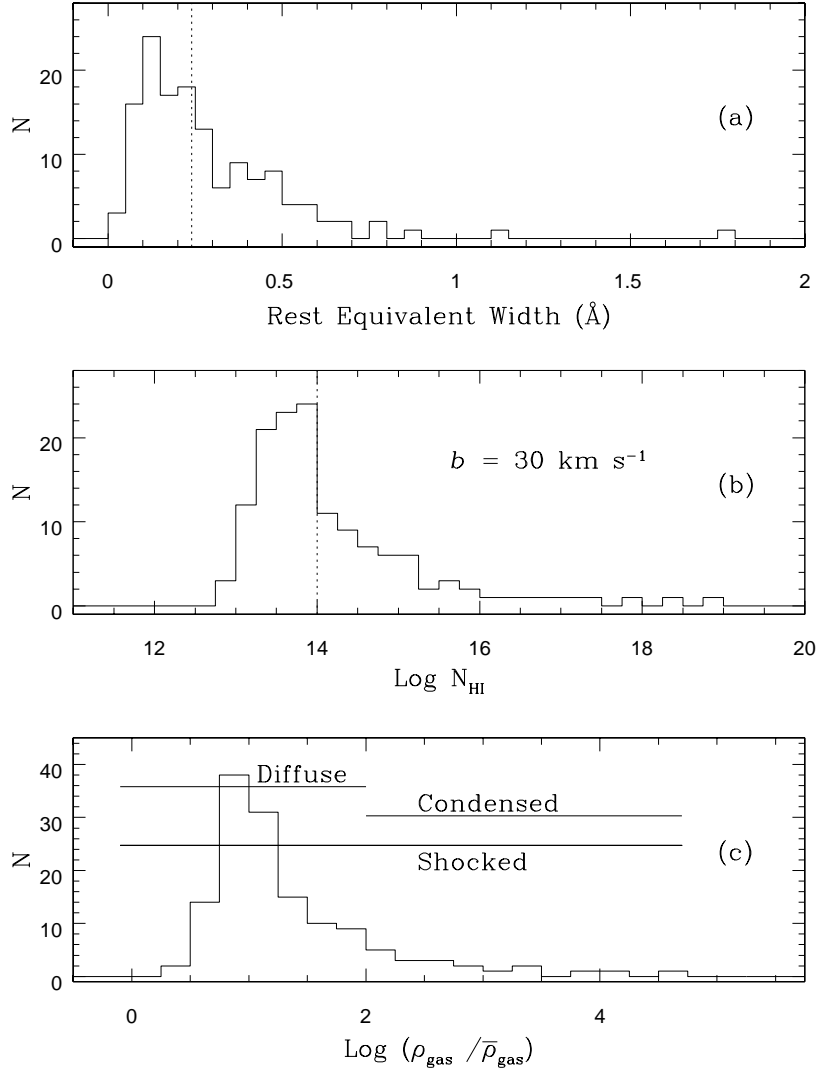}
\epsscale{1.00}
\figcaption{ {\bf (a)}  The distribution of rest equivalent widths for the 
139 \Lya\ absorbers.  The dashed line indicates the completeness limit used
by the Quasar Absorption Line Key Project (Weymann et al. 1998).  
{\bf (b)}  The
distribution of column densities obtained from the equivalent widths assuming
a Doppler parameter of 30 \kms\ and unresolved lines.
The dashed line indicates the column density below which the metallicity
of the absorbers falls sharply.  {\bf (c)}  The 
distribution of gas overdensities ($\rho_{gas} / \bar{\rho}_{gas}$) estimated
from the column density using the relation from Fig. 10 of Dav\'{e} et al. 
(1998).  The bars show the approximate dynamical state of the gas.}
\end{figure}

\clearpage
\begin{figure}
\epsscale{0.85}
\plotone{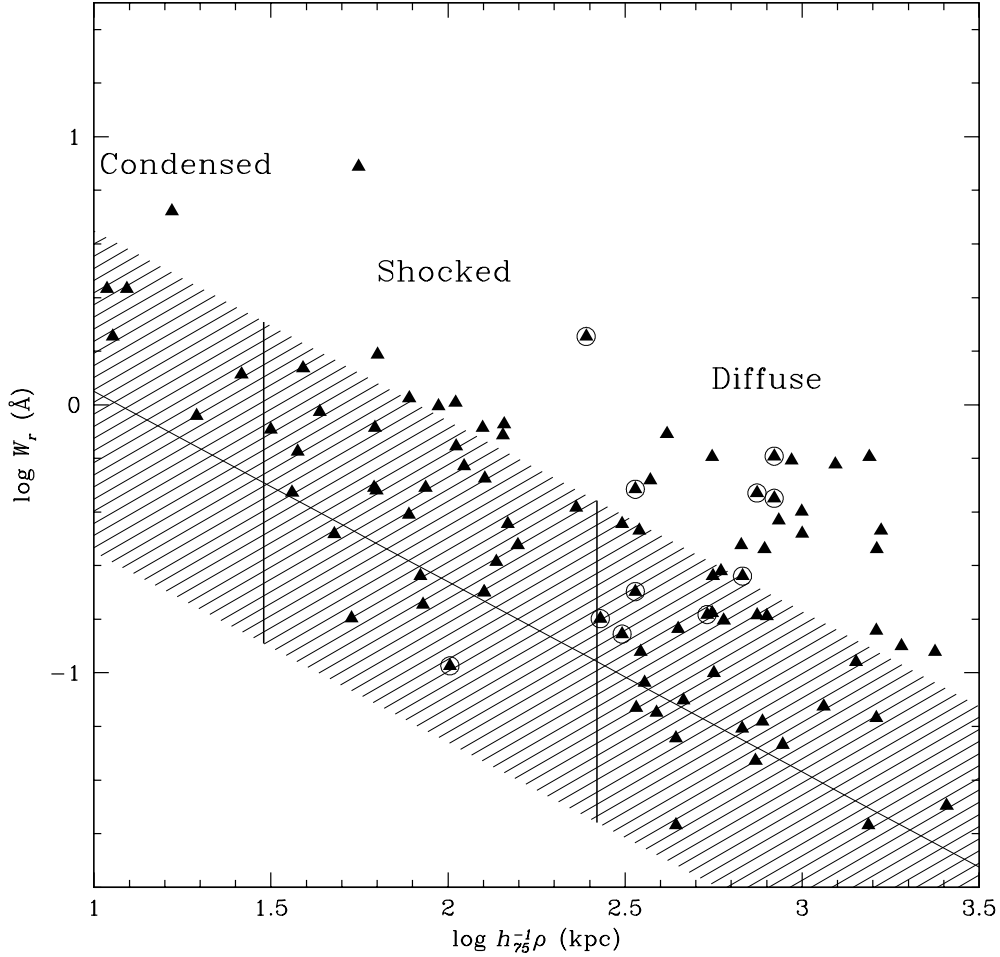}
\epsscale{1.00}
\figcaption{Same as Figure 13, with the sum of all the data
from the literature plus the $L \ge 0.25L^*$ pairs from this
paper indicated uniformly as triangles.  The data from this paper is
also circled. 
The shaded area defines the approximate region that the
simulations of Dav\'{e} et al. (1998) would populate on this diagram for
the galaxies they associate with low column density \Lya\ absorbers in a $z =
0, ~\Lambda$-CDM universe.  The line is their best fit, and 
the two vertical lines roughly denote the 
impact parameters at which the predominant phase of the absorbing gas
changes from cold, condensed gas (smallest $\rho$), to shock heated
gas (intermediate $\rho$), to diffuse gas (high $\rho$).}
\end{figure}

\begin{figure}
\plotone{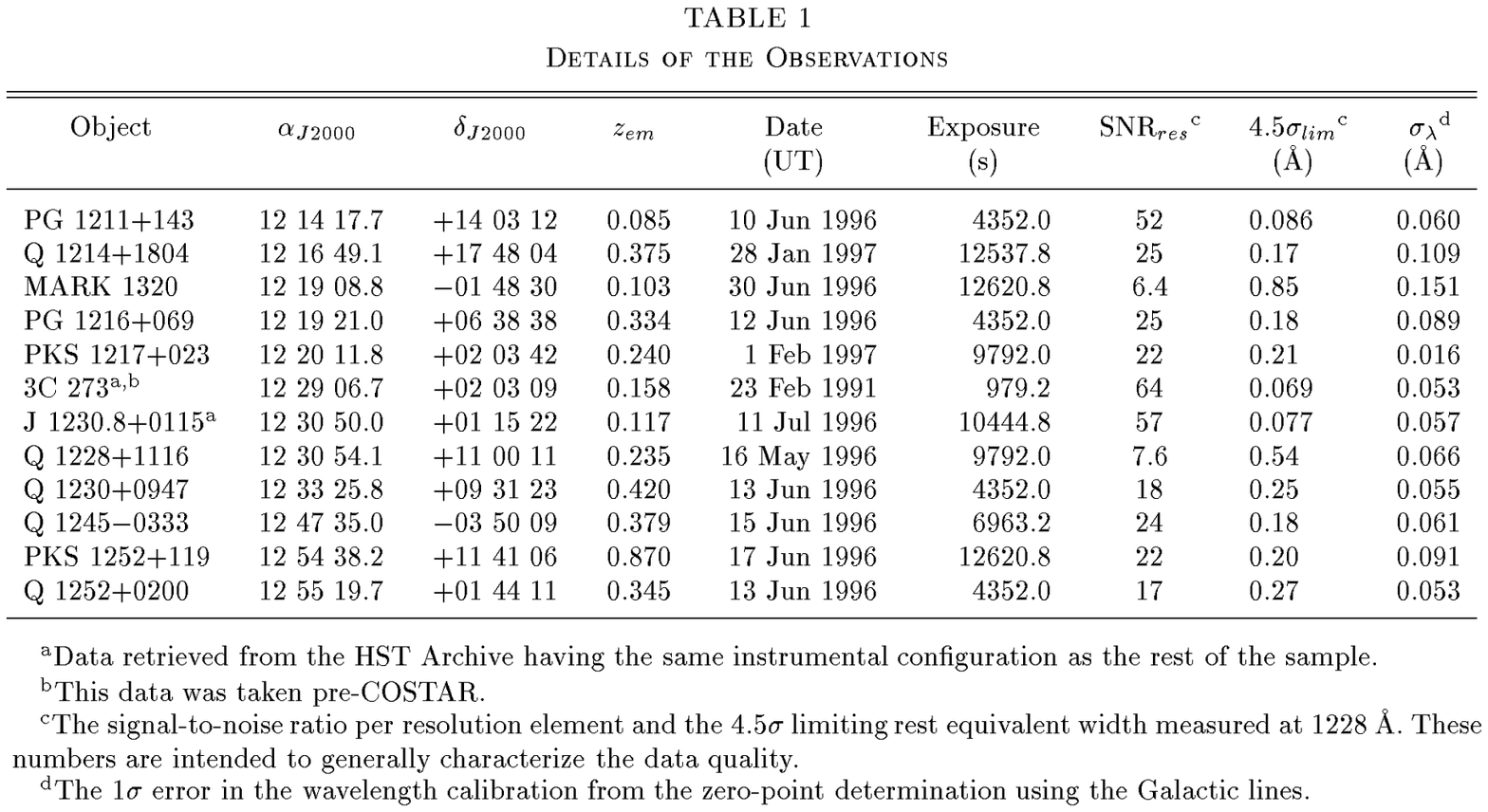}
\end{figure}

\begin{figure}
\plotone{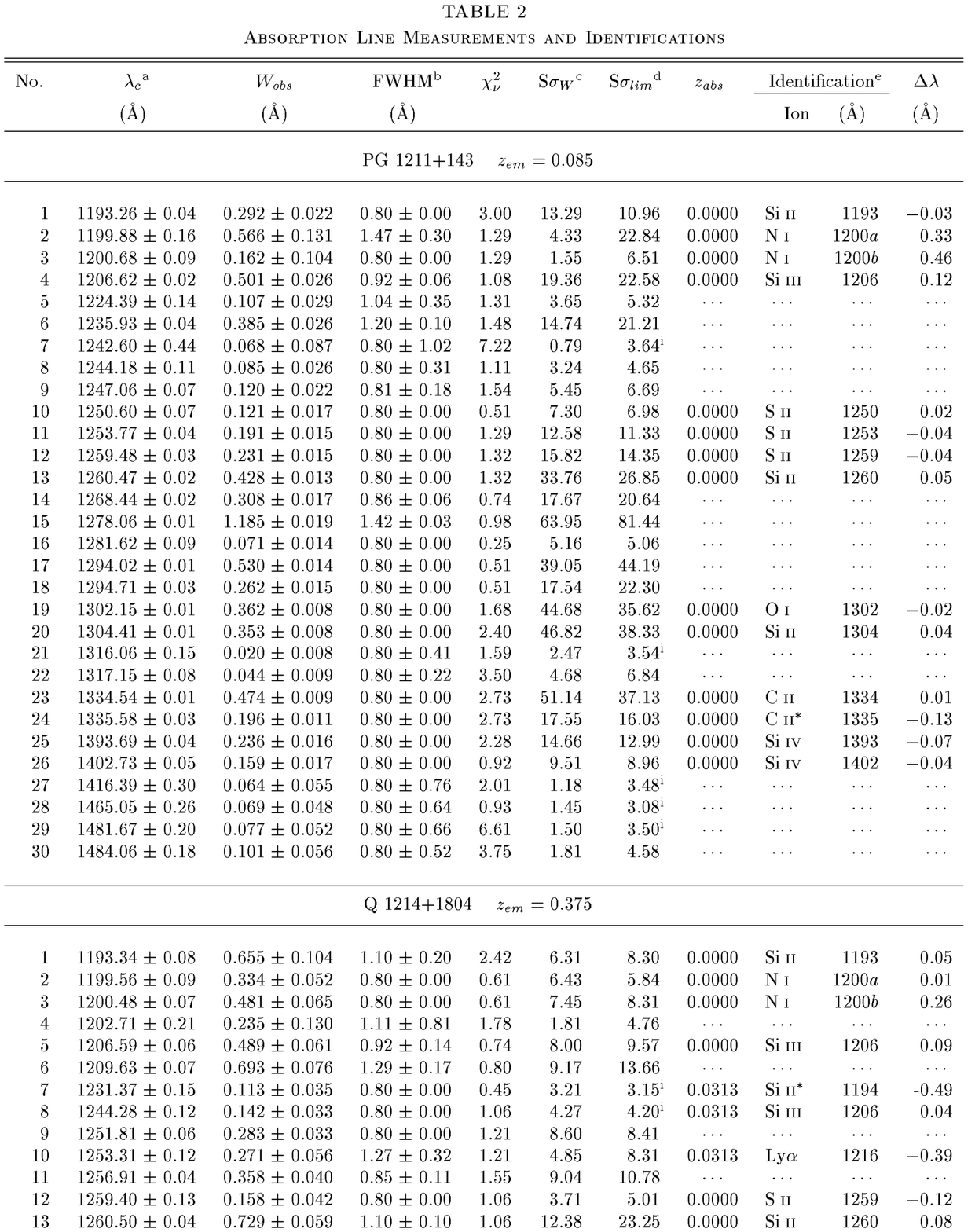}
\end{figure}

\begin{figure}
\plotone{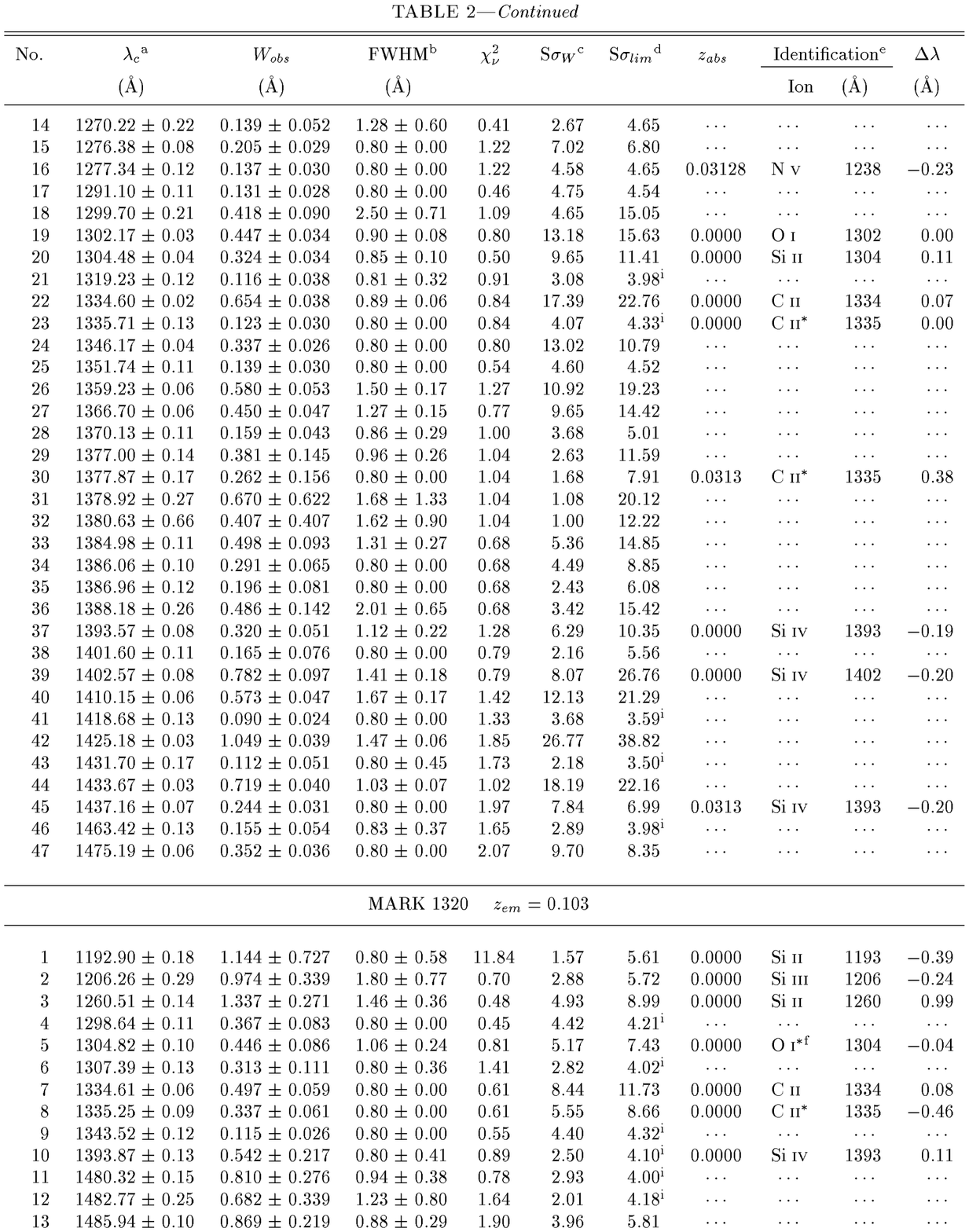}
\end{figure}

\begin{figure}
\plotone{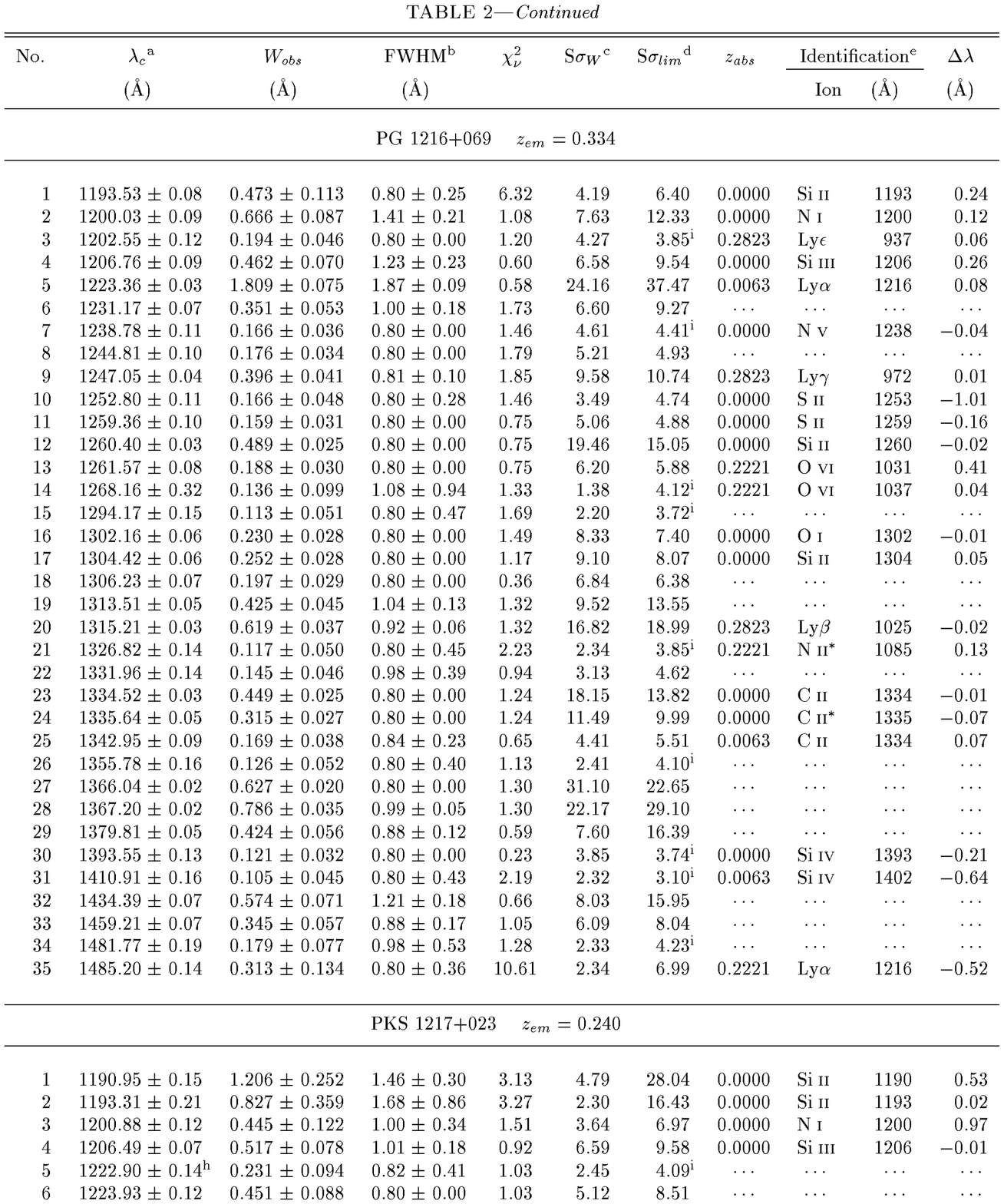}
\end{figure}

\begin{figure}
\plotone{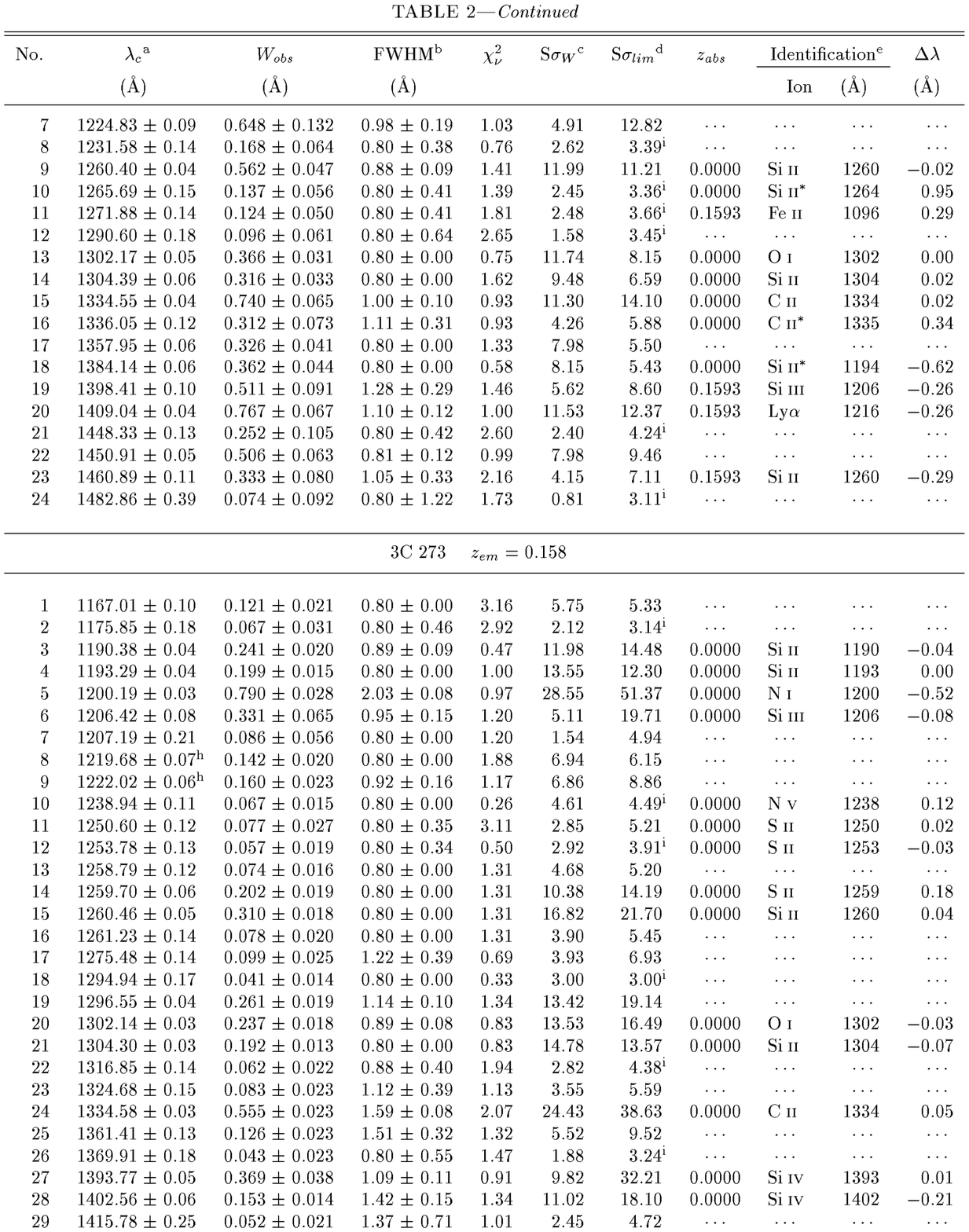}
\end{figure}

\begin{figure}
\plotone{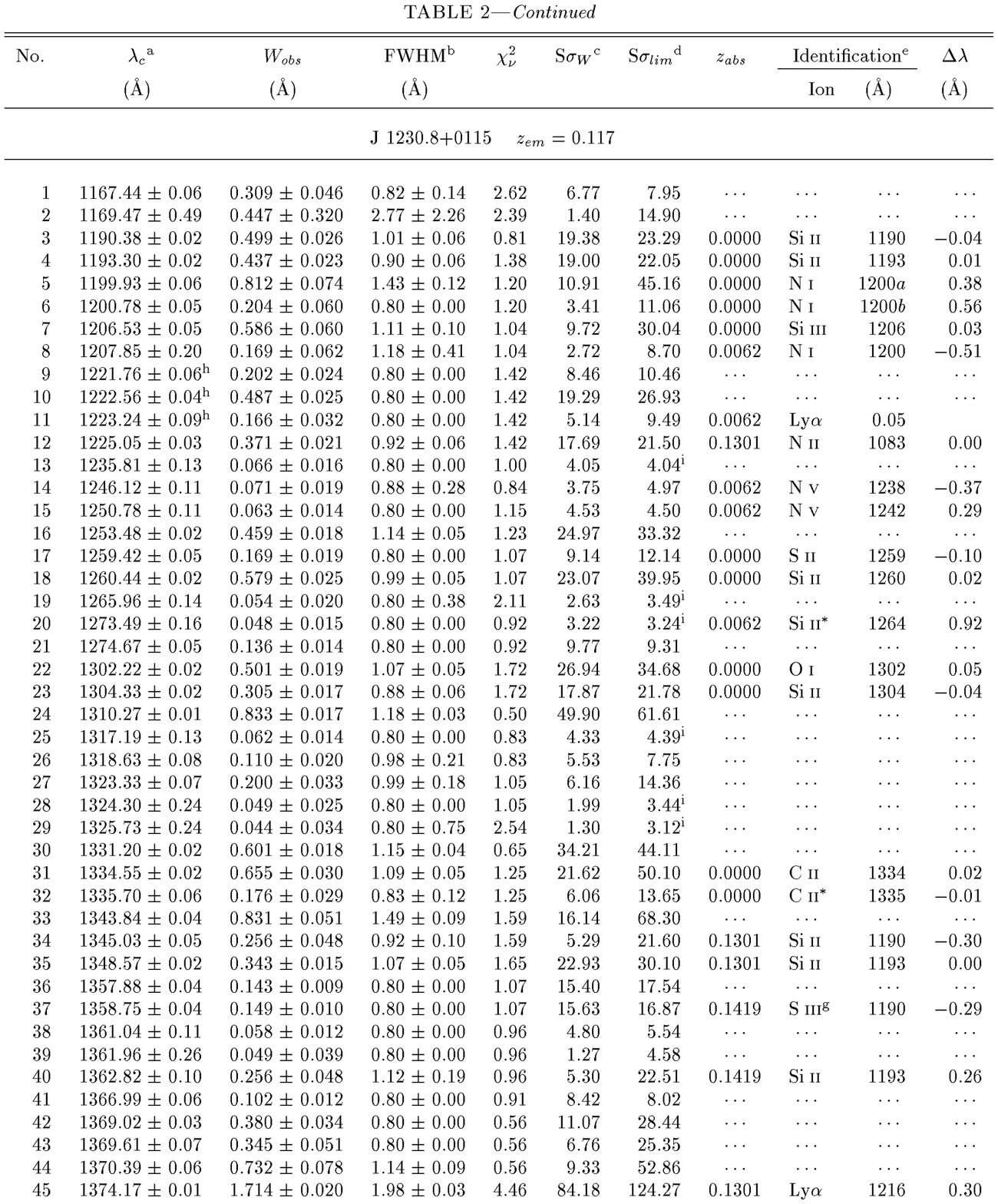}
\end{figure}

\begin{figure}
\plotone{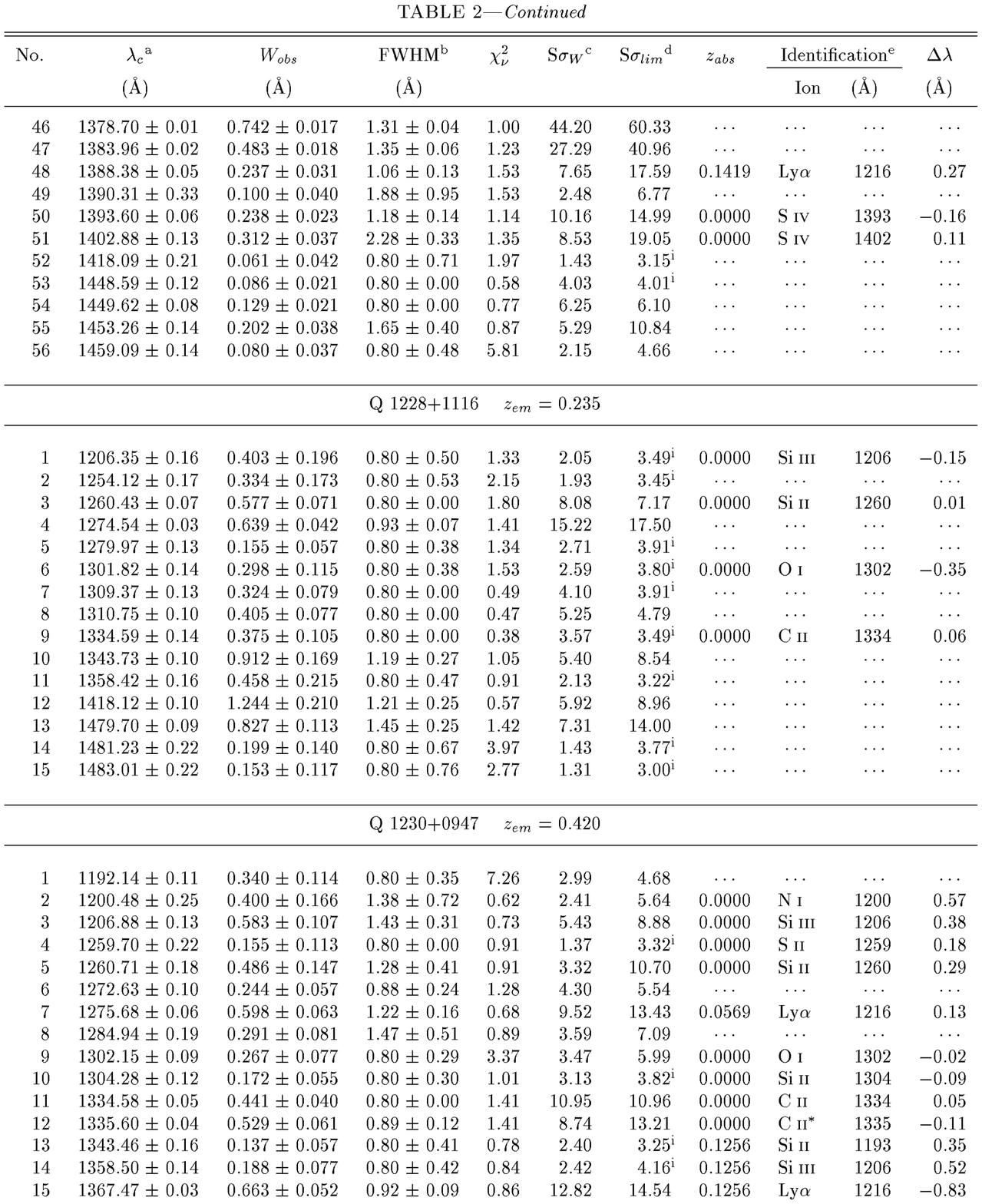}
\end{figure}

\begin{figure}
\plotone{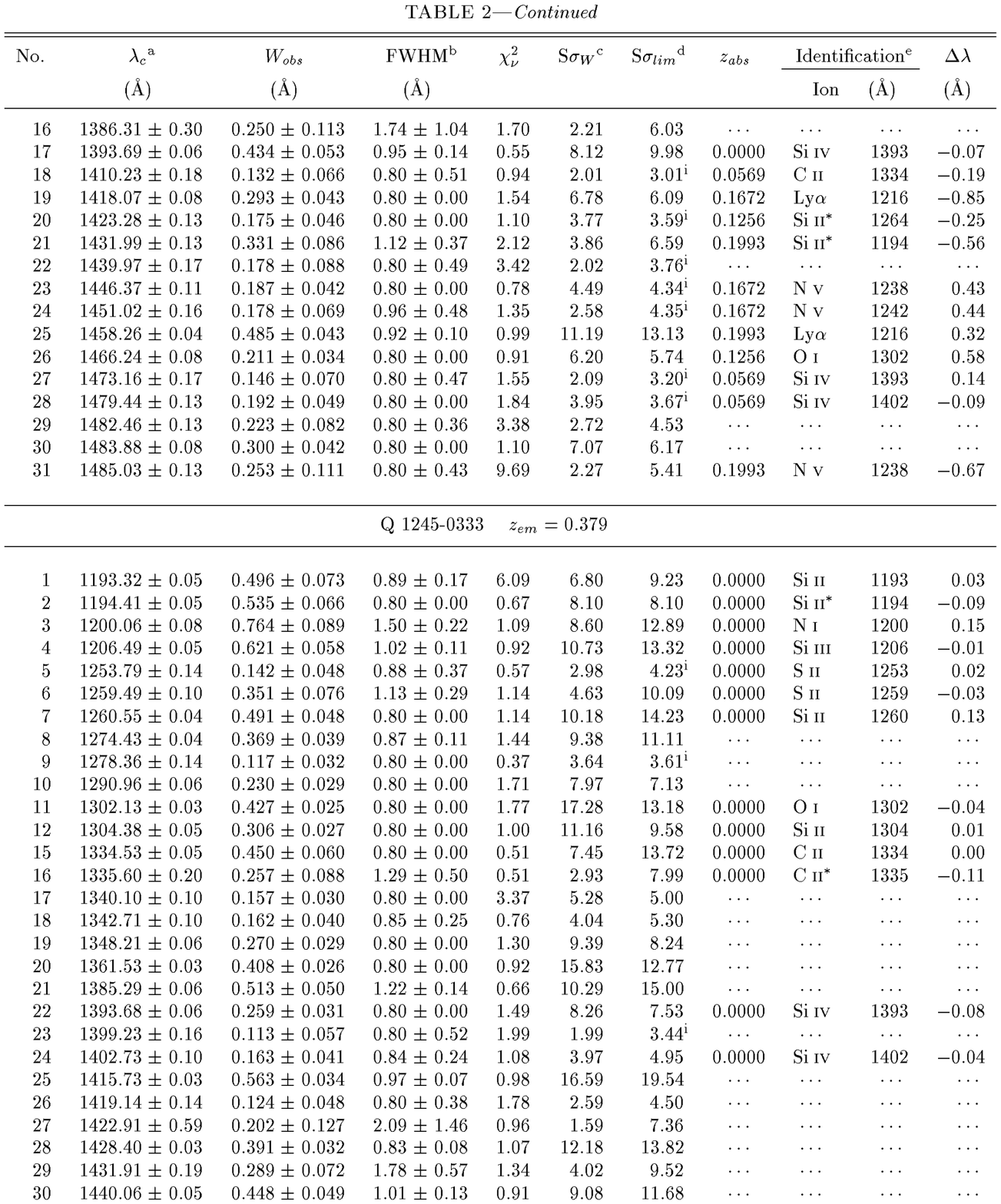}
\end{figure}

\begin{figure}
\plotone{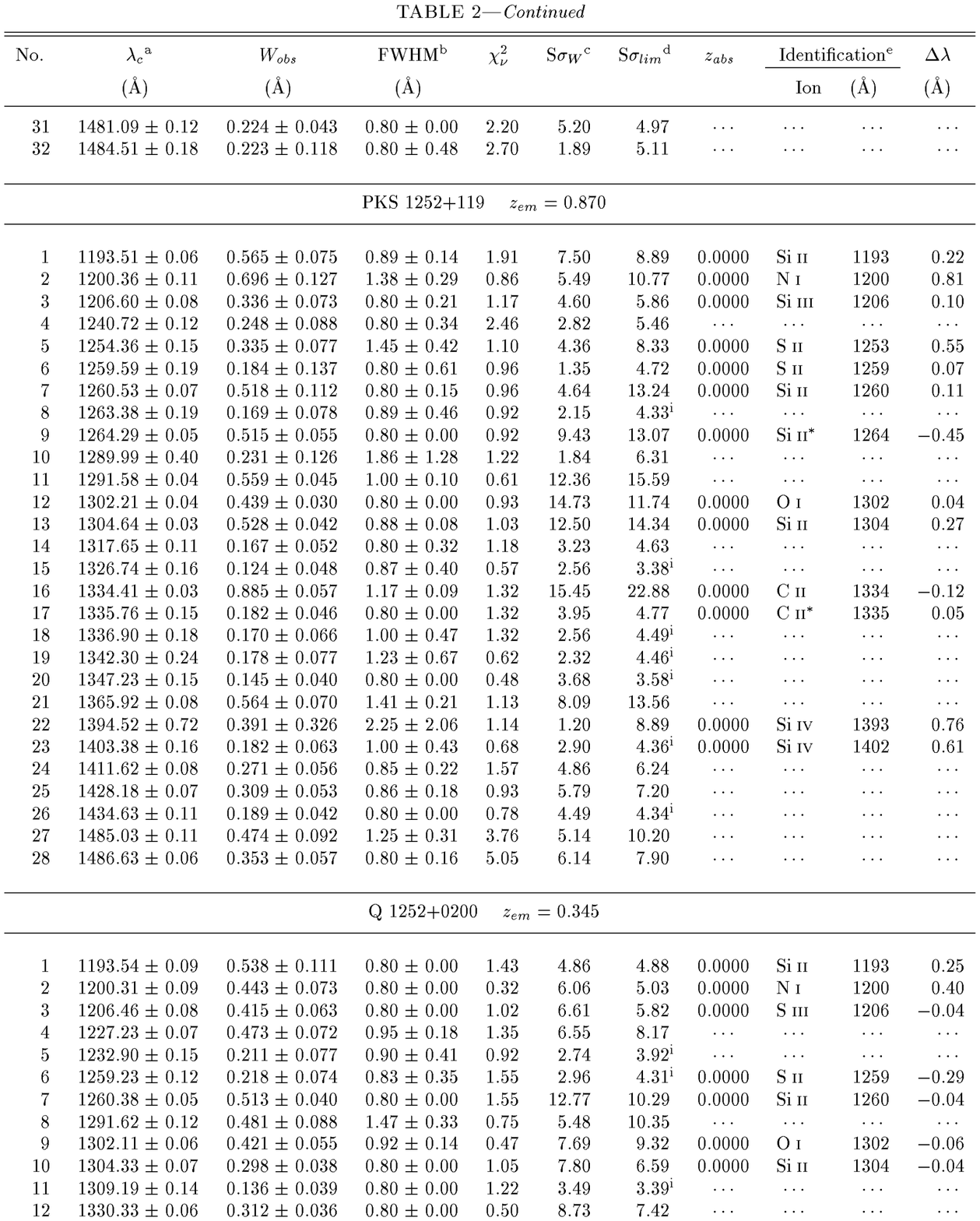}
\end{figure}

\begin{figure}
\plotone{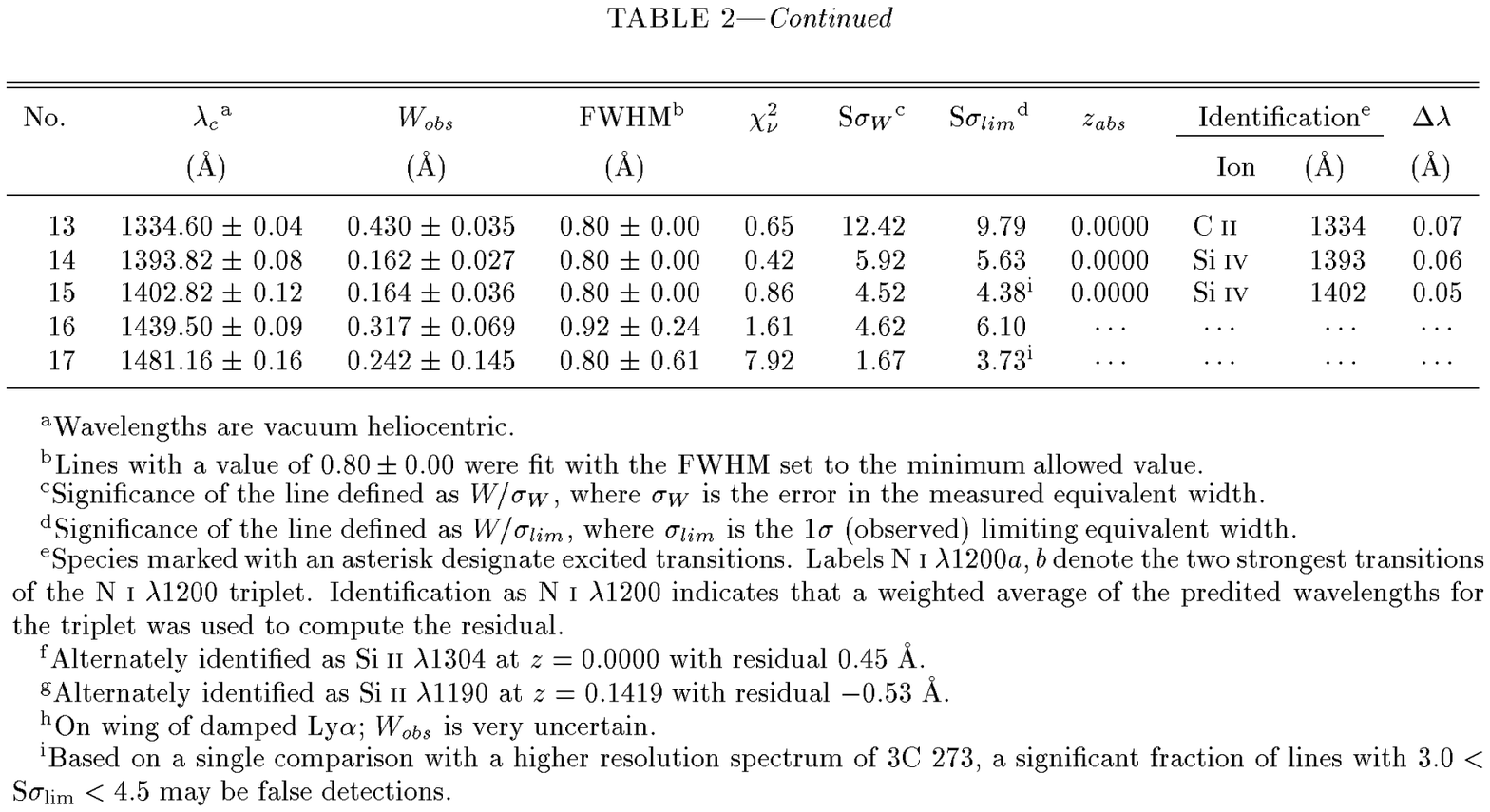}
\end{figure}

\begin{figure}
\plotone{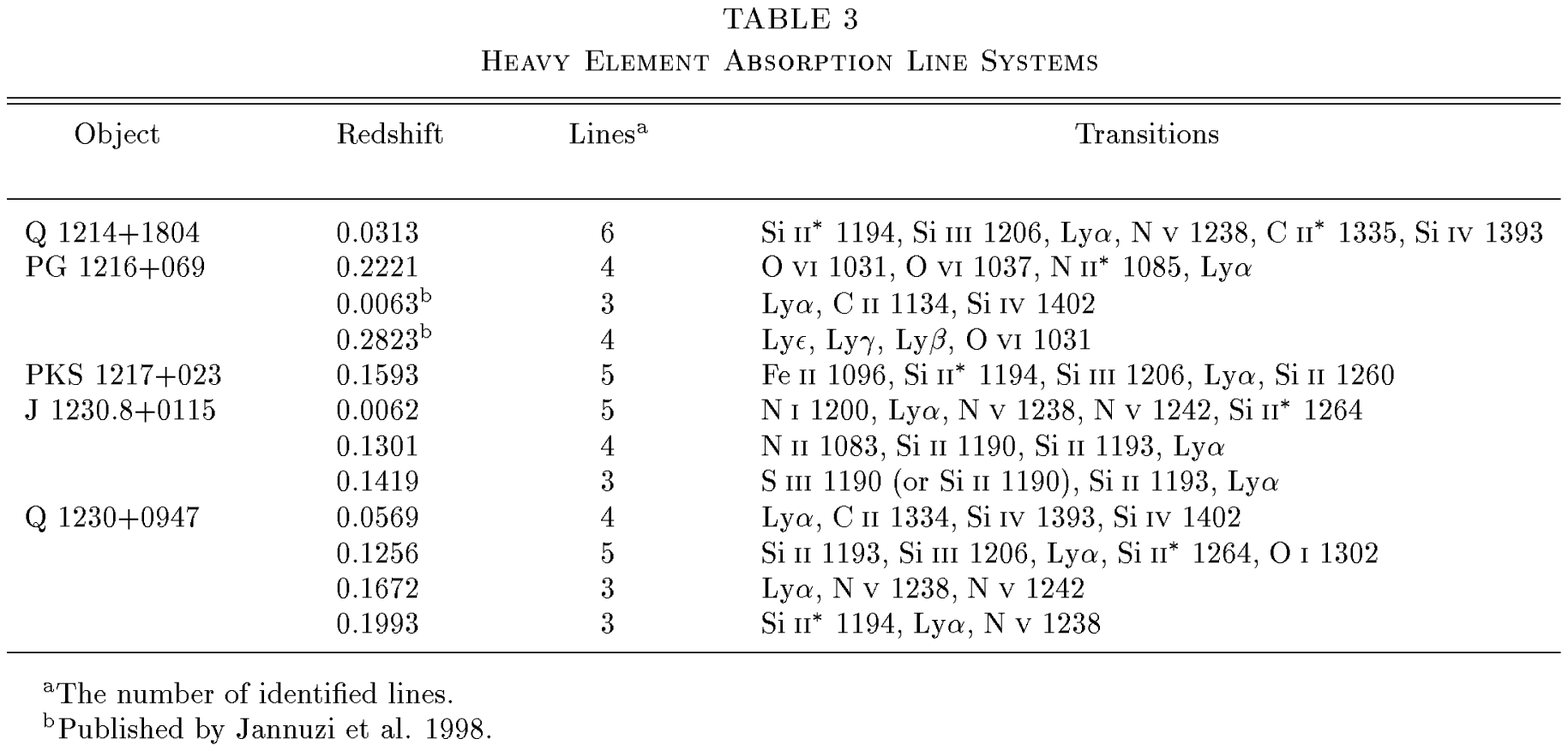}
\end{figure}

\begin{figure}
\plotone{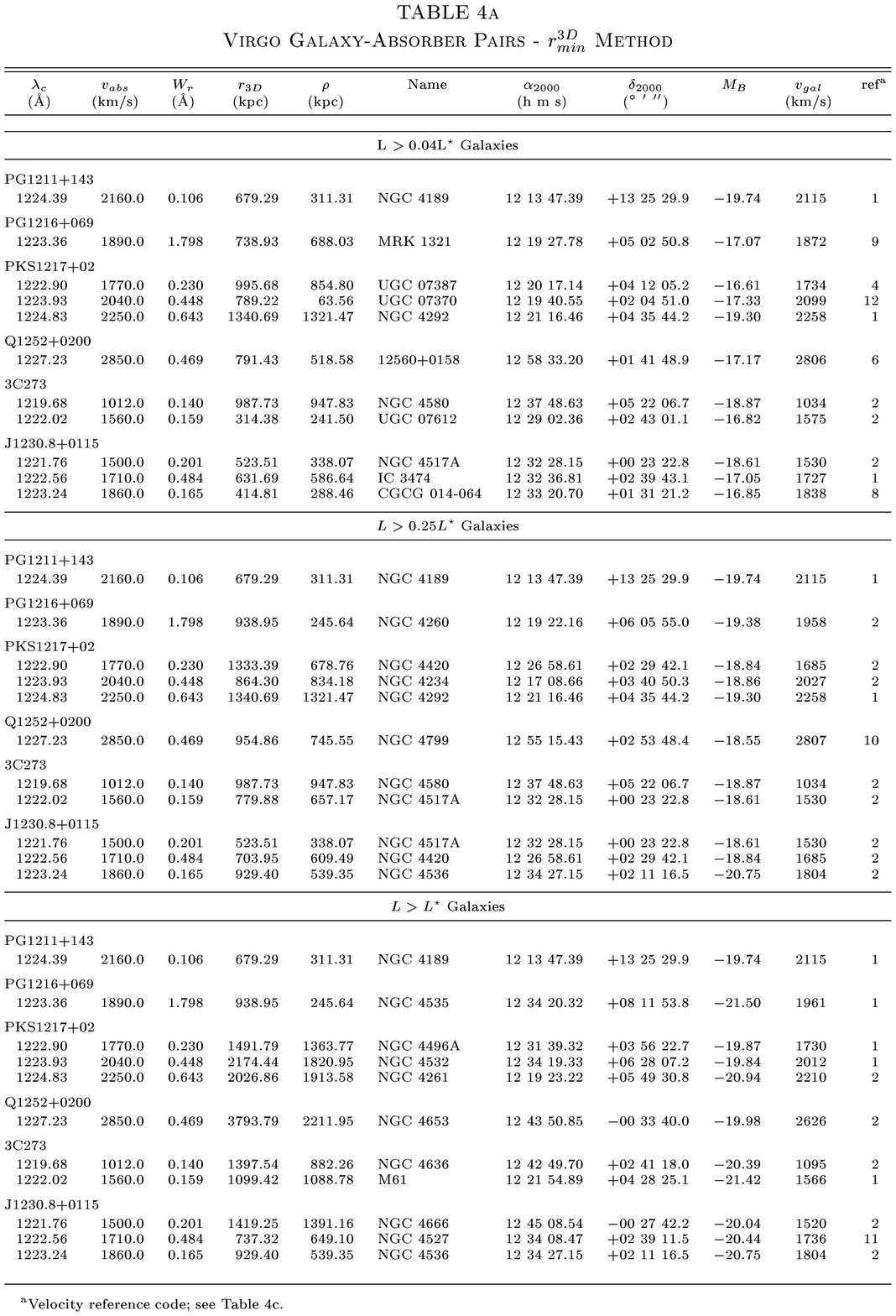}
\end{figure}

\begin{figure}
\plotone{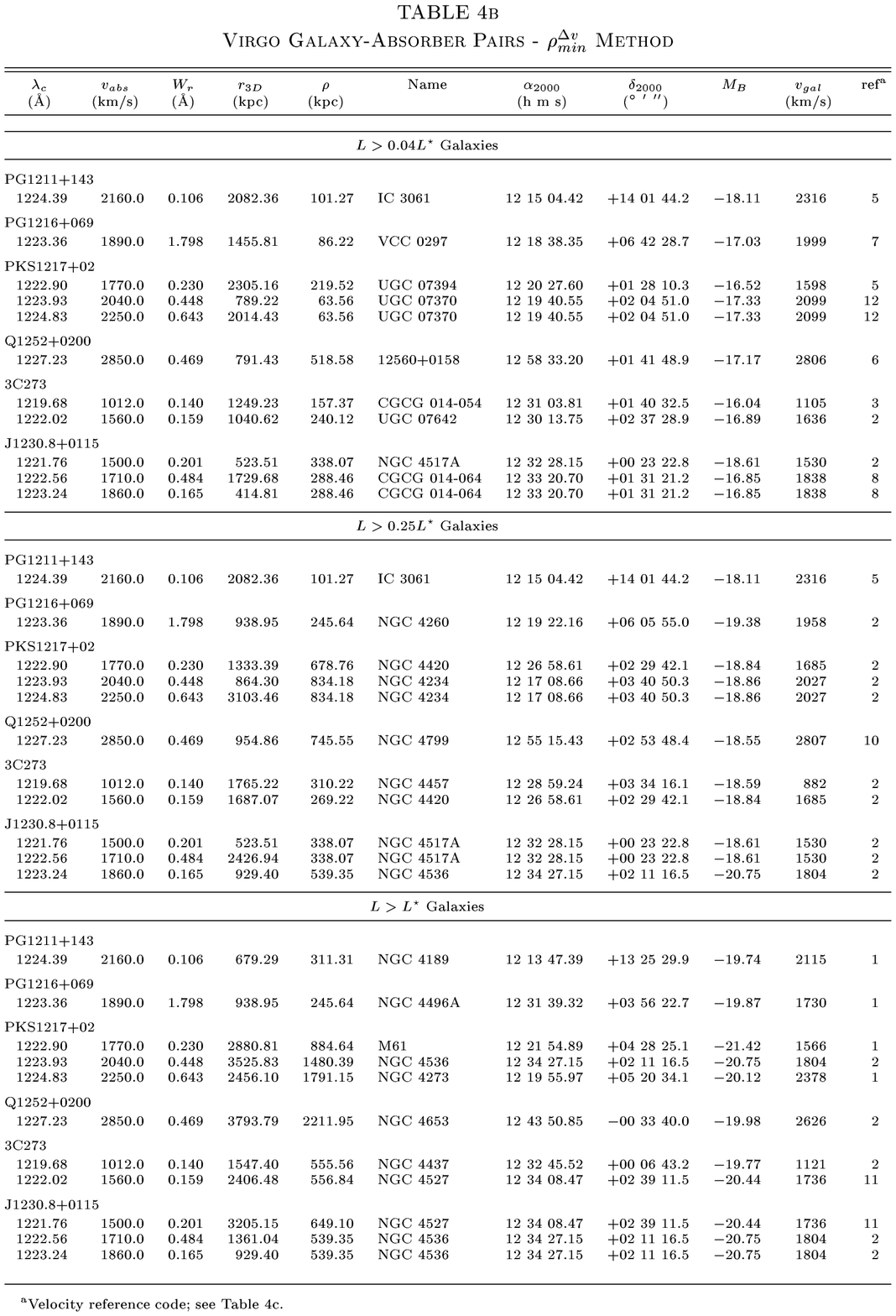}
\end{figure}

\begin{figure}
\plotone{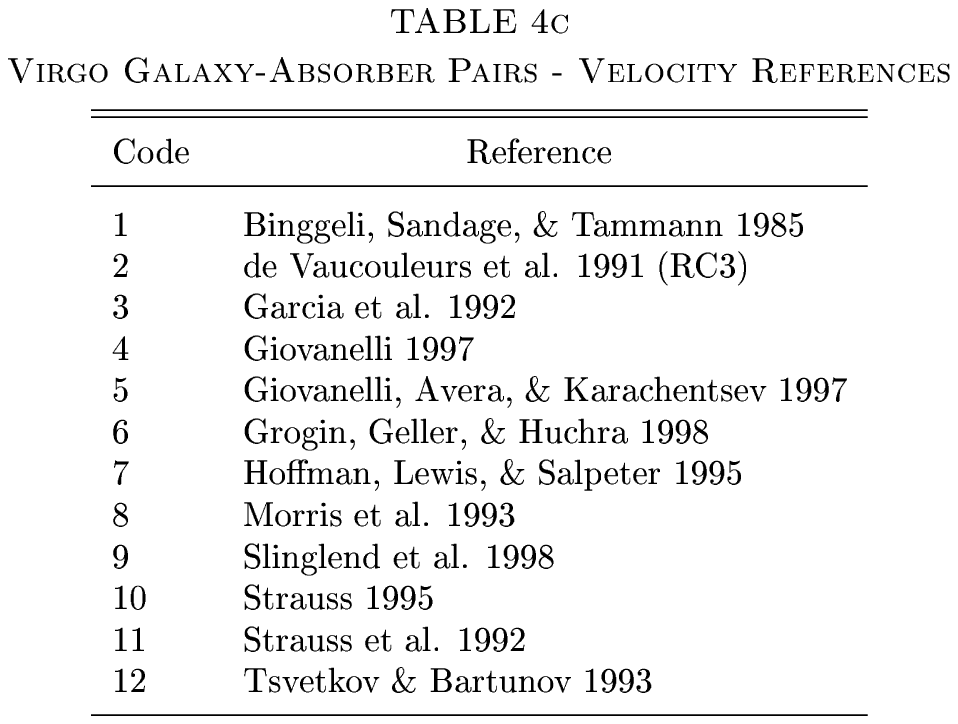}
\end{figure}

\end{document}